\documentclass[pdftexm,12pt]{article}  
\usepackage[dvips]{graphicx}         
\usepackage{pslatex}

\begin{document}

\vskip 0.75in
{\bf \qquad }
\begin{center}{\bf \large On the universality of the distribution 
of the eigenvalues of Wigner random matrices in the bulk of the spectrum.}
\vskip 0.25in{\bf Anastasia A. Ruzmaikina}
\vskip 0.15in
(submitted: October 3, 2005, 1st revised version March 19, 2007)
\vskip 0.01in
(2nd revised version, {\it to be submitted})
\vskip 0.25in
{\bf \large Abstract.}
\end{center} \vskip 0.1in
In this paper we consider real Wigner random matrices -- symmetric $n$ by $n$ 
random matrices whose entries are independent identically distributed real 
random variables. We prove that the probability distribution of one or several 
eigenvalues close to the center of the spectrum does not depend on 
the probability distribution of the entries of the matrix and is the same as 
for the Gaussian Orthogonal Ensemble. We make only mild smoothness assumptions 
on the probability distribution of the entries and assume that the probability 
distribution of the entries decays polynomially with sufficiently large power 
or faster than polynomially.

\section{Introduction.}
Random matrices are an interesting subject in itself and also arise in a number 
of problems of interest in mathematics, physics and mathematical physics.

Random matrices were introduced  by Wigner to model the energy levels in 
large nuclei. In the modeling of the energy levels of large nuclei, one has 
to deal with a very complicated many body problem and probability provides 
effective tools to describe it.  A Hamiltonian of a very large nucleus can be 
represented as a random matrix, in which all the entries are independent 
random variables. 

In addition to nuclear physics, random matrices also found applications 
in the solid state physics to model electronic energy levels of small metallic 
particles, to model ultrasonic resonance frequencies of structural materials 
and to model the energy levels of a spin glass assuming that the entries 
of a random matrix are the strengths of interactions between the different 
spins.

Numerical simulations show that energy levels of certain quantum systems behave 
like the eigenvalues of a random matrix taken from a Gaussian Orthogonal 
Ensemble. 

In number theory, in connection with the Riemann hypothesis, it is conjectured 
and checked analytically in certain regions as well as numerically that 
the spacings between the zeroes of Riemann zeta function on the line ${\rm Re} 
z = \frac{1}{2}$ have the same distribution as the spacings between the nearest 
eigenvalues of a random matrix, and that other local fluctuation properties 
of the eigenvalues of a random matrix and of a Riemann zeta function are also 
identical.
  
These wide applications stimulate the search for general patterns 
in the behaviour of eigenvalues of random matrices. 
In this paper we show that for a general class of systems which can be 
described by random matrices with independent identically distributed random 
entries, the behaviour of eigenvalues is universal, that is it does not depend 
on the distribution of the entries of the matrix for a wide range 
of distributions.  

\subsection{Definitions, and method of the proof developed in this paper.} 

Consider a Wigner ensemble of $n$-dimensional
real random symmetric matrices $A_n = ||a_{ij}||$ with $a_{ij} = a_{ji} =
\frac{\xi_{ij}}{\sqrt{n}}$, for $1 \leq i,j \leq n$. Let $\xi_{ij}$
be independent identically distributed  random
variables with a symmetric distribution such that $E \xi_{ij}^2 =
\frac{1}{4}$.
%

Let $\lambda_1, \ldots \lambda_n$ be the eigenvalues of $A_n$. The limit in
probability of the empirical distribution function of eigenvalues $N_n(\lambda) 
=\frac{1}{n} \# \{k: \lambda_k \leq \lambda\}$ is given by a Wigner semi circle 
theorem (see papers by Wigner \cite{W1}, \cite{W2}, Marchenko \cite{M}, 
Pastur \cite{P1}, \cite{P2}, L.Arnold \cite{A1}, \cite{A2}, Wachter \cite{Wa}, 
Girko \cite{G} and others): 
\begin{equation}\label{wigner}
    \lim_{n \to \infty} N_n(\lambda) = \int_{-\infty}^{\lambda} I_{[-1,1]}
    \frac{2}{\pi} \sqrt{1 - t^2} d t,
\end{equation}
where $I_{[-1,1]} (x) = 1$ for $-1 \leq x \leq 1$ and $=0$ otherwise.

The outstanding problem of probability theory is to prove the universality 
in the bulk for Wigner random matrices, i.e. that in the limit of large $n$ 
the probability distributions of any $k$ eigenvalues of a Wigner random matrix 
$A_n$ converge to a limit which is independent of the distributions of entries 
of $A_n$, and is the same a for the matrices in Gaussian Orthogonal Ensemble 
(or Gaussian Unitary Ensemble).

The question of probability distribution of the individual eigenvalues
has been studied thoroughly for the Gaussian Unitary Ensemble (GUE)
and Gaussian Orthogonal Ensemble (GOE), where GUE is the ensemble of
random hermitian matrices such that Re $a_{ij}$ = Im $a_{ij}$ = $N (0,
\frac{1}{8n})$ for $i \neq j$ and $a_{ii} =   N (0,
\frac{1}{4n})$ and GOE is the ensemble of random real symmetric
matrices such that  $a_{ij}$ = $N (0,
\frac{1}{4n})$ for $i \neq j$ and $a_{ii} =   N (0,
\frac{1}{2n}).$ The joint distribution of the matrix elements can be
written as 
$$ P(d A_n) = {\rm const}_{n,\beta} e^{- \beta n {\rm Trace}(A_n^2)} d
A_n,
$$ where $\beta =2$ for GUE and $\beta =1$ for GOE.

This equation implies the following formula for the distribution of
the eigenvalues for GUE and GOE:
$$ d P(\lambda_1, \ldots, \lambda_n) = P_{n,\beta}(\lambda_1, \ldots,  
\lambda_n) d\lambda_1 \ldots d\lambda_n,$$
where
$$ P_{n,\beta}(\lambda_1, \ldots,  
\lambda_n) = {\rm const}_{n,\beta} \Pi_{1 \leq i \leq j \leq n}
|\lambda_i - \lambda_j|^{\beta} e^{-\beta n (\lambda_1^2+ \ldots+\lambda_n^2)}.$$

In \cite{TW1}, \cite{TW2}, \cite{F} Tracy-Widom, and  Forrester studied 
the distribution of the
first few largest eigenvalues in GUE and GOE.  They showed that the
distribution of first $k$ rescaled eigenvalues $(\lambda_1^{(n)} -1)\cdot
2n^{2/3}, (\lambda_2^{(n)} -1)\cdot 2n^{2/3}, \ldots, (\lambda_k^{(n)} -1)\cdot
2n^{2/3}$ has a limit as $n \rightarrow \infty$:
$$ F_{\beta, k} (s_1, \ldots, s_k) = \lim_{n \rightarrow \infty}
P_{\beta} (\lambda_i^{(n)} \leq 1 + \frac{s_i}{2 n^{2/3}}, i=1,
\ldots, k).$$
The limiting $k$-dimensional distribution function can be expressed in
terms of the solutions of a completely integrable PDE.
 The distribution of the rescaled maximal eigenvalue $\lambda_1^{(n)}$ is given by
\begin{equation}\label{tracywidom}
F(x) = \lim_{n \to \infty} P(\lambda_1^{(n)} \leq 1 + \frac{x}{2 n^{2/3}}) =
 \exp\left( - \int_{x}^{\infty} (t - x) u^2 (t) dt \right),
\end{equation}
where $u(t)$ is the solution of the Panleve II equation $u''(t) = t u(t) + 2 u(t)^3$.

Sinai and Soshnikov in \cite{SS, SS1} studied the fluctuations around the semi circle law on the
edges of the spectrum: $\lambda = \pm 1$ for the matrices $A_n$ such that the
distributions of random variables $\xi_{ij}$ decay at least as fast as Gaussian
distributions. The eigenvalues are rescaled by the factor $r_n$ such that $r_n n^{2/3}
\rightarrow_{n \rightarrow \infty} \infty$, so that
\[
    \lambda_k = 1 - r_n \theta_k,
\]and the mass
\[
    \mu_n(\theta_k) = \frac{1}{n r_n^{3/2}}
\]
is placed at each point $\theta_k$. The main theorem in \cite{SS} is
\vskip 0.1in
\noindent{\bf Theorem [SiSo]}
\vskip 0.01in
  {\it    As $n \rightarrow \infty$ the measures $\mu_n$ weakly converge in probability on each finite interval to a measure $\mu$ concentrated on the half-line ${{\rm R}^+} $ and absolutely continuous with respect to the Lebesque measure. The density $\frac{d \mu}{d x}$ has the form $\frac{2 \sqrt{2}}{\pi}\sqrt{x}$ for $x\geq 0$ and is $0$ for $x < 0$. If $r_n > n^{\varepsilon-2/3}$ for some $\varepsilon > 0$, then the measures $\mu_n$ weakly converge to $\mu$ with probability $1$. }
The consequences of the Theorem [SiSo] give the distribution of the maximal
eigenvalue $\lambda_{{\rm max}}$ of $A_n$ and the number of eigenvalues on
$[1+n^{-2/3} x, \infty]$ for $x > 0$, see \cite{SS}.
 
In \cite{Ru} I generalize the Theorem [SiSo] to the matrices $A_n$ with
distribution of $\xi_{ij}$ decaying only polynomially fast at infinity. The main
result can be stated as the following Theorem.
\vskip 0.1in
\noindent{\bf Theorem [Ru1]}
\vskip 0.01in
{\it Let the distribution of $\xi_{ij}$ satisfy the assumption 
\begin{equation}\label{decay}
    P(|\xi_{i,j}| \geq x) \leq \frac{1}{x^p},
\end{equation} 
 with $p \geq 18$.
Then Theorem [SiSo] holds.
}

The idea of the proof is to introduce a large cutoff $\Lambda_n > 0$, such that the
set of $\omega$ on which at least one of the $\xi_{ij}$ exceeds $\Lambda_n$ has a
small measure. On the compliment of this set we approximate the distribution of
$\xi_{ij}$ by the distribution with a cutoff. Then we show, following the argument in
\cite{SS} with the necessary modifications (due to the fact that the cutoff
$\Lambda_n$ is an increasing function of $n$), that

\vskip 0.1in
{\bf Theorem [Ru2]}
\vskip 0.01in
{\it Let $p_n \rightarrow_{n \rightarrow \infty} \infty$ so that $p_n = o(n^{2/3})$. Then
\begin{equation}\label{traceconv1}
    {\rm E}({\rm Tr} A_n^{p_n} \frac{p_n^{3/2}}{n}) \rightarrow_{n \rightarrow \infty} \frac{2^{3/2}}{{\pi}^{1/2}}
     \ \ {\it in~ probability.}
\end{equation}}

In order to prove the universality at the edge of the spectrum and to
find the distribution of 1st, 2nd, 3rd eigenvalues, one needs to
consider the rescaling by the factor $r_n$ such that $r_n n^{2/3}
\rightarrow 1$.

The main result proved by Soshnikov  in \cite{S}  states 
\vskip 0.1in
{\bf Theorem [So]}
\vskip 0.01in
{\it For Wigner symmetric $n$ by $n$ random matrices with elements $a_{ij} = \frac{\xi_{ij}}{\sqrt n}$, where $\xi_{ij}$ are i.i.d, with distributions of entries which are  symmetric, which decay faster than a Gaussian at infinity and such that ${\rm E} \xi_{ij}^2 = \frac{1}{4}$, the joint distribution function of
k-dimensional random vector with the components $(\lambda_1^{(n)} - 1)
2 n^{2/3}$, \ldots, $(\lambda_k^{(n)} - 1) 2 n^{2/3}$ has a weak limit
as $n \rightarrow \infty$ which coincides with the case of Gaussian
Unitary Ensemble. }

 In \cite{Ru} I generalize this theorem  to the
matrices $A_n$ with distribution of $\xi_{ij}$ decaying only
polynomially fast at infinity:

\vskip 0.1in
{\bf Theorem [Ru3]}
\vskip 0.01in
 {\it Consider a Wigner ensemble of $n$-dimensional
real random symmetric matrices $A_n = ||a_{ij}||$ with $a_{ij} = a_{ji} = \frac{\xi_{ij}}{\sqrt{n}}$, for $1 \leq
i,j \leq n$. Let $\xi_{ij}$ be i.i.d. random variables with a symmetric distribution such that ${\rm E} \xi_{ij}^2 =
\frac{1}{4}$ and such that the distributions of $\xi_{ij}$ satisfy the assumption $P(|\xi_{ij}| \geq x) \leq \frac{1}{x^p}$ with $p \geq 18$.

Then the joint distribution function of
k-dimensional random vector with the components $(\lambda_1^{(n)} - 1)
2 n^{2/3}$, \ldots, $(\lambda_k^{(n)} - 1) 2 n^{2/3}$ has a weak limit
as $n \rightarrow \infty$ which coincides with the case of Gaussian
Unitary Ensemble. 
}

 The limiting distribution of the largest Gaussian Unitary Ensemble eigenvalue appeared in the paper by Baik, Deift and Johansson \cite{BDJ} as a limit of the rescaled distribution of the length of the longest increasing subsequence $l_{N}(\sigma)$ of a random permutation $\sigma$ of $N$ letters.
Okunkov   in {\cite{Ok}} generalized the Baik-Deift-Johannson result for an arbitrary number of rows of partitions of $N$. 
The limiting distribution of the largest Gaussian Unitary Ensemble eigenvalue appeared also in the paper by Johansson \cite{Jo2} on shape fluctuations in certain random growth models in two dimensions.

To understand the joint distribution of several eigenvalues in the bulk we
introduce 
the $k$ point correlation functions are defined as
$$ \rho_{n,\beta,k}(\lambda_1, \ldots, \lambda_k) = \frac{n!}{(n-k)!} 
\int_{{\rm R}^{n-k}} P_{n, \beta} (\lambda_1, \ldots, \lambda_n) d
\lambda_{k+1} \ldots d \lambda_{n}.$$
Let $\nu_{n,I}$ be the number of eigenvalues in a set $I$: $\nu_{n,I} =
\#\{\lambda_i^{(n)}: \lambda_i^{(n)} \in I, i =1,2, \ldots, n\}.$ It
can be shown that ${\rm E} \nu_{n,I} = \int_I \rho_{n, \beta, 1} (x) dx,$
i.e. in order to understand the distribution of several eigenvalues
around the point $x$, one should look at the intervals $(x - \frac{\rm
  const}{\rho_{n, \beta, 1} (x)},  x + \frac{\rm
  const}{\rho_{n, \beta, 1} (x)}).$
Therefore consider a rescaling
$$ \lambda_i = x + \frac{y}{\rho_{n, \beta,1}(x)}, \qquad i=1,2, \ldots,
k. $$
The rescaled $k$-point correlation function will be defined as:
$$R_{n, \beta,k} (y_1, \ldots, y_k) = \rho_{n, \beta,1}(x)^{-k} \cdot  \rho_{n,
  \beta, k} (\lambda_1, \ldots, \lambda_k).$$
It can be shown that
$$ {\rm E} \nu_{n,I_n} (\nu_{n,I_n} - 1) \ldots (\nu_{n,I_n} -k +1) =
\int_{c_1, c_2}^k R_{n, \beta, k} (y_1, \ldots, y_k) dy_1 \ldots dy_k.$$

In the cases of GUE and GOE one can derive the limiting distribution
of $\nu_{n, I_n}$ which gives the distribution of
eigenvalues for in the bulk of the spectrum (see \cite{Meh}).
 
For GUE it was shown that the rescaled $k$-point
correlation functions have the following limit on the compact subsets
of $R^k$,
 $$ \lim_{n \rightarrow \infty} R_{n,2,k} (y_1, \ldots, y_k) =
R_{2,k}(y_1, \ldots, y_k) = {\rm det} (K(y_i,y_j))_{i,j=1}^k,$$
where $K(y,z) = \frac{{\rm sin} (\pi (y-z))}{y-z}$.
In the case of GOE $K(y,z) = \frac{{\rm sin}(\pi (y-z))}{(y-z)} + \frac{{\rm sin}(\pi (y+z))}{(y+z)}$.

In \cite{Gu} is derived the distribution of one or several eigenvalues of GUE.
Gustavsson (\cite{Gu}) proves the following theorems:
\vskip 0.1in
\noindent{\bf Theorem [Gu1]} {\rm (The bulk).}
\vskip 0.1in
{\it Set 
$$
 G(t) = \frac{2}{\pi} \int_{-1}^{t} \sqrt{1 - x^2} {\rm dx} \qquad -1 \leq t \leq 1
$$
and $t = t(k,n) = G^{-1} (k/n)$ where $k = k(n)$ is such that $k/n \rightarrow a \in (0,1)$ as $n \rightarrow \infty$. If $x_k$ denotes the $k$:th eigenvalue in the GUE then it holds that as $n \rightarrow \infty$
$$ 
\frac{x_k - t \sqrt{2n}}{\left( \frac{{\rm log} n}{4 (1 - t^2) n} \right)^{1/2}} \rightarrow {\rm N(0,1)}
$$
in distribution.
}
\vskip 0.1in
\noindent{\bf Theorem [Gu2]} {\rm (The edge).}
\vskip 0.01in
{\it Let $k$ be such that $k \rightarrow \infty$ but $\frac{k}{n} \rightarrow 0$ as $n \rightarrow \infty$ and define $x_{n-k}$ as eigenvalue number $n-k$ in the GUE. Then it holds that as $n \rightarrow \infty$,
$$
\frac{x_{n-k} - \sqrt{2 n} \left(1 - \left(\frac{3 \pi k}{4 \sqrt 2 n}\right)^{2/3} \right)}{\left(\left(\frac{1}{12 \pi}\right)^{2/3} \frac{{\rm log} k}{n^{1/3} k^{2/3}} \right)^{1/2}} \rightarrow {\rm N(0,1)}
$$
in distribution.
}
\vskip 0.1in
\noindent{\bf Theorem [Gu3]} {\rm (The bulk).}
\vskip 0.01in
{\it Let $\{ x_{k_i} \}_1^m$ be eigenvalues of the GUE such that $0 < k_{i+1} - k_i \sim n^{\theta_i}$, $0 < \theta_i \leq 1$, and $k_i/n \rightarrow a_i$, where $a_i \in (0,1)$ as $n \rightarrow \infty$. Define $s_i = s_i(k_i,n) = G^{-1}(k_i/n)$ and set
$$
X_i = \frac{x_{k_i} - s_i \sqrt{2 n}}{ \left( \frac{{\rm log} n}{4 (1 - s_i^2) n} \right)^{1/2}} \qquad i = 1, \ldots, m.
$$
Then as $n \rightarrow \infty$
$$
{\bf P}[X_1 \leq x_1, \ldots, X_m \leq x_m] \rightarrow \Phi_{\Lambda} (x_1, \ldots, x_m)
$$
where $\Lambda$ is the $m \times m$ correlation matrix with $\Lambda_{ij} 
= 1 - {\rm max}_{i \leq k < j < m} \theta_k$, and $\Phi_{\Lambda}$ is the cumulative distribution function for the normalized $m$-dimensional Normal Distribution with correlation matrix $\Lambda$.
}
\vskip 0.1in
\noindent{\bf Theorem [Gu4]} {\rm (The edge).}
\vskip 0.01in
{\it Let $\{x_{n-k_i} \}_1^m$ be eigenvalues of the GUE such that $k_1 \sim n^{\gamma}$ where $0 < \gamma < 1$ and $0 < k_{i+1} - k_i \sim n^{\theta_i}$ , $0 < \theta_i < \gamma$. Set
$$
X_i = \frac{x_{n-k_i} - \sqrt{2n}\left( 1 - \left(\frac{3 \pi k_i}{4 \sqrt 2 n}\right)^{2/3}\right)}{\left(\left(\frac{1}{12 \pi}\right)^{2/3} \frac{{\rm log} k_i}{n^{1/3} k_i^{2/3}} \right)^{1/2}} \qquad i = 1, \ldots, m
$$
then as $n \rightarrow \infty$
$$
{\bf P}[X_1 \leq x_1, \ldots, X_m \leq x_m] \rightarrow \Phi_{\Lambda} (x_1, \ldots, x_m)
$$
where $\Lambda$ is the $m \times m$ correlation matrix with $\Lambda_{i,j} = 1 - \frac{1}{\gamma} {\rm max}_{i \leq k < j < m} \theta_k$, and $\Phi_{\Lambda}$ is the cdf for the normalized $m$-dimensional Normal Distribution with correlation matrix $\Lambda$.
}

In \cite{BZ}, \cite{PS}, \cite{DIZ}, \cite{BI} and \cite{DKMVZ} the result of the universality in the bulk of the spectrum
has been generalized to the certain classes of unitary invariant
ensembles of hermitian random matrices, when
$$ P(d A_n) = {\rm const}_n e^{-n {\rm Trace} V(A_n)} d A_n,$$
where $V$ is for example a polynomial of even degree with a positive
leading coefficient. If $V$ is not a quadratic polynomial, the matrix
elements of $A$ are strongly correlated (therefore these results do
not apply to Wigner random matrices.)

Johansson in \cite{Jo1} shows the universality of rescaled two point correlation functions in the bulk of the spectrum for quite a general class of hermitian Wigner matrices, with elements whose distributions are convolutions of Gaussians with other distributions.

The goal of this paper is to prove the universality in the bulk of Wigner random matrices for a general class of matrices.
In other terms, we intent to prove that for any sufficiently 
smooth, in general case non-gaussian, $g(x)$  
\begin{eqnarray} \nonumber
 P\left(\frac{b\sqrt{{\rm ln}n}}{\sqrt{n}} \leq x_i \leq \frac{c\sqrt{\rm ln}{n}}{\sqrt{n}} \right) 
= \frac{{\rm Vol}\left(\frac{b\sqrt{{\rm ln}n}}{\sqrt{n}} \leq x_i \leq \frac{c\sqrt{{\rm ln}n}}{\sqrt{n}}, S_{m_i,m_j} \right)}{{\rm Vol} (S_{m_i,m_j})} (1+\varepsilon_{m_i,m_j}).
\end{eqnarray}
The proof is cmplicated by the fact that level surfaces $\prod{g(x_j)}= {\rm const}$ are not spherical, and are not necessary  rescalings of each other, unless 
$g(x)$ can be closely approximated by the relation
\begin{equation} \label{rescaling4g}
 g(x)= {\rm Const}\; e^{|x|^{\alpha}} .
\end{equation}
In this case to obtain level surfaces which are rescalings of each other, we subdivide the space of  $\{x_{i,j}\}_{i,j=1,...n}$, i.e.
${\bf R}^{\frac{n(n+1)}{2}}$ into cones. these cones are narrow enough that level surface $\prod{ g(x)} = {\rm const}$ can be approximated by quaratic surface by Taylor expansion. We prove the result of Proposition 3 separately in each cone, but the cones $C_m$ have to be selectedd specially so that

\begin{equation} \label{partition4cones}
\frac{{\rm Vol}\left(\frac{b\sqrt{{\rm ln}n}}{\sqrt{n}} \leq x_i \leq \frac{c\sqrt{{\rm ln}n}}{\sqrt{n}}, C_m, S_{m_1,m_2} \right)}{{\rm Vol} (C_m,S_{m_1,m_2})}
=  \frac{{\rm Vol}\left(\frac{b\sqrt{{\rm ln}n}}{\sqrt{n}} \leq x_i \leq \frac{c\sqrt{{\rm ln}n}}{\sqrt{n}}, S_{m_1,m_2} \right)}{{\rm Vol} (S_{m_1,m_2})} .
\end{equation}

The partitioning of the space ${\bf R}^{\frac{n(n+1)}{2}}$ into cones $C_m$ satisfying (\ref{partition4cones}) is discussed in details in Lemma 6, in which we also show that the subset of the space not in $\cup C_m$ has neglidible probability.

This approach allows to prove  in this paper that the probability distribution for the eigenvalues in the bulk is the same for different  probability densities $g(x)$ for the entries $x_{i,j}$, and that it
does not depend on the distribution of the entries of the random matrix.

To this end we take Gaussian random matrices, i.e. those with 
$$g(x)=\frac{e^{-\frac{{x^2}}{2}}}{\sqrt{2 \pi}}, \;\; -\infty <x< \infty.$$
We show in Proposition 2 that probability distribution of eigenvalues $x_i \in (-1,1)$ can be written as 
\begin{eqnarray}
 P\left(\frac{b\sqrt{{\rm ln}n}}{\sqrt{n}} \leq x_i \leq \frac{c\sqrt{\rm ln}{n}}{\sqrt{n}} \right) 
=  \frac{{\rm Vol}\left(\frac{b\sqrt{{\rm ln}n}}{\sqrt{n}} \leq x_i \leq \frac{c\sqrt{{\rm ln}n}}{\sqrt{n}}, S_{m_1,m_2} \right)}{{\rm Vol} (S_{m_1,m_2})}(1+\varepsilon_{m_1,m_2}), 
\end{eqnarray}
where $S_{m_1,m_2}$ is a set of probability 
$$
 S_{m_1, m_2} =\{x_:\frac{n(n+1)}{8} - m_1\; n\leq\sum{x^2_{ij}}\leq \frac{n(n+1)}{8} - m_2\; n \}.
$$
By the Central Limit Theorem, it is equal $1-\varepsilon_n$.
Proposition 2 is valid because for the gaussian random value level surfaces are spherical and look exactly the same on each level surface up to a rescaling, and therefore $P(A)$ is equal to 
$ \frac{{\rm Vol}\left(A, S_{m_1,m_2} \right)}{{\rm Vol} (S_{m_1,m_2})},$
for every level surface $\prod{ g(x_{i,j})} ={\rm const}\, \Leftarrow\Rightarrow \, \sum{x^2_j} = r^2$ up to rescaling. Once we find the cones $C_m$, we show that level surfaces are close to rescaling of one another inside each cone. Therefore probability
can be replaced by ratio of volumes.
\vskip 0.15in
\section{Main result.}
The main result of the paper is the following Theorem:
\vskip 0.15in
\noindent{\bf Theorem 1.1}
\vskip 0.05in
{\it 
Consider a symmetric $n$ by $n$ Wigner random matrix, whose entries $x_{ij}$ are independent random variables with the probability distribution $g$, with ${\rm E} x_{ij}^2 = \frac{1}{4}$ and  such that $f(x) = -{\rm ln} g(x)$ satisfies the following assumptions:

\noindent for all $\Lambda$ large enough ${\rm max}_{[0, \Lambda]} f(x) \leq C {\rm max}_{[\Lambda, \Lambda + 1]} f(x),$


\noindent $ |f''(x)|, |f'''(x)| \leq{\rm max}( C |f(x)|, C).$ 


Let $x_i$ be an eigenvalue of a Wigner random matrix such that the expected position of $x_i$ is within distance $O(\frac{1}{\sqrt n})$ from $0$. Then for large $n$
$$
P(\frac{b\sqrt{ {\rm ln} n}}{\sqrt n} \leq x_i \leq \frac{c \sqrt{{\rm ln} n}}{\sqrt n}) 
$$
is independent (up to negligible corrections) of the distribution $g$ of the entries of the Wigner random matrix and is the same as for GOE.
}
\vskip 0.05in
\noindent{\bf Remark.} \hskip 0.1 in

The result of the Theorem 1.1 can also be rephrased equivalently as:

{\it Let $x_{-\frac{n}{2}} \leq \ldots \leq x_{\frac{n}{2}}$ be the eigenvalues of the matrix. Then for $x_i$ with $|i| \leq {\rm const}$, 
$$ 
P(\frac{b \sqrt{{\rm ln} n}}{\sqrt n} \leq x_i - \frac{i}{\sqrt n} \leq \frac{c\sqrt{ {\rm ln} n}}{\sqrt n})$$
is the same as for GOE up to negligible corrections.}

The proof of Theorem 1.1 consists of a number of Propositions and Lemmas given below. In Proposition 1 we show that it is sufficient to consider the problem on the set $\frac{n(n+1)}{8} - Mn \leq \sum_{ij} x_{ij}^2 \leq \frac{n(n+1)}{8} + Mn$. In Proposition 2 we explain the idea of the proof on the simple case of $g$ gaussian. In Proposition 3 we prove the main Theorem 1.1 for a general $g$, by subdividing the space into small cones and on each cone conducting an argument similar to the argument outlined in Proposition 2. Lemmas 4 -- 11 contain technical statements needed in the proof of Proposition 3.

The proof of the Theorem 1.1 is the same up to obvious modifications as the proof of the following more general Theorem:
\vskip 0.1in
\noindent{\bf Theorem 1.2} 
\vskip 0.05in
{\it Consider a symmetric $n$ by $n$ Wigner random matrix, whose entries $x_{ij}$ are independent random variables with the probability distribution $g$, with ${\rm E} x_{ij}^2 = \frac{1}{4}$ and such that $f(x) =- {\rm ln} g(x)$ satisfies the following assumptions:

\noindent for all $\Lambda$ large enough ${\rm max}_{[0, \Lambda]} f(x) \leq C {\rm max}_{[\Lambda, \Lambda+1]} f(x),$


\noindent $ |f''(x)|, |f'''(x)| \leq{\rm max}( C |f(x)|, C).$ 


Let $x_{i_1}$, $x_{i_2}$, \ldots $x_{i_k}$ be eigenvalues of a Wigner random matrix, whose expected positions are within $O(\frac{1}{\sqrt n})$ from $0$.
Then for large $n$,
$$
P(\frac{b_1 \sqrt{{\rm ln} n}}{\sqrt n} \leq x_{i_1} \leq \frac{c_1 \sqrt{{\rm ln} n}}{\sqrt n}, \ldots , \frac{b_k \sqrt{{\rm ln} n}}{\sqrt n} \leq x_{i_k} \leq \frac{c_k \sqrt{{\rm ln} n}}{\sqrt n}) 
$$
 does not depend (up to negligible corrections) on the distribution of the entries of the matrix and is the same as for GOE.}

\vskip 0.1in
\noindent{\bf Remark:}
\hskip 0.1in The assumptions that $${\rm max}_{[0, \Lambda]} f(x) \leq C {\rm max}_{[\Lambda, \Lambda+1]} f(x)\;\;\;  {\rm and\;\; that}\;\;\; |f''(x)|, |f'''(x)| \leq {\rm max}(C |f(x)|, C)$$
 were made for simplicity of explanation and can be relaxed significantly. Some assumption that $|f''|$ and $|f'''|$ are not too large (less than $n^{\varepsilon}$ on the set of measure $1- \varepsilon_n$) is needed for Lemma 10 to hold.

\section{Proof of the main result.}
 
We consider each matrix $(x_{ij})$ as a point in the $\frac{n(n+1)}{2}$ dimensional space ${\bf R^{\frac{n(n+1)}{2}}}$ with coordinates $(x_{ij})$.
In what follows we shall assume that $\varepsilon_M$ is a small positive or negative number and that $\varepsilon_M \rightarrow_{M \rightarrow \infty} 0$ (or $\varepsilon_n \rightarrow_{n \rightarrow \infty} 0$).  Sometimes we shall use the notation $\varepsilon_{M,n}$ to emphasize $n$ dependence.

We shall denote by $S_M$ (or by $S_K$) a region between two spheres in ${\bf R^{\frac{n(n+1)}{2}}}$  such that 
$$S_M = \{x_{ij} :  \frac{n(n+1)}{8} - Mn \leq \sum_{ij} x_{ij}^2 \leq \frac{n(n+1)}{8} + Mn \}.$$

We shall denote by $S_{M_1, M_2}$ a region between two spheres in ${\bf R^{\frac{n(n+1)}{2}}}$  such that
$$S_{M_1, M_2} = \{x_{ij} :  \frac{n(n+1)}{8} - M_2 n \leq \sum_{ij} x_{ij}^2 \leq \frac{n(n+1)}{8} + M_1 n \}.$$


We shall denote by $S_r $ a spherical shell $\sum_{ij} x_{ij}^2 = r^2$.
 In the rest of the paper we shall use the following notation:
by ${\rm Vol} (S_M)$ we shall mean the volume of the surface, by ${\rm Vol} (\sum_{ij} x_{ij}^2 = r^2)$ we shall mean the surface area of the surface.

\vskip 0.1in
\noindent{\bf Proposition 1:}
\vskip 0.05in
{\it For all $g$ such that Central Limit Theorem holds, 
$$
\int \prod_{ij} g(x_{ij}) 1_{\frac{b \sqrt{{\rm ln} n}}{\sqrt n} \leq x_i \leq \frac{c \sqrt{{\rm ln} n}}{\sqrt n}} \prod_{ij} d x_{ij} = 
 \int \prod_{ij} g(x_{ij}) 1_{\frac{b \sqrt{{\rm ln} n}}{\sqrt n} \leq x_i \leq \frac{c \sqrt{{\rm ln} n}}{\sqrt n}, S_M} \prod_{ij} d x_{ij} + \varepsilon_M 
$$
where $\varepsilon_M \rightarrow 0$ as $M \rightarrow \infty$.
}
\vskip 0.1in
\noindent{\bf Proof:} 
\hskip 0.05in
For any measurable set $C$ in ${\bf R^{\frac{n(n+1)}{2}}}$,
$$ 
\int \prod_{ij} g(x_{ij}) 1_C \prod_{ij} d x_{ij} = P(x_{ij} \in C, i, j = 1, \ldots n).$$
Since, by the Central Limit theorem, 
$$
P(S_M) = P(\frac{n(n+1)}{8} - Mn \leq \sum_{ij} x_{ij}^2 \leq \frac{n(n+1)}{8} + Mn) = 1 + \varepsilon_M \rightarrow_{M \rightarrow \infty} 1
$$
\vskip 0.01in
Proposition 1 is proved.
\vskip 0.05in
The following Proposition is important in the subsequent results and also illustrates the method used to prove the main Theorem.
\vskip 0.1in
\noindent{\bf Proposition 2:}
\vskip 0.05in
\noindent{\it For a gaussian $g$,
\begin{eqnarray}
&&P (\frac{b \sqrt{{\rm ln} n}}{\sqrt n} \leq x_i \leq \frac{c \sqrt{{\rm ln} n}}{\sqrt n})\qquad\qquad\cr&=&\frac{{\rm Vol} (\frac{b \sqrt{{\rm ln} n}}{\sqrt n} \leq x_i \leq \frac{c \sqrt{{\rm ln} n}}{\sqrt n}, \frac{n(n+1)}{8} - M_2 n \leq \sum x_{ij}^2 \leq \frac{n(n+1)}{8} + M_1 n)}{{\rm Vol}( \frac{n(n+1)}{8} - M_2 n  \leq  \sum x_{ij}^2 \leq \frac{n(n+1)}{8} + M_1 n)}\; (1 + \varepsilon_{M}),\qquad
\end{eqnarray}
where $\varepsilon_M \rightarrow_{M \rightarrow \infty} 0$.
}
\vskip 0.1in
\noindent{\bf Proof:}
\begin{eqnarray}{\label{21}}
&& P(\frac{b \sqrt{{\rm ln} n}}{\sqrt n} \leq x_i \leq \frac{c \sqrt{{\rm ln} n}}{\sqrt n}) = \int e^{- \sum_{ij} x_{ij}^2 } 1_{\frac{b \sqrt{{\rm ln} n}}{\sqrt n} \leq  x_i \leq \frac{c \sqrt{{\rm ln} n}}{\sqrt n}} \prod_{ij} d x_{ij} \cr  && \qquad = \int e^{-r^2} {\rm Vol}(\sum_{ij} x_{ij}^2= r^2, \frac{b \sqrt{{\rm ln} n}}{\sqrt n} \leq x_i \leq \frac{c \sqrt{{\rm ln} n}}{\sqrt n}, S_{M_1, M_2})dr + \varepsilon_M. 
\end{eqnarray}
Since the hypersurface $\sum_{ij} x_{ij}^2 = r'^2$ is a rescaling 
of the hypersurface $\sum_{ij} x_{ij}^2 = r^2$ by a factor $\frac{r'}{r}$, we obtain that to each point $(x_{ij})$ on the surface $\sum_{ij} x_{ij}^2 = r^2$ there corresponds a point $(x'_{ij}) = (x_{ij} \frac{r'}{r})$, on the surface $\sum_{ij} x_{ij}^2 = r'^2$ on the same radial line, we have that the eigenvalues $(x_i)$ of the matrix $(x_{ij})$ correspond to the eigenvalues $(x_i' = x_i \frac{r'}{r})$ of the matrix $(x'_{ij})$. Since one hypersurface is the rescaling of the other one
\begin{eqnarray}\label{22}
&&\frac{{\rm Vol} ( \frac{b \sqrt{{\rm ln} n}}{\sqrt n} \leq x_i \leq \frac{c \sqrt{{\rm ln} n}}{\sqrt n}, \sum_{ij} x_{ij}^2 = r^2, \frac{n(n+1)}{8} - M_2 n \leq \sum_{ij} x_{ij}^2 \leq \frac{n(n+1)}{8} + M_1n)}{{\rm Vol}
(\sum_{ij} x_{ij}^2 = r^2, \frac{n(n+1)}{8} - M_2n \leq \sum_{ij} x_{ij}^2 \leq \frac{n(n+1)}{8} + M_1n)} \cr &=& \frac {{\rm Vol} (\frac{b \sqrt{{\rm ln} n}}{\sqrt n} \frac{r'}{r} \leq x_i \leq \frac{c \sqrt{{\rm ln} n}}{\sqrt n} \frac{r'}{r}, \sum_{ij} x_{ij}^2 = r'^2, \frac{n(n+1)}{8} - M_2n \leq \sum_{ij} x_{ij}^2 \leq \frac{n(n+1)}{8} + M_1n)}{{\rm Vol} (\sum_{ij} x_{ij}^2 = r'^2, \frac{n(n+1)}{8} - M_2n \leq \sum_{ij} x_{ij}^2 \leq \frac{n(n+1)}{8} + M_1n)} \qquad  
\end{eqnarray}
i.e. relative volumes stay the same under a rescaling.

Using the fact that $\frac{n(n+1)}{8} - M_2n \leq \sum_{ij} x_{ij}^2 \leq \frac{n(n+1)}{8} + M_1n$ we obtain that $\frac{r'}{r}=1 + O(\frac{M}{n}),$ where $M = {\rm max}(M_1, M_2)$ here and subsequently. By Lemma 4 we write 
\begin{eqnarray}\label{23}
&&{\rm Vol} (\frac{b \sqrt{{\rm ln} n}}{\sqrt n} (1 + O(\frac{M}{n})) \leq x_i \leq \frac{c \sqrt{{\rm ln} n}}{\sqrt n}(1 + O(\frac{M}{n})), \sum_{ij} x_{ij}^2 = r'^2, S_{M_1, M_2})\qquad  \cr &&= 
{\rm Vol} (\frac{b \sqrt{{\rm ln} n}}{\sqrt n} \leq x_i \leq \frac {c \sqrt{{\rm ln} n}}{\sqrt n}, \sum_{ij} x_{ij}^2 = r'^2, S_{M_1, M_2})(1 + \varepsilon_M).
\end{eqnarray}
Thus
\begin{eqnarray}\label{24}
 &&\int e^{-r^2} {\rm Vol} ( \sum_{ij} x_{ij}^2 = r^2, \frac{b \sqrt{{\rm ln} n} }{\sqrt n} \leq x_i \leq \frac{c \sqrt{{\rm ln} n}}{\sqrt n}, S_{M_1, M_2}) dr   \cr&&  = 
\frac{{\rm Vol} (\sum_{ij} x_{ij}^2 = r_0^2,\frac{b \sqrt{{\rm ln} n}}{\sqrt n} \leq x_i \leq \frac{c \sqrt{{\rm ln}n}}{\sqrt n}, \frac{n(n+1)}{8} - M_2n \leq \sum_{ij} x_{ij}^2 \leq \frac{n(n+1)}{8} + M_1n)}{{\rm Vol}
  (\sum_{ij} x_{ij}^2 = r_0^2, \frac{n(n+1)}{8} - M_2n \le \sum_{ij} x_{ij}^2 \leq \frac{n(n+1)}{8} + M_1n)} \cr && \qquad  \int e^{-r^2} {\rm Vol} (\sum_{ij} x_{ij}^2 = r^2, \frac{n(n+1)}{8} -M_2n \leq \sum_{ij} x_{ij}^2 \leq \frac{n(n+1)}{8} + M_1n) dr (1 + \varepsilon_M) \cr &&  = \frac{{\rm Vol} (\sum_{ij} x_{ij}^2 = r_0^2, \frac{b \sqrt{{\rm ln} n}}{\sqrt n} \leq x_i \leq \frac{c \sqrt{{\rm ln} n}}{\sqrt n}, \frac{n(n+1)}{8} - M_2n \leq \sum_{ij} x_{ij}^2 \leq \frac{n(n+1)}{8} + M_1n)}{{\rm Vol} (\sum_{ij} x_{ij}^2 = r_0^2, \frac{n(n+1)}{8} -M_2n \leq \sum_{ij} x_{ij}^2 \leq \frac{n(n+1)}{8} + M_1n)} \cr && \qquad   P(\frac{n(n+1)}{8} -M_2n   \leq \sum_{ij} x_{ij}^2  \leq \frac{n(n+1)}{8} + M_1 n)(1+\varepsilon_M).\qquad \qquad \qquad
\end{eqnarray}
By the Central Limit Theorem $P( S_{M_1, M_2})=1 + \varepsilon_M.$
Now, observe that using the same arguments
\begin{eqnarray}\label{25}
  &&{\rm Vol}(\frac{b \sqrt{{\rm ln} n}}{\sqrt n} \leq x_i \leq \frac{c \sqrt{{\rm ln} n}}{\sqrt n}, \frac{n(n+1)}{8} - M_2n \leq \sum_{ij} x_{ij}^2 \leq \frac{n(n+1)}{8} + M_1n)  \cr && = \int  {\rm Vol}(\sum_{ij} x_{ij}^2 = r^2, \frac{b \sqrt{{\rm ln} n}}{\sqrt n} \leq x_i \leq \frac{c \sqrt{{\rm ln} n}}{\sqrt n}, S_{M_1, M_2}) dr \qquad
 \cr && = \frac{{\rm Vol} (\sum_{ij} x_{ij}^2 = r_0^2, \frac{b \sqrt{{\rm ln} n}}{\sqrt n}  \leq x_i \leq \frac{c \sqrt{{\rm ln} n}}{\sqrt n}, \frac{n(n+1)}{8} -M_2n  \leq \sum_{ij} x_{ij}^2 \leq \frac{n(n+1)}{8} + M_1n)}{{\rm Vol} (\sum_{ij} x_{ij}^2 = r_0^2, \frac{n(n+1)}{8} - M_2n \leq \sum_{ij} x_{ij}^2 \leq \frac{n(n+1)}{8} + M_1n)} \cr &&  \qquad\int{\rm Vol} (\sum_{ij} x_{ij}^2 = r^2, \frac{n(n+1)}{8}-M_2n  \leq   \sum_{ij} x_{ij}^2 \leq \frac{n(n+1)}{8} + M_1n) dr (1+\varepsilon_M).\qquad  \end{eqnarray}
Therefore,
\begin{eqnarray}\label{26}
  &&\frac{{\rm Vol}(\sum_{ij} x_{ij}^2 = r_0^2, \frac{b \sqrt{{\rm ln} n}}{\sqrt n}  \leq x_i \leq \frac{c \sqrt{{\rm ln} n}}{\sqrt n}, \frac{n(n+1)}{8}-M_2n   \leq  \sum_{ij} x_{ij}^2 \leq \frac{n(n+1)}{8} + M_1n)}{{\rm Vol} (\sum_{ij} x_{ij}^2 = r_0^2, \frac{n(n+1)}{8} - M_2n \leq \sum_{ij} x_{ij}^2 \leq \frac{n(n+1)}{8} + M_1n)}\cr && = \frac{{\rm Vol} ( \frac{b \sqrt{{\rm ln} n}}{\sqrt n} \leq x_i \leq \frac{c \sqrt{{\rm ln} n}}{\sqrt n}, \frac{n(n+1)}{8} - M_2n \leq \sum_{ij} x_{ij}^2 \leq \frac{n(n+1)}{8} + M_1n)}{{\rm Vol} (\frac{n(n+1)}{8} - M_2n  \leq  \sum_{ij} x_{ij}^2 \leq \frac{n(n+1)}{8} + M_1 n)} 
(1+\varepsilon_M).\qquad 
\end{eqnarray}
Thus the Proposition 2 is proved.
\vskip 0.05in
\noindent{In the following Proposition, and  thereafter, we use the notation:}
$$f(x) = - {\rm ln}~ g(x) $$
\vskip 0.1in
{\bf Proposition 3:}
\vskip 0.05in
\noindent{\it For any $f =- {\rm ln}~ g$ such that, for all $\Lambda$ large enough $${\rm max}_{[0, \Lambda]}f(x) \leq C {\rm max}_{[\Lambda, \Lambda+1]} f(x),$$ and 
$$ 
|f''(x)|, |f'''(x)| \leq {\rm max} (C |f(x)|, C)
$$
we have that
\begin{eqnarray}\label{statementlemma31}
&& \int \prod_{ij} g(x_{ij}) 1_{\frac{b \sqrt{{\rm ln} n}}{\sqrt n} \leq x_i \leq \frac{c \sqrt{{\rm ln} n}}{\sqrt n}} \prod_{ij} d x_{ij}  \cr && = 
\frac{{\rm Vol}(\frac{b \sqrt{{\rm ln} n}}{\sqrt n} \leq x_i \leq \frac{c \sqrt{{\rm ln} n}}{\sqrt n}, \frac{n(n+1)}{8} - Mn \leq \sum_{ij} x_{ij}^2 \leq \frac{n(n+1)}{8} + Mn)}{{\rm Vol} ( \frac{n(n+1)}{8} - Mn \leq \sum_{ij} x_{ij}^2 \leq \frac{n(n+1)}{8} + Mn)} (1+\varepsilon_M). \qquad\qquad\qquad 
\end{eqnarray}
}
\vskip 0.1in
{\bf Proof:}
\hskip 0.05in
{Let us call  a set $C$ a cone, if $(x_{ij}) \in C$ implies that $(\lambda x_{ij}) \in C$ 
for all $\lambda \geq 0$.
Consider partitioning the set ${\bf R^{\frac{n(n+1)}{2}}}$ into a finite number 
of small cones $C_m$, $m = 1, \ldots, m_n$ and a conical set denoted by $B$, such that the set $B$ has negligible probability.}
Most importantly we assume that  all $C_m$ are such that
\begin{eqnarray}\label{3111}
&&\frac{{\rm Vol} (\frac{b \sqrt{{\rm ln} n}}{\sqrt n} \leq x_i \leq \frac{c \sqrt{{\rm ln} n}}{\sqrt n},\frac{n(n+1)}{8} - Mn \leq \sum_{ij} x_{ij}^2 \leq \frac{n(n+1)}{8} + Mn, C_m)}{{\rm Vol} ( \frac{n(n+1)}{8} - Mn \leq \sum_{ij} x_{ij}^2 \leq \frac{n(n+1)}{8} + Mn, C_m  )} \cr  &&=    \frac{{\rm Vol} (\frac{b \sqrt{{\rm ln} n}}{\sqrt n}\leq x_i \leq \frac{c \sqrt{{\rm ln} n}}{\sqrt n}, \frac{n(n+1)}{8}- Mn \leq \sum_{ij} x_{ij}^2 \leq \frac{n(n+1)}{8} + Mn)}{{\rm Vol} (\frac{n(n+1)}{8} - Mn \leq \sum_{ij} x_{ij}^2 \leq \frac{n(n+1)}{8} + Mn  )}(1+ \varepsilon_{M,n}).\qquad
\end{eqnarray}
A procedure of construction on the cones $C_m$ satisfying these assumptions and thus the existence of partition into the cones $C_m$ is proved in Lemmas 6, 7, 8a, 8b, 8c.

Below we  use the following notation. Let $V_m$ be the area inside $C_m$ between the surfaces $\prod_{ij} g(x_{ij}) = z_0$ and $\prod_{ij} g(x_{ij}) = z_1$ which encloses all the points such that $\frac{n(n+1)}{8} - Mn \leq \sum_{ij} x_{ij}^2 \leq \frac{n(n+1)}{8} + Mn$. 

We can rewrite $P(\frac{b \sqrt{{\rm ln} n}}{\sqrt n} \leq x_i \leq \frac{c \sqrt{{\rm ln} n}}{\sqrt n})$ as
\begin{eqnarray}
&& \int \prod_{ij} g(x_{ij}) 1_{\frac{b \sqrt{{\rm ln} n}}{\sqrt n} \leq x_i \leq \frac{c \sqrt{{\rm ln} n}}{\sqrt n}} \prod_{ij} d x_{ij} \cr  && = 
\sum_m \int z {\rm Vol} (\prod_{ij} g(x_{ij})=z, \frac{b \sqrt{{\rm ln} n}}{\sqrt n} \leq x_i \leq \frac{c \sqrt{{\rm ln} n}}{\sqrt n}, V_m, C_m) dh + \varepsilon_{M,n}, \qquad\qquad
\end{eqnarray}
where $d h$ is the distance between the surfaces $\prod_{ij} g(x_{ij}) = z$ and $\prod_{ij} g(x_{ij}) = z + dz$.

First we  consider the example $f (x) = x^{\alpha}$. In this case the surfaces $\sum_{ij} x_{ij}^{\alpha} = z$ and $\sum_{ij} x_{ij}^{\alpha} = z'$ are rescalings of each other via a factor $(\frac{z'}{z})^{\frac{1}{\alpha}}.$ So we obtain that
\begin{eqnarray}\label{31111}
&& \frac{{\rm Vol} (\frac{b \sqrt{{\rm ln} n}}{\sqrt n} \leq x_i \leq \frac{c \sqrt{{\rm ln} n}}{\sqrt n}, \sum_{ij} f(x_{ij}) = - {\rm ln} z, V_m, C_m)}{{\rm Vol} (\sum_{ij} f(x_{ij}) = - {\rm ln} z, V_m, C_m)}   \cr &&=  \frac{{\rm Vol}
(\frac{b \sqrt{{\rm ln} n}}{\sqrt n} \frac{r'}{r} \leq x_i \leq \frac{c \sqrt{{\rm ln} n}}{\sqrt n} \frac{r'}{r}, \sum_{ij} f(x_{ij}) = - {\rm ln} z', V_m, C_m)}{{\rm Vol} ( \sum_{ij} f(x_{ij}) = - {\rm ln} z', V_m, C_m)}. \qquad\qquad
\end{eqnarray}
To prove (\ref{31111}) in the case of general $g$ we first approximate the
level surfaces $\sum_{ij} f(x_{ij}) =- {\rm ln}~ z$ by quadratic surfaces and prove the analog (\ref{31111}) for the quadratic surfaces and then show that  (\ref{31111}) holds, since the quadratic surfaces approximate the level surfaces closely. 
For this we use Lemmas 4 and 10 which are formulated and proved further in the paper, and  are summarized as follow:

\noindent{In Lemma 10 it is proved that for general $g$
\begin{eqnarray}\label{331}
&&\frac{{\rm Vol} (\frac{b \sqrt{{\rm ln} n}}{\sqrt n} \leq x_i \leq \frac{c \sqrt{{\rm ln} n}}{\sqrt n}, \prod_{ij} g(x_{ij}) = z, V_m, C_m)}{{\rm Vol} ( \prod_
{ij} g(x_{ij}) = z, V_m, C_m)}\cr &&  = \frac{{\rm Vol} (\frac{b \sqrt{{\rm ln} n}}{\sqrt n}\frac{r'}{r}  \leq x_i \leq \frac{c \sqrt{{\rm ln} n}}{\sqrt n}\frac{r'}{r}, \prod_{ij} g(x_{ij}) = z', V_m, C_m)}{{\rm Vol} ( \prod_{ij} g(x_{ij})=z', V_m, C_m)}(1+\varepsilon_{M,n}). \qquad
\end{eqnarray}}
By Lemma 4 we obtain,
\begin{eqnarray}
&&{\rm Vol} (\prod_{ij} g(x_{ij}) = z', \frac{b \sqrt{{\rm ln} n}}{\sqrt n} \frac{r'}{r} \leq x_i \leq \frac{c \sqrt{{\rm ln} n}}{\sqrt n} \frac{r'}{r}, V_m, C_m)\cr && = {\rm Vol} (\prod_{ij} g(x_{ij}) = z', \frac{b \sqrt{{\rm ln} n}}{\sqrt n} \leq x_i \leq \frac{c \sqrt{{\rm ln} n}}{\sqrt n}, V_m, C_m) (1+\varepsilon_M).\qquad\qquad
\end{eqnarray}
Therefore,
\begin{eqnarray}
&& \frac{{\rm Vol} (\prod_{ij} g(x_{ij}) = z, \frac{b \sqrt{{\rm ln} n}}{\sqrt n}\leq x_i \leq \frac{c \sqrt{{\rm ln} n}}{\sqrt n}, V_m, C_m)}{{\rm Vol} ( \prod_{ij} g(x_{ij}) = z, V_m, C_m)}\cr  && = \frac {{\rm Vol} (\prod_{ij} g(x_{ij}) = z', \frac{b \sqrt{{\rm ln} n}}{\sqrt n} \leq x_i \leq \frac{c \sqrt{{\rm ln} n}}{\sqrt n}, V_m, C_m)}{{\rm Vol} ( \prod_{ij} g(x_{ij}) = z', V_m, C_m)}(1+\varepsilon_M).\qquad\qquad
\end{eqnarray}
And we obtain
\begin{eqnarray}
&& P(\frac{b \sqrt{{\rm ln} n}}{\sqrt n} \leq x_i \leq \frac{c \sqrt{{\rm ln}n}}{\sqrt n}, V_m, C_m) \cr  &=& \frac{{\rm Vol}(\prod_{ij} g(x_{ij}) = z_0, \frac{b \sqrt{{\rm ln} n}}{\sqrt n} \leq x_i \leq \frac{c \sqrt{{\rm ln} n}}{\sqrt n}, V_m, C_m)}{{\rm Vol}(\prod_{ij} g(x_{ij}) = z_0, V_m, C_m)} \cr && \qquad \qquad \qquad \qquad \qquad \cdot \int \prod_{ij} g(x_{ij}) {\rm Vol}(\prod_{ij} g(x_{ij}) = z, V_m, C_m) dh
(1 + \varepsilon_n)  \cr &=& \frac{{\rm Vol}(\prod_{ij} g(x_{ij}) = z_0, \frac{ b \sqrt{{\rm ln} n}}{\sqrt n} \leq x_i \leq \frac{c \sqrt{{\rm ln} n}}{\sqrt n}, V_m, C_m)}{{\rm Vol}(\prod_{ij} g(x_{ij}) = z_0, V_m, C_m)} P(C_m, V_m) (1 + \varepsilon_n).\qquad\qquad
\end{eqnarray}
By the same methods we obtain that
\begin{eqnarray}\label{34}
&&{\rm Vol} (\frac{b \sqrt{{\rm ln} n}}{\sqrt n} \leq x_i \leq \frac{c \sqrt{{\rm ln} n}}{\sqrt n}, V_m, C_m)\cr  && = \int {\rm Vol} (\prod_{ij} g(x_{ij}) =z, \frac{b \sqrt{{\rm ln} n}}{\sqrt n}\leq x_i \leq \frac{c \sqrt{{\rm ln} n}}{\sqrt n}, V_m, C_m)dh \cr &&= \frac{ {\rm Vol} (\prod_{ij} g(x_{ij}) = z, \frac{b \sqrt{{\rm ln} n}}{\sqrt n} \leq x_i  \leq \frac{c \sqrt{{\rm ln} n}}{\sqrt n},  V_m, C_m)}{{\rm Vol} ( \prod_{ij} g(x_{ij}) =z, V_m, C_m)} \int  {\rm Vol} (\prod_{ij} g(x_{ij}) = z, V_m, C_m) dh \cr && \qquad \qquad \qquad \qquad \qquad \qquad \qquad \qquad \qquad \qquad (1 + \varepsilon_{M,n}).\qquad\qquad
\end{eqnarray}
Therefore,
$$
\frac{{\rm Vol} ( \frac{b \sqrt{{\rm ln} n}}{\sqrt n}\leq x_i \leq \frac{c \sqrt{{\rm ln} n}}{\sqrt n}, V_m, C_m)}{{\rm Vol}( V_m, C_m)} = \frac{{\rm Vol} (\frac{b \sqrt{{\rm ln} n}}{\sqrt n}  \leq x_i \leq \frac{c \sqrt{{\rm ln} n}}{\sqrt n}, \prod_{ij} g(x_{ij}) = z_0, V_m, C_m)} {{\rm Vol} ( \prod_{ij} g(x_{ij}) = z_0, V_m, C_m)}(1 + \varepsilon_{M,n}).
$$
Thus we obtain,
$$
\int \prod_{ij} g(x_{ij}) 1_{\frac{b \sqrt{{\rm ln} n}}{\sqrt n} \leq x_i \leq \frac{c \sqrt{{\rm ln} n}}{\sqrt n}} \prod_{ij} d x_{ij}=
\sum_{m} \frac{{\rm Vol}(\frac{b \sqrt{{\rm ln} n}}{\sqrt n} \leq  x_i \leq \frac{c \sqrt{{\rm ln} n}}{\sqrt n}, V_m, C_m)}{{\rm Vol} (V_m, C_m)} P(C_m) (1+\varepsilon_{M,n}).
$$
By Lemma 9 and by the assumptions on $C_m$,
\begin{eqnarray}\label{33}
  &&\frac{{\rm Vol} (\frac{b \sqrt{{\rm ln} n}}{\sqrt n} \leq   x_i \leq \frac{c \sqrt{{\rm ln} n}}{\sqrt n}, V_m, C_m)}{{\rm Vol} ( V_m, C_m)}\cr && = \frac{{\rm Vol}( \frac{b \sqrt{{\rm ln} n}}{\sqrt n}  \leq x_i \leq \frac{c \sqrt{{\rm ln} n}}{\sqrt n}, \frac{n(n+1)}{8} - Mn \leq \sum_{ij} x_{ij}^2 \leq \frac{n(n+1)}{8} + Mn, C_m)}{{\rm Vol}( \frac{n(n+1)}{8} - Mn \leq \sum_{ij} x_{ij}^2 \leq \frac{n(n+1)}{8} + Mn, C_m)} (1 + \varepsilon_{M,n})\qquad \cr&& = \frac{{\rm Vol} (\frac{b \sqrt{{\rm ln} n}}{\sqrt n} \leq x_i \leq \frac{c \sqrt{{\rm ln} n}}{\sqrt n}, \frac{n(n+1)}{8} - Mn \leq \sum_{ij} x_{ij}^2 \leq \frac{n(n+1)}{8} + Mn)}{{\rm Vol} (\frac{n(n+1)}{8} - Mn \leq \sum_{ij} x_{ij}^2 \leq \frac{n(n+1)}{8} + Mn)}(1+\varepsilon_{M,n}).\qquad\qquad
\end{eqnarray}
Therefore we derive,
\begin{eqnarray}
&& \int 1_{\frac{b \sqrt{{\rm ln} n}}{\sqrt n} \leq x_i \leq \frac{c \sqrt{{\rm ln} n}}{\sqrt n}} \prod_{ij} g(x_{ij}) \prod_{ij} dx_{ij} \cr && = \frac{{\rm Vol}(\frac{b \sqrt{{\rm ln} n}}{\sqrt n} \leq x_i \leq \frac{c \sqrt{{\rm ln} n}}{\sqrt n}, \frac{n(n+1)}{8} - Mn \leq \sum_{ij} x_{ij}^2 \leq \frac{n(n+1)}{8} + Mn)} {{\rm Vol} (\frac{n(n+1)}{8} - Mn \leq \sum_{ij} x_{ij}^2 \leq \frac{n(n+1)}{8} +Mn)} \sum_m P(C_m)(1+\varepsilon_{M,n}) \cr  && = \frac{{\rm Vol}(\frac{b \sqrt{{\rm ln} n}}{\sqrt n} \leq x_i \leq \frac{c \sqrt{{\rm ln} n}}{\sqrt n}, \frac{n(n+1)}{8} - Mn \leq \sum_{ij} x_{ij}^2 \leq \frac{n(n+1)}{8} + Mn)} {{\rm Vol} (\frac{n(n+1)}{8} - Mn \leq \sum_{ij} x_{ij}^2 \leq \frac{n(n+1)}{8} +Mn)}(1+\varepsilon_{M,n}), \qquad \qquad \qquad 
\end{eqnarray}
Where we use the fact that $\sum_m P(C_m) = 1+\varepsilon_{M,n}$.
\vskip 0.01 in
\noindent{The Proposition 3 is proved.}
\vskip 0.1in
\noindent{\bf Lemma 4:} 
\vskip 0.05 in
\noindent{\it For all $C_m$,
\begin{eqnarray}
&& {\rm Vol} (\prod_{ij} g(x_{ij}) = z,\frac{b \sqrt{{\rm ln} n}}{\sqrt n}(1+O(\frac{M}{n})) \leq  x_i \leq \frac{c \sqrt{{\rm ln} n}}{\sqrt n} (1 + O (\frac{M}{n})), V_m, C_m) \cr && =  {\rm Vol} (\prod_{ij} g(x_{ij}) =z, \frac{b \sqrt{{\rm ln} n}}{\sqrt n} \leq x_i \leq \frac{c \sqrt{{\rm ln} n}}{\sqrt n}, V_m, C_m)(1+\varepsilon_{M,n}).\qquad \qquad
\end{eqnarray}
Also, 
\begin{eqnarray}
&& {\rm Vol} (\prod_{ij} g(x_{ij}) = z, \frac{b \sqrt{{\rm ln} n}}{\sqrt n}(1+O(\frac{M}{n})) \leq x_i \leq \frac{c \sqrt{{\rm ln} n}}{\sqrt n} (1 + O (\frac{M}{n})), V_m, S_M)\cr && =  {\rm Vol} (\prod_{ij} g(x_{ij}) =z, \frac{b \sqrt{{\rm ln} n}}{\sqrt n} \leq x_i \leq \frac{c \sqrt{{\rm ln} n}}{\sqrt n}, V_m, S_M)(1+\varepsilon_{M,n}).\qquad \qquad 
\end{eqnarray}
}
\vskip 0.1in
\noindent{\bf Proof:}
\hskip 0.01in
\noindent We note that
$$
\prod_{ij} d x_{ij} = \prod_{ij} |x_i - x_j| \prod_i d x_i \prod_{ij} O_{ij} \prod_{ij} d \xi_{ij},
$$
where $\prod_{ij} O_{ij} \prod_{ij} d \xi_{ij}$ represents the angular part. 

\noindent{Everywhere in the proof of Lemma 4 we shall assume that $x_i = O({\sqrt{\frac{{\rm ln}\, n}{ n}}}).$} 
$$
\prod_{k \neq j} |x_k - x_j| = \prod_{k,j\neq i} |x_k - x_j| \prod_j |x_i - x_j
 - O(\frac{M}{n}\, \sqrt{\frac{{\rm ln}\, n}{n}}\,)| = \prod_{k \neq j} |x_k - x_j|
\, e^{- O(\frac{M}{n}\, \sqrt{\frac{{\rm ln}\, n}{n}}\,)\, \sum_j \frac{1}{|x_i - x_j|}.}
$$
By Lemma 5, on a set of probability $(1 - \varepsilon_K) P(C_k)$ we have that 
$$ \frac{M \sqrt{{\rm ln} n}}{n^{\frac{3}{2}} } \sum_{j \neq i} \frac{1}{|x_i - x_j|} \leq \varepsilon_n. $$
Take the set $V_m \cap C_m$ and subdivide it into small conical pieces $W_l$, $l=1, \ldots, L$.
Replacing $x_i = \frac{c \sqrt{{\rm ln} n}}{\sqrt n}(1 + O(\frac{M}{n}))$ with $x_i = \frac{c \sqrt{{\rm ln} n}}{\sqrt n}$ and writing $x_{kj} = \sum_i x_i \xi_{ik} \xi_{ij}$ where $\xi_{ik}$ and $\xi_{ij}$ are the angular variables such that $\sum_i \xi_{ij}^2 =1$ and $\sum_i \xi_{ij} \xi_{ik} = \delta_{jk}$, we obtain that
\begin{eqnarray}\label{442}
&&{\rm Vol}(\sum_{ij} f(x_{ij}) = -{\rm ln}~ z, \frac{b \sqrt{{\rm ln} n}}{\sqrt n}(1+O(\frac{M}{n})) \leq x_i \leq \frac{c \sqrt{{\rm ln} n}}{\sqrt n}(1+O(\frac{M}{n})), W_l, V_m, C_m)\quad \cr && = {\rm Vol}(\sum_{kj} f(x_{kj} + O(\frac{M}{n^{3/2}}) \xi_{ik}\xi_{ij}) = -{\rm ln}~ z, \frac{b \sqrt{{\rm ln} n}}{\sqrt n} \leq x_i \leq \frac{c \sqrt{{\rm ln} n}}{\sqrt n}, W_l, V_m, C_m)(1 + \varepsilon_n).\qquad
\end{eqnarray}
The surface $\sum_{kj} f(x_{kj} + O(\frac{M}{n^{3/2}}) \xi_{ik} \xi_{ij}) = - {\rm ln}~ z$ is close to the surface $\sum_{kj} f(x_{kj}) = - {\rm ln}~ z.$

Let $r^2=\sum x_j^2 $ and $(r')^2 = {\sum_{k \neq j} x_k^2 + (x_i + O(\frac{M}{n^{\frac{3}{2}}}))^2}$.
Then 
\begin{eqnarray}\label{lemma41}
\frac{r'}{r} = \sqrt{\frac{r^2 + O(\frac{M}{n^{\frac{3}{2}}})\sum_{kj}\xi_{ik}\xi_{ij}x_{kj}}{r^2}}= \sqrt{\frac{\sum_{k \neq j} x_k^2 + (x_i + O(\frac{M}{n^{\frac{3}{2}}}))^2}{r^2}} = 1 + O(\frac{M}{n^{\frac{5}{2}}}).\qquad
\end{eqnarray}
Since the surface $\sum_{kj} f(x_{kj} + O(\frac{M}{n^{3/2}}) \xi_{ik}\xi_{ij}) = - {\rm ln}~ z$ and the surface $\sum_{kj} f(x_{kj}) = - {\rm ln}~ z$ are less than
$ O(\frac{M}{n^{3/2}})  $ apart radially; and due to the fact that cones are very small, the change of volume $\sum_{kj} f(x_{kj} + O(\frac{M}{n^{3/2}}) \xi_{ik}\xi_{ij}) = - {\rm ln}~ z$ is replaced by $\sum_{kj} f(x_{kj}) = - {\rm ln}~ z$ in 
the equation (\ref{442}) is
$(\frac{r'}{r})^{\frac{n(n+1)}{2}} = (1 + O(\frac{M}{n^{\frac{5}{2}}}))^{\frac{n(n+1)}{2}} = 1 + O(\frac{M}{n^{\frac{1}{2}}})$
we obtain
 \begin{eqnarray}\label{445}
&& {\rm Vol}(\sum_{ij} f(x_{ij}) = - {\rm ln}~ z, \frac{b \sqrt{{\rm ln} n}}{\sqrt n}(1+O(\frac{M}{n})) \leq x_i \leq \frac{c \sqrt{{\rm ln} n}}{\sqrt n}(1+O(\frac{M}{n})), W_l, V_m, C_m) \cr  && = {\rm Vol}(\sum_{kj} f(x'_{kj})  = - {\rm ln}~ z, \frac{b \sqrt{{\rm ln} n}}{\sqrt n} \leq x_i \leq \frac{c \sqrt{{\rm ln} n}}{\sqrt n}, W_l, V_m, C_m)(1 + O(\varepsilon_n)). \qquad
\end{eqnarray} 
 \vskip 0.1in
 \noindent{\bf Lemma 5:}
\vskip 0.05 in
\noindent{\it
For every cone $C_k$ defined in Proposition 3, on the subset $C$ of the cone $C_k$ of probability $(1 - \varepsilon_K) P(C_k)$
\begin{equation}\label{sumestimate}
\frac{M \sqrt{{\rm ln} n}}{n^{\frac{3}{2}}} \sum_{j \neq i} \frac{1}{|x_i - x_j|} \leq \varepsilon_n.
\end{equation}
Also, (\ref{sumestimate}) holds on the subset of $C_k \cap \frac{b \sqrt{{\rm ln} n}}{\sqrt n} \leq x_i \leq \frac{c \sqrt{{\rm ln} n}}{\sqrt n}$ of probability \break $(1 - \varepsilon_K) P(C_k \cap \frac{b \sqrt{{\rm ln} n}}{\sqrt n} \leq x_i \leq \frac{c \sqrt{{\rm ln} n}}{\sqrt n})$.
Consequently on the set of probability  \break   $(1 - \varepsilon_K) P(C_k \cap \frac{b \sqrt{{\rm ln} n}}{\sqrt n} \leq x_i \leq \frac{c \sqrt{{\rm ln} n}}{\sqrt n})$
$$
\frac{{\rm Vol}(\prod_{i,j} g(x_{ij}) = z, \frac{b \sqrt{{\rm ln} n}}{\sqrt n} \leq x_i \leq \frac{c \sqrt{{\rm ln} n}}{\sqrt n}, C^c, C_k)}{{\rm Vol}(\prod_{i,j} g(x_{ij}) = z, \frac{b \sqrt{{\rm ln} n}}{\sqrt n} \leq x_i \leq \frac{c \sqrt{{\rm ln} n}}{\sqrt n}, C_k)} \leq \varepsilon_n.$$ 
}
\vskip 0.1in
\noindent{\bf Proof:}
\hskip 0.05in
{By Lemma 8a, on the set of probability $1 - \varepsilon_K$, the relation $|x_j - \frac{j}{\sqrt n} + \frac{\sqrt n}{2}| \leq \frac{{\rm ln}^4 n}{\sqrt n}$ is satisfied for all $j =1, \ldots, n$.}
Then for the cones $C_k$, except for the cones in the set of probability $\sqrt{\varepsilon_K}$, we also have that  $|x_{j} - \frac{j}{\sqrt n} + \frac{\sqrt n}{2}| \leq \frac{{\rm ln}^4 n}{\sqrt n}$ (for all $j$)
 on the set of probability $(1 - \sqrt{\varepsilon_K}) P(C_k)$,.

For a given cone $C_k$ on a set of probability $(1 - \sqrt{\varepsilon_K})P(C_k)$ we have that
$$
\frac{M \sqrt{{\rm ln} n}}{n^{\frac{3}{2}}} \sum_{|j-i| \geq 3 {\rm ln}^4 n} \frac{1}{|x_i - x_j|} \leq o(\frac{1}{n}).
$$
Now,  consider the eigenvalues $x_j$ for $|j-i| \leq 3 {\rm ln}^4 n$, and 
consider a subset $C_{(i)}^{(0)}$ of $C_k$ such that 
$$
C_{(i)}^{(0)} = \{x_j: |x_i - x_{i+1} | \leq \frac{M\,  {{\rm ln}^{2} n}    }{n^{\frac{3}{2}}   }, |x_{j} - x_{j+1}| \geq \frac{M {\rm ln}^8 n}{n^{\frac{3}{2}}} ~{\rm for} ~ |j-i| \leq 3 {\rm ln}^4 n, x_j \in C_k\}, 
$$
as well as a subset $C_{(i)}$ of $C_k$ such that
$$
C_{(i)} = \{ x_j: |x_i - x_{i+1} | \leq \frac{M {\rm ln}^8 n}{n^{\frac{3}{2}}}, |x_j - x_{j+1}| \geq \frac{M {\rm ln}^8 n}{n^{\frac{3}{2}}} ~{\rm for} ~ |j-i| \leq 3 {\rm ln}^4 n, x_j \in C_k \}.
$$
Divide $C_{(i)}^{(0)}$ into disjoint neighborhoods $\delta(C_{(i)}^{(0)})$ such that in each $\delta(C_{(i)}^{(0)})$, $\Delta x_i = \Delta x_{i+1} = \Delta \cdot \frac{M\,{\rm ln}^{2} n}  {n^{\frac{3}{2}}} $ and $\Delta x_j = \Delta \cdot \frac{\delta}{n^{\frac{1}{2} + \varepsilon}}$, with $\Delta = O(\frac{1}{n^{2 + \varepsilon}})$ and $\delta \leq \frac{1}{{\rm ln} n}$.

Similarly divide $C_{(i)}$ into disjoint neighborhoods $\delta(C_{(i)})$ such that in each neighborhood, $\Delta x_i = \Delta x_{i+1} = \Delta \cdot \frac{M {\rm ln}^8 n}{n^{\frac{3}{2}}}$ and $\Delta x_j = \Delta \cdot \frac{\delta}{n^{\frac{1}{2} + \varepsilon}}$.
Notice that this way have equal number of neighborhoods $\delta(C_{(i)}^{(0)})$ and $\delta(C_{(i)})$ and a natural correspondence between them. We can see then 
$$
{\rm Vol}(\delta(C_{(i)}^{(0)})) = O\left(\frac{M\,  {{\rm ln}^{2} n}}{n^{\frac{3}{2}}}\right) \prod_{j \neq i} |x'_i - x_j| \prod_{j,k \neq i} |x_j - x_k| \prod_{j\neq i} \Delta x_j \Delta x'_i,
$$
and
$$
{\rm Vol}(\delta(C_{(j)})) = O(\frac{M {\rm ln}^8 n}{n^{\frac{3}{2}}}) \prod_{j \neq i} |x_i - x_j| \prod_{j,k \neq i} |x_j -x_k| \prod_{j \neq i} \Delta x_j \Delta x_i.
$$
(We used $x_i'$ in the product $\prod_{j,k \neq i} |x_i' - x_j|$ to emphasize that the values $x_i$ are different in $\delta(C_{(i)}^{(0)})$ and $\delta(C_{(i)})$.)
\noindent{We observe that} 
$$
\prod_{j \neq i} |x_j' - x_i| = e^{O\left(\frac{M {\rm ln}^8 n}{n^{\frac{3}{2}}}\right) \sum_{|j - i | \leq 3 {\rm ln}^4 n} \frac{1}{|x_i - x_j|}} 
\prod_{ j \neq i} |x_i - x_j| \leq
O({\rm ln}^4 n) \prod_{j \neq i} |x_i - x_j|.
$$
Therefore we obtain that
$$
{\rm Vol}(\delta(C_{(i)}^{(0)})) \leq \frac{1}{{\rm ln}^{12} n} {\rm Vol}(\delta(C_{(i)})).
$$
Summing over $\delta(C_{(i)})$ we obtain that
$$
{\rm Vol} (C_{(i)}^{(0)}) \leq \frac{1}{{\rm ln}^{12} n} {\rm Vol} (C_{(i)}).$$
Since $\sum_{ij} f(x_{ij})$ is the same on $C_{(i)}^{(0)}$ and on $C_{(i)}$ up to $\varepsilon_n$ (see the arguments in the proof of Lemma 8b), we obtain that
\begin{eqnarray}
&& P(C_{(i)}^{(0)}) = \sum P(\delta(C_{(i)}^{(0)})) = \sum \int_{\delta(C_{(i)}^{(0)})} e^{- \sum_{ij} f(x_{ij})} d {\rm Vol} \cr 
&& \leq \frac{1}{{\rm ln}^{12} n} \sum \int_{\delta(C_{(i)})} e^{- \sum_{ij} f(x_{ij})} d {\rm Vol}  \leq \frac{1}{{\rm ln}^{12} n} P(C_{(i)}). \qquad\qquad
\end{eqnarray}
Now similarly consider the set $C_{(i)}^{(1)}$ such that $|x_i - x_{i+1}| \leq \frac{M\,{{\rm ln}^{2} n}    }{n^{\frac{3}{2}}  },$ and there exists $x_{j'}$ with $|j' - i| \leq 3 {\rm ln}^4 n$ such that $\frac{M\, {{\rm ln}^{2} n}  }{n^{\frac{3}{2}} } \leq |x_{j'} - x_{j'+1}| \leq \frac{M {\rm ln}^8 n}{n^{\frac{3}{2}}}$ and for other $x_j$, $|x_{j} - x_{j+1}| \geq \frac{M {\rm ln}^8 n}{n^{\frac{3}{2}}}$.

Consider a small neighborhood $\delta(C_{(i)}^{(1)})$ inside $C_{(i)}^{(1)}$, such that $\Delta x_i = \Delta x_{i+1} \leq \Delta \cdot \frac{M\,{{\rm ln}^{2} n}}{n^{\frac{3}{2}} }$, $\Delta x_{j'} = \Delta x_{j'+1} \leq \Delta \cdot \frac{M {\rm ln}^8 n}{n^{\frac{3}{2}}}$, $\Delta x_j \leq \Delta \cdot \frac{\delta}{n^{\frac{1}{2}+ \varepsilon}}$.
 Consider a corresponding small neighborhood $\delta(C_{(i)})$.
Then we obtain that
$$
{\rm Vol}(\delta(C_{(i)}^{(1)})) \leq 3 {\rm ln}^4 n
 \frac{M\, {{\rm ln}^{2} n} }{n^{\frac{3}{2}}}
\frac{M {\rm ln}^8 n}{n^{\frac{3}{2}}} 
\prod_{j \neq i} |x_j  - x'_i| \prod_{j \neq j'} |x_{j} - x'_{j'}| \prod_{j,k \neq j', i} |x_j - x_k| \prod_{j \neq i,j'} \Delta x_j \Delta x'_i \Delta x'_{j'}.$$
Here the notation  $x'_i$ and $x'_{j'}$ is used in $\delta(C_{(i)}^{(1)})$ to emphasize that $x_i$ and $x_{j'}$ are different from those in $\delta(C_{(i)})$.
Then we obtain
$$ \prod_j |x_j - x'_{i'}| = e^{O(\varepsilon_n)} \prod_j |x_j - x_i| $$

$$\prod_{j \neq i, j'} |x_j - x'_{j'}| = e^{O(\frac{M {\rm ln}^8 n}{n^{\frac{3}{2}}}) \sum{\frac{1}{|x_j - x_{j'}|}}} \prod_{j \neq i, j'} |x_j - x_{j'}| \leq O({\rm ln}^4 n) \prod_{j \neq i, j'} |x_j - x_{j'}|.$$
Therefore
$$
{\rm Vol}(\delta(C_{(i)}^{(1)})) \leq M {\rm ln}^{14}n \frac{1}{\delta n^{1-\varepsilon}} {\rm Vol}(\delta(C_{(i)})). 
$$
Similarly to the arguments above we obtain
$$
P(C_{(i)}^{(1)}) \leq \frac{M {\rm ln}^{14}n}{\delta n^{1 - \varepsilon}} P(C_{(i)}).
$$
Repeating such arguments for the sets 
$C_{(i)}^{(2)}$, $C_{(i)}^{(3)}$ etc. (where set $C_{(i)}^{(k)}$ 
is a set such that $|x_i - x_{i+1}| \leq \frac{M {{\rm ln}^2 n}}{n^{\frac{3}{2}}}$ 
and there exist $k$ eigenvalues $x_{j'}$ with $|j' - i| \leq 3 {\rm ln}^4 n$ 
such that $\frac{M\, {\rm ln}^{2} n}{n^{\frac{3}{2}} } \leq |x_{j'} - x_{j'+1}| 
\leq \frac{M {\rm ln}^8 n}{n^{\frac{3}{2}}}$ and for other $j$ such that 
$|j - i| \leq 3 {\rm ln}^4 n$, $|x_j - x_{j+1}| \geq \frac{M {\rm ln}^8 n}{n^{\frac{3}{2}}}$), 
we obtain
$$
P(C_{(i)}^{(0)} \cup C_{(i)}^{(1)} \cup C_{(i)}^{(2)} \cup \ldots ) \leq \frac{{\rm const}}{{\rm ln}^{10} n} P(C_{(i)}).
$$
(With a little more work,
 we can show that the condition $ \frac{M\,{{\rm ln}^{2} n}}{n^{\frac{3}{2}} }\leq |x_{j'} - x_{j'+1}| $ can be removed).
Therefore probability that $|x_i - x_{i+1}| \leq \frac{M\,{{\rm ln}^2 n}}{n^{\frac{3}{2}}}$ is less than $\varepsilon_n P(C_k)$.
The arguments above go through without any changes on the set $\frac{b \sqrt{{\rm ln} n}}{\sqrt n} \leq x_i \leq \frac{c \sqrt{{\rm ln} n}}{\sqrt n} \cap C_k$.

We derive then that $P\left(\exists \,\, i \, \,{\rm such \, that}\, |x_i - x_{i+1}| \leq \frac{M\,{{\rm ln}^{2} n}}{n^{\frac{3}{2}}}\right) \leq \varepsilon_n.$
Therefore
$$ P\left(C^c\right) = P\left( \sum {\frac{M \sqrt{{\rm ln} n}}{n^{\frac{3}{2}}} \, \frac{1}{|x_i - x_j|} \geq \varepsilon_n}\right)\,\, \leq \,\, \sum_k {\frac{M \,\sqrt{{\rm ln}\,n}}{n^{\frac{3}{2}}}\,\frac{n^{\frac{3}{2}}}{M\,{k {\rm ln}^2 n}}= \frac{\sqrt{{\rm ln}\,n}\,{{\rm ln} n}}{{{\rm ln}^2 n}}\approx \frac{1}{\sqrt{{\rm ln}\,n}}} \leq \varepsilon_n $$  
Using the results of Lemma 10 we derive that for some $z_1$ determined by mean value theorem 
\begin{eqnarray}
&& \frac{{\rm Vol}(\frac{b \sqrt{{\rm ln} n}}{\sqrt n} \leq x_i \leq \frac{c \sqrt{{\rm ln} n}}{\sqrt n}, C^c, V_k, C_k)}{{\rm Vol}(\frac{b \sqrt{{\rm ln} n}}{\sqrt n} \leq x_i \leq \frac{c \sqrt{{\rm ln} n}}{\sqrt n}, V_k, C_k)} \cr 
&& = \frac{{\rm Vol}(\prod_{i,j} g(x_{ij}) = z_1, \frac{b \sqrt{{\rm ln} n}}{\sqrt n} \leq  x_i \leq  \frac{c \sqrt{{\rm ln} n}}{\sqrt n}, C^c, V_k, C_k)}{{\rm Vol}(\prod_{i,j} g(x_{ij}) = z_1, \frac{b \sqrt{{\rm ln} n}}{\sqrt n} \leq x_i \leq \frac{c \sqrt{{\rm ln} n}}{\sqrt n}, V_k, C_k)}\;(1+\varepsilon_n).
\end{eqnarray}

\newpage
Since by Lemma 10, 
\begin{eqnarray}
&& \frac{{\rm Vol}(\prod_{i,j} g(x_{ij}) = z_1, \frac{b \sqrt{{\rm ln} n}}{\sqrt n}\frac{r_1}{r} \leq x_i \leq \frac{c \sqrt{{\rm ln} n}}{\sqrt n}\frac{r_1}{r}, C^c, V_k, C_k)}{{\rm Vol}(\prod_{i,j} g(x_{ij}) = z_1, \frac{b \sqrt{{\rm ln} n}}{\sqrt n}\frac{r_1}{r} \leq x_i \leq \frac{c \sqrt{{\rm ln} n}}{\sqrt n}\frac{r_1}{r}, V_k, C_k)}
\cr && =
\frac{{\rm Vol}(\prod_{i,j} g(x_{ij}) = z, \frac{b \sqrt{{\rm ln} n}}{\sqrt n} \leq x_i \leq \frac{c \sqrt{{\rm ln} n}}{\sqrt n}, C^c, V_k, C_k)}{{\rm Vol}(\prod_{i,j} g(x_{ij}) = z, \frac{b \sqrt{{\rm ln} n}}{\sqrt n} \leq x_i \leq \frac{c \sqrt{{\rm ln} n}}{\sqrt n}, V_k, C_k)}(1+\varepsilon_n),
\end{eqnarray}
\noindent{we obtain that}
$$
\frac{{\rm Vol}(\prod_{i,j} g(x_{ij}) = z, \frac{b \sqrt{{\rm ln} n}}{\sqrt n} \leq x_i \leq \frac{c \sqrt{{\rm ln} n}}{\sqrt n}, C^c, V_k, C_k)}{{\rm Vol}(\prod_{i,j} g(x_{ij}) = z, \frac{b \sqrt{{\rm ln} n}}{\sqrt n} \leq x_i \leq \frac{c \sqrt{{\rm ln} n}}{\sqrt n}, V_k, C_k)} \leq \varepsilon_n.
$$
\noindent{Lemma is proved.}
\vskip 0.05in
\noindent{\bf Remark:} \hskip 0.1in 
Before proceeding  with the 
 partitioning of the space of $\{x_{i,j}\}_{i,j=1, \dots n}$ into the cones $C_m$, let us introduce additional  relations which are used in the Lemmas that follow:\hskip 0.1in
For the matrix with elements $x_{ij}$,  we can represent each element as 
$$
 x_{ij} = \sum_{k =1}^{n} x_k \xi_{ki} \xi_{kj}
$$
where $\xi_{ij}$ are the $\frac{n(n-1)}{2}$ angular variables, such that $\sum_k \xi_{ki}^2 =1$ and $\sum_k \xi_{ki} \xi_{kj} = \delta_{ij}$.
As is shown in Mehta (see \cite{Meh}) 
$$ \prod_{ij} d x_{ij} = \prod_{ij} |x_i - x_j| \prod_{j} d x_j \prod_{ij} O_{ij} \prod_{ij} d \xi_{ij}, $$
where $O_{ij}$ depends only on the angular variables.
 
In order to construct the partition of ${\bf R}^{\frac{n(n+1)}{2}}$ into cones $C_m$ (and a set $B$ of negligible probability) such that  
\begin{eqnarray}\label{conevolumeratio}
\frac{{\rm Vol}(\frac{b \sqrt{{\rm ln} n}}{\sqrt n} \leq x_i \leq \frac{c \sqrt{{\rm ln} n}}{\sqrt n}, S_K, C_m)}{{\rm Vol}(S_K, C_m)} = \frac{{\rm Vol}(\frac{b \sqrt{{\rm ln} n}}{\sqrt n} \leq x_i \leq \frac{c \sqrt{{\rm ln} n}}{\sqrt n}, S_K)}{{\rm Vol}(S_K)} (1 + \varepsilon_n)
\end{eqnarray}
it is sufficient to construct the cones satisfying this condition in the set of eigenvalues $x_1 \leq \ldots \leq x_n$, and then take rectangles in the space of $\xi_{ij}$ (and rescale them by all $\eta > 0$ to make cones) so that in each cone $C_m \cap S_r$ for all $i$ and $j$ each $\Delta \xi_{ij} \leq \frac{1}{n^{3+ \varepsilon}}$ (this can be made smaller if needed), then the ratio of volumes  in (\ref{conevolumeratio}) stays the same, because the $\xi_{ij}$ part in each integral is the same in the numerator and denominator and therefore cancels. 
\vskip 0.1 in 
\noindent{\bf Lemma~6:}
\vskip 0.05 in
{\it
The set $x_1 \leq x_2 \leq \ldots \leq x_n$ can be divided into the cones $C_k$ (and a set $B$ of negligible probability) such that in each cone $x_i$ changes from $-\frac{N \delta}{2 n^{\frac{1}{2} + \varepsilon}}$ to $\frac{N \delta}{2 n^{\frac{1}{2} + \varepsilon}}$ and the length of each $C_k \cap S_r$ is $\frac{N^{\frac{3}{2}} \delta}{2 n^{\frac{1}{2} + \varepsilon}}$ in the long direction and less than $\frac{\delta}{n^{\frac{1}{2} + \varepsilon}}$ in each perpendicular direction, where $N \leq n^{\frac{1}{3}}$ and $\delta \leq \frac{1}{{\rm ln} n}$. Also in each cone $C_k$,
$$
\frac{{\rm Vol}(\frac{b \sqrt{{\rm ln} n}}{\sqrt n} \leq x_i \leq \frac{c \sqrt{{\rm ln} n}}{\sqrt n}, S_K, C_k)}{{\rm Vol}(S_K, C_k)} = \frac{{\rm Vol}(\frac{b \sqrt{{\rm ln} n}}{\sqrt n} \leq x_i \leq \frac{c \sqrt{{\rm ln} n}}{\sqrt n}, S_K)}{{\rm Vol}(S_K)}(1 + \varepsilon_{K,n}).
$$
}
\vskip 0.1 in
\noindent{\bf Proof:}
\hskip 0.05in
Let $|\lambda| \leq \frac{N \delta}{2 n^{\frac{1}{2} + \varepsilon}}$.
Divide the set $x_1 \leq x_2 \leq \ldots \leq x_n$ into cones in the following way. Consider all the points $(x_j)$ on a particular $S_r$ such that $S_r \subset S_K$. If at $x_i = \frac{i}{\sqrt n} - \frac{\sqrt n}{2}$, all $x_j$'s (with $|j-i| \leq \sqrt n$) have the property that $x_{j} - x_{j-1} \geq \frac{1}{n^{\frac{1}{2} + \varepsilon}}$, then draw a line through this point on which take $x_j' = x_j + \lambda(1 - \frac{|j-i|}{N})$ for $|j-i| \leq N$ and $x_j' = x_j$ for $|j-i| > N$.  If for $x_i = \frac{i}{\sqrt n} - \frac{\sqrt n}{2}$ there exist $j$ such that $|x_{j} - x_{j-1}| \leq \frac{1}{n^{\frac{1}{2} + \varepsilon}}$ then draw a line through this point such that for those $x_j$ and $x_{j-1}$, $x_{j}' - x_{j-1}' = x_{j} - x_{j-1}$ and for other $x_j$  we have as before that $x_j' = x_j + \lambda(1 - \frac{|j' - i|}{N})$ where $j'$ is the number of eigenvalue $x_j$ ($j' \geq i$) such that $x_j - x_{j-1} \leq \frac{1}{n^{\frac{1}{2}+\varepsilon}}$ (i.e. omitting all the eigenvalues which distance less than $\frac{1}{n^{\frac{1}{2}+\varepsilon}}$ between the neighboring eigenvalues). Take such lines distance $\frac{\delta}{n^{\frac{1}{2} + \varepsilon}}$ apart in the direction of each $x_j$ at $x_i = \frac{i}{\sqrt n} - \frac{\sqrt n}{2}$ to be the boundaries of the cones ${\tilde C}_k$. To obtain the cone take the rescalings by all possible $\eta > 0$ -- or connect all the points of the parallelepiped just constructed with the origin.  Construct the cones consequently: first construct one cone, then, using its boundaries, construct neighboring cones. For the cones where there exist $j$ such that $x_{j} - x_{j-1}  \leq \frac{1}{n^{\frac{1}{2} + \varepsilon}}$ let the boundaries of the cones be given by the lines such that for those $j$, $x_{j}' - x_{j-1}' = x_{j} - x_{j-1}$ (and therefore these would also be the boundaries of the neighboring cones). Therefore if a cone has boundaries given by non-parallel lines, the smallest distance between the boundaries in any $x_j$ direction is $\frac{\delta}{2 n^{\frac{1}{2} + \varepsilon}}$.
Now in the cones ${\tilde C}_k$ we have that
\begin{eqnarray}\label{ratioCk}
\frac{{\rm Vol}(\frac{b \sqrt{{\rm ln} n}}{\sqrt n} \leq x_i \leq \frac{c \sqrt{{\rm ln} n}}{\sqrt n}, S_r, {\tilde C}_k)}{{\rm Vol}(S_r, {\tilde C}_k)} = 
\frac{\int \ldots \int_{S_r \cap {\tilde C}_k} \prod_{j \neq i} d x_j'  \int_{\frac{b \sqrt{{\rm ln} n}}{\sqrt n}}^{\frac{c \sqrt{{\rm ln} n}}{\sqrt n}} d x_i' \prod_{j \neq k}|x_j' - x_k'|}{\int \ldots \int_{S_r \cap {\tilde C}_k} \prod_{j \neq i} d x_j' \int_{- \frac{N \delta }{2 n^{\frac{1}{2} + \varepsilon}}}^{\frac{N \delta}{n^{\frac{1}{2} + \varepsilon}}} d x_i' \prod_{j \neq k} |x_j' - x_k'|}.
\end{eqnarray}
We shall now show that if $p$ is the number of $x_j$'s such that $x_{j+1} - x_j \leq \frac{1}{n^{\frac{1}{2} + \varepsilon}}$ then for some $|C_n'| \leq {\rm const}$, $|C_n''| \leq {\rm const}$ 
$$
\prod_{j \neq k} |x_j' - x_k'| = \prod_{j \neq k} |x_j - x_k| e^{\lambda (N+p) n^{\frac{1}{2} + \varepsilon} {\rm ln} n C_n ' + \lambda (N +p) n^{\frac{1}{2} + \varepsilon}C_n''\,+\, 6C_n''' \lambda N n^{\frac{1}{2}+\varepsilon}\, {\rm ln}^4 n}.
$$
Indeed, in Lemmas~7, 8a and 8b it is shown that with probability $1 - \varepsilon_K$ the number $p$ of $x_j$ such that $x_{j} - x_{j-1} \leq \frac{1}{n^{\frac{1}{2} + \varepsilon}}$ is less than $n^{\frac{1}{2} - \varepsilon} ~{\rm ln}^5 n$.

(Because we are going to use factors $C_n \leq {\rm const}$ in the estimate, this amounts to that if $x_{j} - x_{j-1} \leq \frac{1}{n^{\frac{1}{2} + \varepsilon}}$ then we can treat them in the estimate as if  they actually coinside, and if $x_{j} - x_{j-1} > \frac{1}{n^{\frac{1}{2} + \varepsilon}}$ then we can treat them in the estimate as if the distance between them is equal to $\frac{1}{n^{\frac{1}{2} + \varepsilon}}$.)

Now by $x_j$ and $x_k$ we shall mean eigenvalues which are distance more than $\frac{1}{n^{\frac{1}{2} + \varepsilon}}$ from the neighboring eigenvalues (for those eigenvalues $x_j' = x_j + \lambda(1-\frac{|j'-i|}{N})$ if $|j'-i| \leq N$), and later in the calculation we shall take into account other $x_j$'s which are distance less than $\frac{1}{n^{\frac{1}{2} + \varepsilon}}$ from the neighboring eigenvalues.

Now suppose $i \leq j' \leq i+N$. If $i \leq k' \leq j' \leq i+N$ then $x_j' - x_k' = x_j - x_k - \frac{\lambda}{N}(j'-k')$. Therefore
$$
|x_j' - x_k'| = |x_j - x_k| e^{\frac{\lambda}{N} \frac{|j'-k'|}{|x_j - x_k|}},
$$ 
If $j' \leq k' \leq i+N$ then $x_j' - x_k' = x_j - x_k + \frac{\lambda}{N}(j'-k')$.
Then
$$
|x_j' - x_k'| = |x_j - x_k| e^{\frac{\lambda}{N} \frac{|j'-k'|}{|x_j - x_k|}} 
$$
If $i-N \leq k' \leq i$ then $x_j' - x_k' = x_j - x_k -\frac{\lambda}{N}(j' + k' - 2i)$. Therefore
$$
|x_j' - x_k'| = |x_j - x_k| e^{-\frac{\lambda}{N}\frac{(j'+k'-2i)}{|x_j-x_k|}}. 
$$
If $i+N \leq k' \leq n$  then $|x_j' - x_k'| = |x_j - x_k + \lambda(1 - \frac{|j'-i|}{N})|$. Therefore
$$
|x_j' - x_k'| = |x_j - x_k| e^{-\frac{\lambda}{N} \frac{(N -j'+i)}{|x_j - x_k|}}.
$$ 
If $1 \leq k' \leq i - N$, then $x_j\,' = x_j + \lambda(1-\frac{j\,'-i}{N})$ and $x_k' = x_k$. Then
$$
|x_j\,'-x_k'| = |x_j - x_k + \lambda(1-\frac{j\,'-i}{N})| = |x_j - x_k| e^{\frac{\lambda}{N}\frac{(N-j\,'+i)}{|x_j - x_k|}}. 
$$
Therefore we obtain
\begin{eqnarray}
&&{\prod_{i \leq j\,' \leq i+N}\prod_{k' \neq j\,'} |x_j'- x_k'|}= {\prod_{i \leq j\,' \leq i+N} \prod_{k' \neq j\,'} |x_j - x_k|e^{\sum_{i \leq j\,' \leq i+N} S'_{n,j},C_n}} , \cr
{\rm where}\;\; |C_n| &\leq&  1,\;\; {\rm and}\;\;  \qquad \qquad \cr 
S'_{n,j} &=& \sum_{i \leq k' \leq i+N}  \frac{\lambda}{N}{\frac{|j\,'-k'|}{|x_{j\,'}-x_{k'}|}} + \sum_{i-N \leq k' \leq i} \frac{\lambda}{N} \frac{|2i - j\,' - k'|}{|x_{j\,'} - x_{k'}|} \cr
 &+& \sum_{k' \leq i - N} \frac{\lambda}{N}  
\frac{|N - j\,' + i|}{|x_{j\,'} - x_{k'}|} 
 + \sum_{k' \geq i+N} \frac{\lambda}{N} 
\frac{|N - j\,' + i|}{|x_{k'} - x_{j\,'}|}.
\end{eqnarray}
Now we shall take into account next to some points $x_j$ there can be several points $x_k \geq x_j$ if $j \geq i$ such that $x_{k} - x_{k-1} < \frac{1}{n^{\frac{1}{2} + \varepsilon}}$ (or $x_k \leq x_j$ if $j < i$ such that $x_{k+1} - x_k \leq \frac{1}{n^{\frac{1}{2} + \varepsilon}}$). Let $p_1$, $p_2$, ..., $p_q$ with $p_1 + \ldots + p_q = p$ be the respective numbers of such ``close'' eigenvalues at different $j$'s.
Suppose that next to the given $x_j$ there are $p_j$ such ``close'' points.  Then we obtain that for these points we can use exactly the same estimates we used for $x_j$. Therefore instead of $S_{n,j}'$ we obtain $S_{n,j}$ such that
\begin{eqnarray}\label{snestimates}
S_{n,j} &=& \sum_{k:i \leq k' \leq i+N} p_k  \frac{\lambda}{N}{\frac{|j'-k'|}{|x_{j'}-x_{k'}|}}
+ \sum_{k:i-N \leq k' \leq i} p_k  \frac{\lambda}{N} \frac{|j'+k'-2i|}{x_{j'} - x_{k'}} 
\cr &+& \sum_{k: k' \leq i - N} p_k \frac{\lambda}{N} \frac{N - j' +i}{|x_{j'} - x_{k'}|} 
 + \sum_{k: k' \geq i+N} p_k \frac{\lambda}{N} 
 \frac{N - j'+i}{|x_{k'} - x_{j'}|}.
\end{eqnarray}
The sums of interest can be estimated as follows:
\begin{eqnarray}
&&\sum_{j,\,k=i}^{i+N} {p_j p_k} {\frac{|j\,'-k'|}{|x_{j\,'}-x_{k'}|}} \cr
&=&\sum_{j\,'=0}^{i+N} p_j \sum_{|j\,' - k'| \leq 3 {\rm ln}^4 n} p_k  {\frac{|j\,'-k'|}{|x_{j\,'}-x_{k'}|}} + 
\sum_{j}\sum_{q \geq 1} \sum_{3 q {\rm ln}^4 n \leq |j - k| \leq 3 (q+1) {\rm ln}^4 n}  {\frac{|j\,'-k'|}{|x_{j\,'}-x_{k'}|}}\cr
&=&
\sum_{j\,'=i}^{i+N} p_j \sum_{|j\,' - k'| \leq 3 {\rm ln}^4 n} {p_k  \frac{|j\,'-k'|}{|x_{j\,'}-x_{k'}|}} + 
\sum_{j} p_j \sum_{q \geq 1} \sum_{3 q {\rm ln}^4 n \leq |j - k| \leq 3 (q-1) {\rm ln}^4 n} {p_k \frac{|j\,'-k'|}{|x_{j'}-x_{k'}|}} \cr
&=& (p+N) N 3{\rm ln}^4n n^{\frac{1}{2}+\varepsilon} + (p+N)
 N 3{\rm ln}^4n \sum_{q \geq 1} {\frac{\sqrt{n}}{(3q-2){\rm ln}^4 n}} \cr
 &=& (p+N) N 3{\rm ln}^4n n^{\frac{1}{2}+\varepsilon} 
+ (p+N) N 3{\rm ln}n {\sqrt{n}}.
\end{eqnarray}
Note that each $p_k \leq 3 {\rm ln}^4 n$, because by Lemma 8a on the set of $\{ x_{ij} \} $ of probability $1-e^{-{\rm ln}^3 n}$ any interval of the length $\frac{1}{\sqrt{n}}$ cannot contain more than $3 {\rm ln}^4 n $ eigenvalues. (Otherwise, eigenvalues will be more than $\frac{{\rm ln}^4 n}{\sqrt n}$ away from their expected positions.)
\vskip 0.01in
\noindent{Also,}
\begin{eqnarray}
&& \frac{\lambda}{N} \sum_{ k' \leq {i - N}} 
{ p_k  \sum_{j\,'=i-N}^{i+N} p_j \frac{N - j\,'+i}{|x_{k'} - x_{j\,'}|}} \cr
 &=& \frac{\lambda}{N}\sum_{j\,'=i-N}^{i+N} {(N - j\,'+i) p_j\;\;
[\sum_{{i-N-3{\rm ln}^4 n} \leq k' \leq {j - N}}\;\frac{p_k}{|x_{k'} - x_{j\,'}|}} \cr
&+&{{\sum_{q \geq 1}}\; {\sum_{i-N-3(q+1){\rm ln}^4 n \leq k' \leq i - N -3 q{\rm ln}^4 n}}}\; 
\frac{p_k}{|x_{k'} - x_{j\,'}|}\;].
\end{eqnarray}

\noindent{Now we can estimate that} 
\begin{eqnarray}
&&\sum_{j'=i-N}^{i+N} {(N - j'+i) p_j}
\sum_{ {|i-N-3{\rm ln}^4 n|} \leq k \leq {|j - N|}} {\frac{p_k}{|x_{k'} - x_{j'}|}} \cr
&=& N (p+2N) C'_N 3{\rm ln}^4 n n^{\frac{1}{2}+\varepsilon} = 3 N (p+2N) C'_N {\rm ln}^4 n n^{\frac{1}{2}+\varepsilon}
\end{eqnarray}
and
\begin{eqnarray}
&&\sum_{j'=i-N}^{i+N} { p_j} {\sum_{q \geq 1}  {(N - j'+i)}   \sum_{ {i-N-3(q+1){\rm ln}^4 n} \leq k \leq {j - N -3 q{\rm ln}^4 n }} 
{\frac{p_k}{|x_k' - x_j'|}}}  \cr
&=&    C''_N N(p+ 2N)  \sum_{q \geq 1} \frac{ 3{\rm ln}^4 n n^{\frac{1}{2}}}{(3q-2) {\rm ln}^4 n } = C''_N N(p+3 N)  {\rm ln}^4 n n^{\frac{1}{2}}
\end{eqnarray}
Therefore
\begin{eqnarray}
\frac{\lambda}{N} \sum_{ k \leq i - N} 
 {p_k  \sum_{j'=i-N}^{i+N} p_j \frac{N - j'+i}{|x_{k'} - x_{j'}|}}=\frac{\lambda}{N} 3 N(p+2N) C'_N {\rm ln}^4 n n^{\frac{1}{2}+\varepsilon} = 3 \lambda (p+2N) C_N {\rm ln}^4 n n^{\frac{1}{2}+\varepsilon}
\end{eqnarray}
And we obtain
\begin{eqnarray}
\prod_{j: i \leq j' \leq i+N}\prod_{k \neq j'} |x_{j'} - x_{k'}|& =& \prod_{k \neq j} |x_j - x_k| 
e^{\sum_{i \leq j \leq i+N} p_j S_{n,j}}. \qquad \qquad 
\end{eqnarray}
Estimating the sum in (\ref{snestimates}) we obtain, 
\begin{eqnarray}\label{prod111}
&&\sum p_jS_{n,j} = \prod_{i \leq j' \leq i+N}\prod_{k' \neq j'} |x_k' - x_j'| \cr
&=& \prod_{i \leq j' \leq i+N} \prod_{k' \neq j'} |x_k - x_j| {e^{{3(p+N)}\, \lambda n^{\frac{1}{2} + \varepsilon} C_n' {\rm ln}^4 n + 3\lambda (p+2 N) C_n'' n^{\frac{1}{2} + \varepsilon} {\rm ln}^4 n}}
e^{ 3C_n'''  N (p+N) n^{\frac{1}{2}}\, {\rm ln} n }
.\qquad 
\end{eqnarray}
Let $i-N \leq j\,' \leq i$. Then we obtain that for $i-N \leq k' \leq i$
$$
|x_j' - x_k'| = |x_j -x_k| e^{\frac{\lambda}{N}\frac{| j\,'- k'|}{|x_{j\,'} - x_{k'}|}}.
$$
For $ i \leq k' \leq i+N$ we obtain
$$
|x_j' - x_k'| = |x_j - x_k|\,
e^{-\frac{\lambda}{N}    \frac{|N+ j\,'-i|}{|x_{j\,'}-x_{k'}|}}. 
$$
For $k' \leq i-N $, we obtain
$$
|x_j' - x_k'| = |x_j - x_k|
e^{\frac{\lambda}{N}\, \frac{|2i- j\,' -k'|}{|x_{k'} - x_{j\,'}|}}.  
$$
For $k' > i+N$, we obtain
$$
|x_j' - x_k'| = |x_j - x_k| e^{-\frac{\lambda}{N}\, \frac{|N+ j\,'-i'|}{|x_{k'}-x_{j\,'}|}}.
$$
Therefore,
$$
\prod_{i-N \leq j' \leq i} \prod_{k' \neq j'} |x_j' - x_k'| = \prod_{i-N \leq j' \leq i}\prod_{k' \neq j'} |x_j - x_k| e^{\sum_{i-N \leq j' \leq i} S_{n,j}' C_n''}
$$ 
where $|C_n''| \leq 1$, and  
\begin{eqnarray}
S_{n,j}' &=& \sum_{k' = i-N}^{i} \frac{\lambda}{N}
\frac{|j\,'-k'|}{|x_{j\,'} -x_{k'}|}
+ \sum_{k' = i+1}^{i+N} \frac{\lambda}{N} \frac{|2i - j\,' -k'|}{|x_{k'}-x_{j\,'}|}  \cr &+& \frac{\lambda}{N} \sum_{k' < i - N} \frac{|N - i + j'|}{|x_j'-x_k'|}  + \frac{\lambda}{N} \sum_{k' > i+N} \frac{|N -i + j\,'|}{|x_{k'} - x_{j\,'}|} \cr & = &
2\,N C_n' \lambda n^{\frac{1}{2} + \varepsilon} +2\, C_n'' \lambda {\rm ln} n n^{\frac{1}{2}}
+ 6C_n''' \lambda N n^{\frac{1}{2}+\varepsilon}\, {\rm ln}^4 n.
\end{eqnarray}
If next to $x_j$ there are $p_j$ ``close'' points, then repeating the estimates as in (\ref{prod111}) we obtain
\begin{eqnarray} \nonumber 
&& \prod_{i-N \leq j' \leq i}\prod_{k' \neq j'} |x_j' - x_k'| \cr
&=&  \prod_{i-N \leq j' \leq i}\prod_{k' \neq j'} |x_j - x_k| e^{6\, (p+N) C_n' \lambda n^{\frac{1}{2} + \varepsilon} {\rm ln} n +  6 (p+N)\lambda C_n'' n^{\frac{1}{2} + \varepsilon}{\rm ln}^4 n + 6C_n''' \lambda N n^{\frac{1}{2}+\varepsilon}{\rm{ln}^4 n}} .
\end{eqnarray}
Suppose $j' < i -N$. Then if $i - N \leq k' \leq i+N$ we obtain
$$
|x_j' - x_k'| = |x_j - x_k| e^{\frac{\lambda}{N} \frac{N - |i-k'|}{|x_{k'}-x_{j'}|}
}.$$
Therefore
$$
\prod_{j' < i - N}\prod_{k'=i-N}^{i+N} |x_j' -x_k'| =\prod_{j' < i-N} \prod_{k' = i - N}^{i+N} |x_j - x_k| e^{\frac{\lambda}{N} \sum_{j' < i-N}p_j \sum_{k' = i-N}^{i+N} p_k\frac{N - |i-k'|}{|x_{k'}-x_{j\,}'|}}.$$
Calculating the exponent we obtain
$$
\frac{\lambda}{N} \sum_{j' < i-N} p_j  \sum_{k'=i-N}^{i+N} p_k \frac{N - |i-k'|}{|x_{k'}-x_{j\,'}|} = 
3C (2N+ p)\lambda n^{\frac{1}{2}} {\rm ln} n +3 C' \lambda (2N
+p) n^{\frac{1}{2}+\varepsilon} {\rm ln}^4n.$$
Suppose $j' > i + N$. Then for $i-N \leq k' \leq i+N$,
$$
\prod_{j' > i+N} \prod_{k' = i -N}^{i+N} |x_j' - x_k'| = \prod_{j' > i+N} \prod_{k' = i - N}^{i+N} |x_j - x_k| e^{\frac{\lambda}{N} \sum_{j' > i + N}\,p_j\, \sum_{k' = i-N}^{i+N}\,p_k\, \frac{N - |i-k'|}{|x_{j\,'}-x_{k'}|}}. 
$$
We estimate the sum in the exponential in the same way as above 
$$
\frac{\lambda}{N} 
\sum_{j' \geq i+N} p_j \sum_{k'=i-N}^{i+N} p_k \frac{N - |i-k'|}{x_j'-x_k'}  = 
C (p+2N) \lambda n^{\frac{1}{2}} {\rm ln}n + 3C_n' \lambda\,(p+2 N) n^{\frac{1}{2}+\varepsilon}\,{\rm ln}^4 n.
$$
Therefore
$$
\prod_{j' \geq i+N} \prod_{k' = i - N}^{i+N} |x_j' - x_k'| = \prod_{j' \geq i+N} \sum_{k' = i - N}^{i+N} |x_j - x_k| 
\,e^{\left(3\, C p \lambda (p+2 N) n^{\frac{1}{2}} {\rm ln}n + 3C_n' \lambda (p+2 N) n^{\frac{1}{2}+\varepsilon}\,{\rm ln}^4 n\right)}.
$$
Combining all of these estimates together we obtain that
\begin{eqnarray}
 \prod_{1 \leq k \leq j \leq n} |x_j' - x_k'| = 
\prod_{1 \leq k \leq j \leq n} |x_j - x_k| e^{S_{n,j}},
\end{eqnarray}
where
\begin{eqnarray}
S_{n,j}&=& 
C_n' \lambda  (p+2N) n^{\frac{1}{2}} {\rm ln} n 
+ 3 C_n''' \lambda (p+2N)  n^{\frac{1}{2}+\varepsilon}{\rm ln}^4 n
\cr &=& C_n \lambda  (p+2N)  n^{\frac{1}{2}+\varepsilon}  {\rm ln}^4 n,\qquad 
\end{eqnarray}
and $C_n$, $C_n'$, $C_n'''$, are $\rm const > 0 $.
If we rename the coordinates inside the cone ${\tilde C}_k$ in terms of $\lambda$ in the following way $x_j' = {\tilde x}_j + \lambda(1 - \frac{|j' - i|}{N})$ for $|j' - i | \leq N$ and $|{\tilde x}_{j} - {\tilde x}_{j-1}| \geq \frac{1}{n^{\frac{1}{2} + \varepsilon}}$ and $x_{j}' - x_{j-1}' = {\tilde x}_{j} - {\tilde x}_{j-1}$ if $|{\tilde x}_{j} - {\tilde x}_{j-1}| \leq \frac{1}{n^{\frac{1}{2} + \varepsilon}}$ (where ${\tilde x}_j$ are the eigenvalues at $\lambda = 0$), we can obtain the coordinates on $C_k \cap S_r$ in the following way
$$
x_j' = ({\tilde x_j} + \lambda(1 - \frac{|j'-i|}{N}))\left(\frac{\sum_j {\tilde x_j}^2}{\sum_{j: |j' - i | > N} {\tilde x_j}^2 + \sum_{j: |j' - i| \leq N} ({\tilde x_j} + \lambda(1 - \frac{|j' - i|}{N}))^2}\right)^{\frac{1}{2}}.
$$
Then we obtain 
\begin{eqnarray}\label{ratioCk1}
&& \frac{{\rm Vol}(\frac{b \sqrt{{\rm ln} n}}{\sqrt n} \leq x_i \leq \frac{c \sqrt{{\rm ln} n}}{\sqrt n}, S_r, {\tilde C}_k)}{{\rm Vol}(S_r, {\tilde C}_k)}=
  \frac{\int \ldots \int_{S_r \cap {\tilde C}_k} \prod_j d x_j' \int_{\frac{b \sqrt{{\rm ln} n}}{\sqrt n}}^{\frac{c \sqrt{{\rm ln} n}}{\sqrt n}} d x_i' \prod_{j,k}|x_j' - x_k'|}{\int \ldots \int_{S_r \cap {\tilde C}_k} \prod_j d x_j' \int_{-\frac{N \delta}{2 n^{\frac{1}{2}+\varepsilon}}}^{\frac{N \delta}{2 n^{\frac{1}{2}+\varepsilon}}} d x_i' \prod_{j,k} |x_j' - x_k'|}~~~~~~ ~ ~ ~\cr 
&=& \frac{\int \ldots \int_{S_r \cap {\tilde C}_k \cap \{\lambda = 0\} } \prod_j d {\tilde x_j} \prod_{j,k}|{\tilde x}_j - {\tilde x}_k| \int_{\frac{b \sqrt{{\rm ln} n}}{\sqrt n}}^{\frac{c \sqrt{{\rm ln} n}}{\sqrt n}} d \lambda~ e^{C_n  \lambda {(p+N)}  n^{\frac{1}{2}+\varepsilon} {\rm ln}^4 n  (1+o(\lambda))}}{\int \ldots \int_{S_r \cap {\tilde C}_k \cap \{\lambda = 0\}} \prod_j d {\tilde x}_j \prod_{j,k} |{\tilde x}_j - {\tilde x}_k| \int_{-\frac{N \delta}{2 n^{\frac{1}{2}+\varepsilon}}}^{\frac{N \delta}{2 n^{\frac{1}{2}+\varepsilon}}} d \lambda~ e^{C_n \lambda{ (p+N)}  n^{\frac{1}{2} + \varepsilon} {\rm ln}^4 n(1+o(\lambda))}}.
\end{eqnarray} 
In order to make this ratio to be equal to 
\begin{equation}\label{ratiosphere}
\frac{{\rm Vol}(\frac{ b \sqrt{{\rm ln} n}}{\sqrt n} \leq x_i \leq \frac{c \sqrt{{\rm ln} n}}{\sqrt n}, S_K)}{{\rm Vol}(S_K)}.
\end{equation}
For each $\lambda = {\rm const}$ that is outside of the interal  $\left[ \frac{b \sqrt{{\rm ln}n}}{\sqrt{n}},  \frac{c \sqrt{{\rm ln}n}}{\sqrt{n}}\right],$
we shall rescale the width of each cone in each 
${\tilde x}_j$ direction for $j$ such that   $|{\tilde x}_j - {\tilde x}_{j - 1}| \geq \frac{1}{n^{\frac{1}{2} + \varepsilon}}$ by a factor 
$$
e^{-\frac{C_n}{n-p} \lambda (p + N) n^{\frac{1}{2} + \varepsilon} {\rm ln}^4 n - \frac{C_{\lambda}}{n-p}}.
$$
Then we claim that the integral in (\ref{ratioCk1}) changes by a factor
$$
e^{-C_n  \lambda  (p+N) n^{\frac{1}{2} + \varepsilon} {\rm ln}^4 n - C_{\lambda}}, 
$$ 
and the ratio in (\ref{ratioCk1}) can be made equal to the ratio in (\ref{ratiosphere}).
(For example, we can take $C_\lambda = 0$ for  
$\lambda {\in} \left[ \frac{b \sqrt{{\rm ln}n}}{\sqrt{n}},  \frac{c \sqrt{{\rm ln}n}}{\sqrt{n}}\right],$
and take $C_\lambda = \alpha'\frac{\sqrt{{\rm ln}n}}{\sqrt{n}}$ for $\lambda$  outside of this interval.
In the latter case, assuming that the ratio (\ref{ratiosphere}) is equal to $g$, we obtain $\alpha'= \frac{(c-b)(1-g)}{g}.$)
Indeed, if each ${\tilde x}_j$ is rescaled, it is shifted by at most 
$$
\frac{\delta}{n^{\frac{1}{2} + \varepsilon}} \frac{C_n  \lambda 
(p+N)  n^{\frac{1}{2} + \varepsilon} {\rm ln}^4 n}{n-p} + \frac{\delta}{n^{\frac{1}{2} + \varepsilon}} \frac{C_{\lambda}}{n-p} = \frac{C_n  \delta \lambda    (p+N){\rm ln}^4 n}{n-p} + \frac{C_{\lambda} \delta}{n^{\frac{1}{2}+\varepsilon} (n-p)}.
$$
By calculations similar to the ones in the beginning of the Lemma we obtain that
\begin{eqnarray}
&& \prod_{j,k}|{\tilde x}_j - {\tilde x}_k + C_n \frac{\delta \lambda  
(p+N) {\rm ln}^4 n}{n-p} + \frac{C_{\lambda} \delta}{n^{\frac{1}{2}+\varepsilon} (n-p)}| \cr &&= \prod_{j,k} |{\tilde x}_j - {\tilde x}_k| e^{ \delta \lambda   (p+N) (n+p)  n^{\frac{1}{2} + \varepsilon} {\rm ln}^5 n \frac{C_n}{n-p}+ (n+p) n^{\frac{1}{2} + \varepsilon} {\rm ln} n \frac{C_{\lambda} \delta}{n^{\frac{1}{2} + \varepsilon} (n-p)}}\cr &&= \prod_{j,k}|x_j - x_k| e^{C_n' \delta \lambda (p+N) n^{\frac{1}{2}+\varepsilon}{\rm ln}^5 n + C_{\lambda} \delta {\rm ln} n}.
\end{eqnarray}
The factor in the exponential is less than 
$$
 \lambda (p+N)  n^{\frac{1}{2} + \varepsilon} {\rm ln}^4 n
$$
provided that
$$ 
\delta {\rm ln} n < 1.
$$

Now to make ratio ({\ref{ratioCk1}}) be equal (\ref{ratiosphere}) we start by 
rescaling a given cone by the factor 
   $$
             e^F = e^{- \frac{C_n}{n-p} (p+N) \lambda n^{\frac{1}{2} + \varepsilon} {\rm ln}^4 n - \frac{C_{\lambda}}{n-p}}
$$
in each $x_j \cap \lambda = {\rm const}$ direction, for $\lambda \notin [\frac{b \sqrt{{\rm ln}n}}{\sqrt n}, \frac{c \sqrt{{\rm ln} n}}{\sqrt n}]$ for $j$ such that $x_j - x_{j-1} \geq \frac{1}{n^{\frac{1}{2} + \varepsilon}}$.
Then after the rescaling (i.e. taking $d x_j = d {\tilde x}_j e^F$ and using the fact discussed above that under this rescaling $\prod_j d x_j \prod_{j,k} |x_j' - x_k'| = e^{(n-p) F} \prod_j d {\tilde x}_j \prod_{j,k} |x_j' - x_k'|$) we obtain
\begin{eqnarray}
&&  \int \ldots \int_{S_r \cap {\tilde C}_k \cap \lambda = {\rm const}} \prod_j d x_j \prod_{j,k} |x_j' - x_k'|\cr && = \int \ldots \int_{S_r \cap {\tilde C}_k \cap \lambda = 0} e^{-C_n \lambda  (p+N)  n^{\frac{1}{2} + \varepsilon} {\rm ln}^4 n  - C_{\lambda}} \prod_j d {\tilde x}_j e^{\delta\, \lambda\, C_n (p+N) n^{\frac{1}{2} + \varepsilon} {\rm ln}^5 n} \prod_{j,k} |{\tilde x}_j - {\tilde x}_k| \cr && = \int \ldots \int_{S_r \cap C_k \cap \lambda = 0} e^{-C_{\lambda}} \prod_{j} d {\tilde x}_j  \prod_{j,k} |{\tilde x}_j - {\tilde x}_k|.
\end{eqnarray}
By choosing $C_{\lambda}$ for each $\lambda$ we can make (\ref{ratioCk1}) equal to (\ref{ratiosphere}). Note that the same $C_{\lambda}$ works in all the cross-sections ${\tilde C}_k \cap S_r$ ($S_r \subset S_K$) of a given cone. We start with one cone ${\tilde C}_k$ and change it as described above. Call the new cone $C_k$. Then using the new boundaries of this cone, we proceed to the neighboring cone and change it as described above until we change all the cones. 
We claim that this is possible because the boundaries 
of each cone as the result change very little.

Indeed, if each cone is rescaled by the factor $e^{ \lambda \frac{p+N}{n-p} {\rm ln}^4 n~ n^{\frac{1}{2} + \varepsilon} {C_n}}$ in the direction of each $x_j \cap \{\lambda = {\rm const}\}$ and if the set $x_1 \leq \ldots \leq x_n$ is divided into $K$ cones of width $\frac{\delta}{n^{\frac{1}{2} + \varepsilon}}$ in the direction $x_j \cap \{\lambda = {\rm const}\}$ the boundary of $K$th cone is shifted in $x_j$ direction by
$$
K \frac{\delta}{n^{\frac{1}{2}+\varepsilon}} (p+N) \lambda {\rm ln}^4n n^{\frac{1}{2} + \varepsilon} \frac{C_n}{n-p}
$$
The requirement that this shift is negligible compared with the width of the cone $\frac{\delta}{n^{\frac{1}{2} + \varepsilon}}$ gives a condition on $N$ and $K$:
$$
 K \lambda (p+N) n^{\frac{1}{2} + \varepsilon}   {\rm ln}^4 n   \frac{C_n}{n-p} << 1.
$$
Taking $|\lambda| \leq \frac{N \delta}{2 n^{\frac{1}{2} + \varepsilon}}$ and $K \leq \frac{\frac{N \delta}{2 n^{\frac{1}{2} + \varepsilon}}}{\frac{\delta}{n^{\frac{1}{2} + \varepsilon}}} = \frac{N}{2}$, we obtain from the previous inequality the condition on $N$:
$$ 
N (p+N) \frac{\delta N}{n^{\frac{1}{2} + \varepsilon}}\, n^{\frac{1}{2} + \varepsilon}\,\frac{C_n {\rm ln}^4 n}{n - p} \leq N^3  \frac{C_n }{n-p}\,{\rm ln}^4 n << 1.
$$
Therefore $ N <<\frac{ n^{\frac{1}{6}}}{{\rm ln}n}$ satisfies the condition. We can take $N \leq \frac{n^{\frac{1}{6}}}{{\rm ln}n}$. For such $N$ Lemma is proved.

\vskip 0.1in
Lemma is proved.
\vskip 0.05 in 
\noindent{\bf Lemma~7:} 
\vskip 0.01 in
{\it For a Gaussian Ensemble with probability $1 - e^{-{\rm ln}^6 n}$, among the eigenvalues $x_{j}$ with $i-\sqrt n \leq j \leq i+\sqrt n$, there are at most $n^{\frac{1}{2}-\frac{\varepsilon}{2}} \frac{2 {\rm ln}^5 n}{\varepsilon}$ eigenvalues such that $x_{j+1} - x_j \leq \frac{1}{n^{\frac{1}{2} + \varepsilon}}$. }
\vskip 0.1 in
\noindent{\bf Proof:}
\hskip 0.05in
Divide the interval $[x_{i-\sqrt n}, x_{i+\sqrt n}]$ into $n^{1 + \varepsilon}$ subintervals $I_k$ of length $\frac{1}{n^{1 + \varepsilon}}$. Then probability that there exist $p$ eigenvalues $x_j$ such that $x_{j+1} \in I_k'$ and $x_j \in I_k'$ for some $k'$  is 
\begin{eqnarray}\label{est}
&& \frac{(n^{1+\varepsilon})!}{p! (n^{1 + \varepsilon} - p)!} \int_{I_1} d x_1 \int_{I_1} d x_2 \ldots \int_{I_{p}} d x_{2p}  {\rm det}(\frac{{\rm sin}(x_i \sqrt n - x_j \sqrt n)}{(x_i \sqrt n - x_j \sqrt n)})_{1 \leq i,j \leq 2p} \nonumber \cr
&\leq & \frac{n^{(1 + \varepsilon)p}}{p!} (2p!) (\frac{1}{n^{2 + 2\varepsilon}})^{p} \leq \frac{\sqrt{2}}{n^{(1+\varepsilon)\, p}}\, \left(\frac{4}{e}\right)^p .
\end{eqnarray} 
Similarly we obtain that for all $1 \leq k \leq n^{\frac{1}{2} - \varepsilon}$ the following holds: for a given $k$, probability that there exist $p$ eigenvalues $x_j$ such that $x_j \in I_k'$ (for some $k'$) and $x_{j+1} \in I_{k' + k}$ can be estimated as
\begin{eqnarray}\label{est'}
&& \frac{(n^{1 + \varepsilon})!}{p! (n^{1 + \varepsilon} - p)!} \int_{I_1} d x_1 \int_{I_{k+1}} d x_2 \ldots \int_{I_{k+p}} d x_{2p} {\rm det} (\frac{{\rm sin}(x_i \sqrt n - x_j \sqrt n)}{(x_i \sqrt n - x_j \sqrt n)})_{1 \leq i, j \leq 2p} \nonumber \cr &\leq & \frac{n^{(1 + \varepsilon) p}}{p!} (2p)! \left(\frac{1}{n^{2 + 2 \varepsilon}}\right)^p \leq \frac{{\sqrt{2}}}{n^{(1\,+\,\varepsilon)\, p}} \, \left(\frac{4}{e}\right)^p.
\end{eqnarray}
Therefore we derive that   
\begin{eqnarray}
P(\exists \;\;{\rm at\;least} \;\;\; p\;\;j\,'s\;\;\;{\rm such\; that}\;\;\;
\frac{k-1}{n^{1 + \varepsilon}}\; \leq \; |x_{j+1} - x_j| \; \leq \; \frac{k+1}{n^{1+\varepsilon}}) \; \leq \; \frac{1}{n^{\varepsilon p}} \leq \frac{\sqrt{2}}{n^{(1+\varepsilon)\, p}}\, \left(\frac{4}{e}\right)^p .\qquad\qquad
\end{eqnarray}
If $p = \frac{2 {\rm ln}^5 n}{\varepsilon}$ then this probability is less than $e^{-2 {\rm ln}^6 n}$. 
If in the set $x_{i - \sqrt n} \leq x_j \leq x_{i + \sqrt n}$ there are at least $\frac{2}{\varepsilon} n^{\frac{1}{2} - \varepsilon} {\rm ln}^5 n$ eigenvalues such that $|x_{j+1} - x_j| \leq \frac{1}{n^{\frac{1}{2} + 2 \varepsilon}}$, then there exists a $k \leq n^{\frac{1}{2} - \varepsilon}-1$ such that for at least $\frac{2}{\varepsilon} {\rm ln}^5 n$ eigenvalues $x_i$ such that
$$\frac{k-1}{n^{1+ \varepsilon}} \leq x_i - x_{i-1} \leq \frac{k+1}{n^{1+\varepsilon}}.$$ 
Thus this set has probability at most $e^{- 2{\rm ln}^6 n}$.
Therefore with probability $1 - e^{-{\rm ln}^6 n}$ there are at most $\frac{2}{\varepsilon} n^{\frac{1}{2}-\varepsilon} {\rm ln}^5 n$  eigenvalues on  $[x_{i-\sqrt n}, x_{i+\sqrt n}]$ with distances $x_{j+1} - x_j \leq \frac{1}{n^{\frac{1}{2} + 2\varepsilon}}$.
The estimates are done here for GUE. However the estimates in (\ref{est}) and in (\ref{est'}) are sufficiently rough to also hold for GOE. 
\vskip 0.1in
Lemma is proved.
\vskip 0.1in
\noindent{\bf Lemma~8a:} 
\hskip 0.1 in
{\it For all $|x_j| << \sqrt {n}$ with probability $1 - \varepsilon_{K,n}$, $|x_j - \frac{j}{\sqrt n} + \frac{\sqrt n}{2}| \leq \frac{{\rm ln}^4 n}{\sqrt n}.$}
\vskip 0.1 in
\noindent{\bf Proof:}
\hskip 0.1in
Consider a particular $x_k$ and the integral
\begin{eqnarray}\nonumber
&&P\left(|x_k- \frac{k}{\sqrt n} + \frac{\sqrt n}{2}| \geq \frac{{\rm ln}^4 n}{\sqrt n}\right)\cr&=&
\int e^{- \sum_{i,j} f(x_{i,j})} {\rm Vol}(\sum_{i,j} x_{ij}^2 = r^2, |x_k - \frac{k}{\sqrt n} + \frac{\sqrt n}{2}| \geq \frac{{\rm ln}^4 n}{\sqrt n}, S_K) d h.
\end{eqnarray} 
For a Gaussian Ensemble it was derived by Gustavsson {\cite{Gu}} that for $|x_k| << \sqrt n$
$$
P_{gaus}(|x_k - \frac{k}{\sqrt n} + \frac{\sqrt n}{2}| \geq \lambda_k) \leq e^{- \frac{c_k \lambda_k^2 n}{2{\rm ln} n}}, \qquad {\rm where}~ c_k \leq 4.
$$
Therefore by the same methods as in Proposition~2 we can show that
$$
\frac{{\rm Vol}(\sum_{i,j} x_{i,j}^2 = r^2, |x_k - \frac{k}{\sqrt n} + \frac{\sqrt n}{2}| \geq \lambda_k, S_K)}{{\rm Vol}(\sum_{i,j} x_{i,j}^2 = r^2, S_K)} \leq e^{-\frac{c_k \lambda_k^2 n}{2 {\rm ln} n}}.
$$
Denote by $x_{i,j}^{(0)}$ the value of $x_{i,j}$ if $x_k = \frac{k}{\sqrt n} -\frac{\sqrt n}{2}$. Then we can find $\sum_{i,j} f(x_{i,j}) - \sum_{i,j} f(x_{i,j}^{(0)}).$

Indeed, if $x_k$ is shifted by $\lambda_k$ then $\sum_{j} x_j^2 = \sum_{j} {x_j^{(0)}}^2 + 2 \lambda_k x_k + \lambda_k^2$. If $\lambda_k$ is sufficiently large then several eigenvalues adjacent to $x_k$ are shifted. 
%
%
Therefore we obtain that the total distance between the initial $x_{j}^{(0)}$ and the final $x_j$ is
$$
\sum_j (x_j - x_j^{(0)})^2 =  {{\sum_k}' \lambda_k^2}
$$ 
where the $\sum'$ is the sum only over the shifted eigenvalues.

By Central Limit Theorem, for the points on the surface $\sum_{i,j} f(x_{i,j}) = z$ in the set of probability $1 - \varepsilon_n$, $\sum_{i,j} x_{i,j}^2 = \frac{n(n+1)}{8} \pm Kn$. Therefore over the distance along the sphere of $O(n^{\frac{2}{3} - \delta})$ the distance between the surface and the sphere increases by $O(1)$. By using similar triangles, we obtain that over a distance along the sphere of order $\lambda$, the distance between the surface $\sum_{i,j} f(x_{i,j}) =z$ and the sphere increases at most by $\frac{\lambda}{n^{\frac{2}{3} - \delta}}$ (See Lemma~8c for more details). Let $x_{ij}'$ be the point on the same $S_r$ as the point $x_{ij}^{(0)}$ and on the same radial line as $x_{ij}$. Therefore the surface $\sum_{i,j} f(x_{i,j}) = z'$ which goes through the point $x_{i,j}'$ and the surface $\sum_{i,j} f(x_{i,j}^{(0)}) = z_0$ which goes through the point $x_{i,j}^{(0)}$ cross the radial line which goes through $x^{(0)}_{i,j}$ at distances $r'$ at $x_{ij}$ and $r_0$ at $x_{ij}'$ such that $\frac{r'}{r_0} = 1 + \frac{\sqrt{{\sum_k}'\lambda_k^2}}{n^{\frac{5}{3} - \delta}}.$
Then on this radial line
\begin{eqnarray}\label{radial}
 z_0 = \sum_{i,j} f(x'_{i,j} \frac{r'}{r_0}) & =& \sum_{i,j} f(x'_{i,j}) + \sum_{i,j} f'(x'_{i,j}) x'_{i,j} (\frac{r'}{r_0} -1) + \frac{1}{2} \sum_{i,j} f''({\tilde x}_{i,j}) {\tilde x}_{i,j}^2 (\frac{r'}{r_0} -1 )^2 \cr & =& z' + \sqrt{{\sum_k}' \lambda_k^2} n^{\frac{1}{3} + \delta} + O(\varepsilon_n).
\end{eqnarray}
%
If we split the integral into the integrals over very thin cones $C_{x_j}$ such that in each cone all $x_j$ don't change except for those eigenvalues which move when $x_k$ is shifted by $\lambda_k$. The way these cones are constructed is not important, but for example these cones are constructed as follows. Through each point on the set $x_k = \frac{k}{\sqrt n} - \frac{\sqrt n}{2}$ draw a line such that on this line all the eigenvalues on $[x_k - \frac{{\rm ln}^9 n}{\sqrt n}, x_k + \frac{{\rm ln}^9 n}{\sqrt n}]$ shift together with $x_k$ so that if $x_k$ shifts by $\frac{\alpha {\rm ln}^4 n}{\sqrt n}$ to the right (for $|\alpha| < {\rm ln}^4 n$), the distance between $x_j$ and $x_k$ such that $x_j > x_k$ decreases by a factor $1 - \frac{1}{{\rm ln}^5 n}$ and the distance between $x_j$ and $x_k$ such that $x_{j} < x_k$ increases by a factor $1 + \frac{1}{{\rm ln}^5 n}$. Then the distances between adjacent eigenvalues $x_j$ and $x_{j+1}$ increases by a factor $1 + \frac{1}{{\rm ln}^5 n}$ for $j < k$. Take parallelepipeds on $x_k = \frac{k}{\sqrt n} - \frac{\sqrt n}{2} \cap S_r$ such that in each parallelepiped, each $x_j$ changes by at most $\frac{1}{n^{2 + \varepsilon}}$. Through the boundaries of each parallelepiped draw lines as described above. Then rescale each resulting figure by all $\lambda > 0$ making a cone, make the cones one after another using the boundaries of the previous cone in the construction of the next cone. 

 Below we show that on the set of measure $1 - e^{-\frac{{\rm ln}^7 n}{2}}$
$$
\frac{{\rm Vol}(|x_j -\frac{j}{\sqrt n} + \frac{\sqrt n}{2}| \geq \frac{{\rm ln}^4 n}{\sqrt n}, C_ {x_j}, S_K)}{{\rm Vol}(C_{x_j}, S_K)} \leq e^{- \frac{{\rm ln}^7 n}{2}}.$$
Similarly from Gustavson's result we derive 
\begin{equation} \label{Gusvol1} 
\frac{{\rm Vol}{(|x_k-\frac{k}{\sqrt n}+ \frac{\sqrt n}{2}|
= {\lambda_k}, {S_k}, \sum{x^2_j=r^2}})}{{\rm Vol}(S_k, \sum{x^2_j=r^2})} = e^{-C_k\,\frac{\lambda^2_k \,n}{2 {\rm ln}\,n}}.
\end{equation}
Now we can write 
\begin{eqnarray} \nonumber
&&\frac{{\rm Vol}{(|x_k-\frac{k}{\sqrt n}+ \frac{\sqrt n}{2}|  
= {\lambda_k}, {S_k}, \sum{x^2_j=r^2}})}{{\rm Vol}(S_k, \sum{x^2_j=r^2})} \cr 
&&=\sum_{C_{x_j   }}} 
{\frac{{\rm Vol}{(|x_k-\frac{k}{\sqrt n}+ \frac{\sqrt n}{2}|= {\lambda_k}, {S_k},{C_{x_j}}, \sum{x^2_j=r^2}})}{{\rm Vol}(S_k,{C_{x_j}}, \sum{x^2_j=r^2})}
{\frac{{\rm Vol}(S_k,C_{x_j}, \sum{x^2_j=r^2})}{{\rm Vol}(S_k, \sum{x^2_j=r^2})}}. 
\end{eqnarray}
Let $B'_{\lambda_k}= \cup'\,C_{x_j}$ be a set such that for any $C_{x_j} \in B'$
\begin{equation} \label{GvolB} 
\frac{{\rm Vol}{(|x_k-\frac{k}{\sqrt n}+ \frac{\sqrt n}{2}|= {\lambda_k}, {S_k},{C_{x_j}}, \sum{x^2_j=r^2}})}{{\rm Vol}({C_{x_j}}, S_k, \sum{x^2_j=r^2})} \geq e^{-C_k\,\frac{\lambda^2_k \,n}{4 {\rm ln}\,n}},
\end{equation}
and $B'$ contains all $C_{x_j}$ for which (\ref{GvolB}) is true, then we must have that 
\begin{equation} 
\frac{{\rm Vol}{(\cup'\, C_{x_j}, {S_k},  \sum{x^2_j=r^2}})}{{\rm Vol}( S_k, \sum{x^2_j=r^2})} \leq e^{-C_k\,\frac{\lambda^2_k \,n}{4 {\rm ln}\,{n}}}.
\end{equation}
Integrating over all values of $\lambda_k \geq \lambda$ we obtain that
\begin{equation} 
\frac{{\rm Vol}{(\cup_{\lambda_k \geq \lambda  }B'_{\lambda_k}, {S_k},  \sum{x^2_j=r^2}})}{{\rm Vol}( S_k, \sum{x^2_j=r^2})} \leq \int_{\lambda}^{\infty}\, d\,\lambda_{k} \, e^{-C_k\,\frac{\lambda^2_k \,n}{4 {\rm ln}\,{n}}}\leq e^{-C_k\,\frac{\lambda^2 \,n}{4 {\rm ln}\,{n}}}  ,
\end{equation}

We can show that
on the cones except in a set of a small probability, there are less than $n^{\frac{4}{3} - \delta}$ eigenvalues on the interval $[x_k - \frac{{\rm ln}^9 n}{\sqrt n}, x_k + \frac{{\rm ln}^9 n}{\sqrt n}]$. Therefore $$\frac{c_k \lambda_k^2 n}{2 {\rm ln} n} > 2\sqrt{{\sum_k}' \lambda_k^2} n^{\frac{1}{3} - \delta}.$$
Indeed, from Gustavsson's result it follows that the relative volume of such set where there are more than $n^{\frac{4}{3} - \delta}$ eigenvalues on $[x_k - \frac{{\rm ln}^9 n}{\sqrt n}, x_k + \frac{{\rm ln}^9 n}{\sqrt n}]$ is less than $e^{-\frac{1}{2 {\rm ln} n} n^{\frac{8}{3} - 2 \delta}}$ and by arguments such as in the Lemma 8c it follows that the probability of this set is less than $\varepsilon_K$.

 Probability of each such cone $C_{x_j}$ (outside of the set of probability $\varepsilon_K$) is therefore
\begin{eqnarray}
  && P(C_{x_j}) = \int_{C_{x_j}} \int_{\lambda_k} d \lambda_k e^{-\sum_{ij} f(x_{ij})} {\rm Vol}(\sum_{ij} x_{ij}^2 = r^2, S_K, x_k = \lambda_k) dh \cr && \leq \int_{\lambda_k} d \lambda_k \int_{C_{x_j}} e^{-\frac{c_k \lambda_k^2 n}{2 {\rm ln} n} + \sqrt{{\sum_k}' \lambda_k^2} n^{\frac{1}{3}+\delta}} e^{-\sum_{ij} f(x_{ij}^{(0)})} {\rm Vol}(\sum_{ij} x_{ij}^2 = r^2, S_K) dh \cr 
&&= \int_{C_{x_j}} e^{-\sum_{ij} f(x_{ij}^{(0)}) + o(1)} {\rm Vol}(\sum_{ij} x_{ij}^2 = r^2, S_K) dh.
\end{eqnarray}

Substituting this into the integral and using the fact that $\sum P(C_{x_j}) = 1$, we obtain
\begin{eqnarray}
&&P(|x_k - \frac{k}{\sqrt n} + \frac{\sqrt n}{2}| \geq \frac{{\rm ln}^4 n}{\sqrt n}) \leq \cr && \leq \sum_{C_{x_j}} \int_{\frac{{\rm ln}^4 n}{\sqrt n}}^{\infty} d \lambda_k \int e^{-\frac{ c_k \lambda_k^2 n}{2 {\rm ln} n}} e^{\sqrt{{\sum_k}'\lambda^2_k} n^{\frac{1}{3} + \delta}}   e^{-\sum_{i,j} f(x_{i,j}^{(0)})} {\rm Vol}(\sum_{i,j} x_{i,j}^2 = r^2, S_K) d h \cr && \leq e^{-{\rm ln}^7 n} \sum_{C_{x_j}} \int e^{- \sum_{i,j} f(x_{ij}^{(0)}) + o(1)} {\rm Vol} (\sum_{i,j} x_{ij}^2 = r^2, S_K) dh \cr && \leq e^{- {\rm ln}^7 n}.
\end{eqnarray}

Now we shall prove that on the set of cones $C_{x_j}$ such that there exists $k$ such that
$$
 \frac{{\rm Vol}(|x_k - \frac{k}{\sqrt n} + \frac{\sqrt n}{2} | \geq \frac{{\rm ln}^4 n}{\sqrt n}, C_{x_j}, S_K)}{{\rm Vol}(C_{x_j}, S_K)} \geq e^{- \frac{1}{2}{\rm ln}^7 n}
$$
has a small probability.
Indeed, from Gustavsson's result it follows that the volume of the set $B$ such that 
$$
\frac{{\rm Vol}(|x_k - \frac{k}{\sqrt n} + \frac{\sqrt n}{2} | \geq \frac{{\rm ln}^4 n}{\sqrt n}, B, S_K)}{{\rm Vol}(B, S_K)} \geq e^{-\frac{1}{2} {\rm ln}^7 n}$$
is less than $e^{- \frac{1}{2} {\rm ln}^7 n}$.

%

Consider the cones $C_{x_j} \subset B$ such that 
$$
\frac{{\rm Vol}(|x_k - \frac{k}{\sqrt n} + \frac{\sqrt n}{2}| \geq \frac{{\rm ln}^4 n}{\sqrt n}, C_{x_j}, S_K)}{{\rm Vol}(
C_{x_j}, S_K)} \geq e^{- \frac{1}{2}{\rm ln}^7 n}.
$$
Let $B_{\alpha}$ be a subset of a set $B$ consisting of several cones which share a common boundary 
with the neighboring cone in $ B_\alpha$, i.e. $B_\alpha$ is a larger  cone.
Now make a cone $\tilde{B_\alpha}$ by taking the cone $B_\alpha$ and rescale each boundary in $x_j$
direction by $\frac{{\rm ln}^4\,n}{n^2}\,x_j$ for each $i,j$. Then 
$${\rm Vol}( \tilde{B_\alpha} \, \cap \, S_R)= e^{{\rm ln}^4\,n}\,{\rm Vol}({B_\alpha} \, \cap \, S_R)$$
Compare $B_\alpha \, \cap \, S_R$ and $\tilde{B_\alpha} \, \cap \, S_R $.
Divide $B_\alpha \, \cap \, S_R$ into the nested parallepipeds such that each bigger one $B_\alpha^{k+1}$ is the rescaling by $1+\frac{{\rm ln}^2\,n}{n^2}$ of the previous $B_\alpha^k$.
Then we claim that on the bigger set ${\tilde B}_{\alpha}$ the sum $\sum_{i,j} f(x_{ij})$ differs by at most $\varepsilon_n$ on the set of probability $(1 - \varepsilon_K)P({\tilde B}_{\alpha})$. 

Indeed, consider small rectangles on the boundary of the set $B_{\alpha}$ such that each small rectangle is centered on the point in $B_{\alpha}$, the small rectangles are non-overlapping, and all together they cover all the points in ${\tilde B}_{\alpha} \cap B_{\alpha}^c$. Then in such rectangle, expand $\sum_{i,j} f(x_{ij})$ about its center $x_{ij}^{(1)}$ (where by the center we mean the point such that ${\rm E} (x_{ij} - x_{ij}^{(1)}) = 0$ for all $i,j$ with respect to the probability conditioned on this rectangle):
$$
\sum_{i,j} f(x_{ij}) = \sum_{i,j} f(x_{ij}^{(1)}) + \sum_{i,j} f'(x_{ij}^{(1)}) (x_{ij} - x_{ij}^{(1)}) + \varepsilon_n.
$$  
Consider the probability measure conditioned on this rectangle. Then by the Central Limit Theorem on the set of probability $1 - \varepsilon_n$ with respect to the conditional measure and therefore on the set of probability $(1 - \varepsilon_n) P({\tilde B}_{\alpha})$ with respect to the original probability measure $$\sum_{i,j} f'(x_{ij}^{(1)}) (x_{ij} - x_{ij}^{(1)}) = O(n) {\rm max }(x_{ij} - x_{ij}^{(1)}) = O(n) \frac{{\rm ln}^4 n}{n^2} \leq \varepsilon_n.$$
Repeating the same argument for each $B_\alpha^k$ we obtain same result that $\sum f(x_j)$ on $B_\alpha^{k+1}$ differs from $\sum f(x_j)$ on $B_\alpha^{k}$ by at most $2\,\varepsilon_k$.

Therefore we obtain that 
$$
P(B_{\alpha}) \leq e^{-{\rm ln}^4 n} P({\tilde B}_{\alpha}).
$$
Since the constructed sets ${\tilde B}_{\alpha}$ do not overlap, we obtain that
$$
P(B) \leq e^{- {\rm ln}^4 n} P({\tilde B}) \leq e^{- {\rm ln}^4 n}.
$$
Lemma is proved.
\vskip 0.1in
\noindent{\bf Lemma~8b:} 
\vskip 0.01in
{\it On the set of probability $1 - \varepsilon_{K,n}$,  out of the eigenvalues $x_j$ with $|j-i| \leq \sqrt n$ there are at most $n^{\frac{1}{2} - \frac{\varepsilon}{2}} {\rm ln}^5 n$ eigenvalues $x_j$ such that $x_{j+1} - x_j \leq \frac{1}{n^{\frac{1}{2} + \varepsilon}}$.}
\vskip 0.1in
\noindent{\bf Proof:}
\hskip 0.1in
Suppose on $[-1,1]$ there are exactly $p_1 n^{1/2-\varepsilon/2}$ (with $p_1 \in {\bf N}$) intervals $[x_{j_k}, x_{j_{k}+1}]$ $k = 1, \ldots p_1 n^{\frac{1}{2}-\frac{\varepsilon}{2}}$ such that $x_{j_k+1} - x_{j_k} \leq \frac{1}{n^{\frac{1}{2}+\varepsilon}}$ and for $j \neq j_k$ $x_{j+1}-x_j > \frac{1}{n^{\frac{1}{2} + \varepsilon}}$.

We start the construction by finding $p_1 n^{\frac{1}{2}-\frac{\varepsilon}{2}}$ intervals $[x_{i_{k'}}, x_{i_{k'+1}}]$ such that $x_{i_{k'}} > x_{j_k+1}$ and $[x_{i_{k'}}, x_{i_{k'}+1}]$ is the first interval to the right of $[x_{j_k}, x_{j_k+1}]$ such that $x_{i_{k'}+1}-x_{i_{k'}} \geq \frac{\varepsilon}{n^{\frac{1}{2}}}.$ (It is easy to show that on the set of probability $1-\varepsilon_n$ all such intervals have the property that $i_{k'}-j_k \leq 3 {\rm ln}^4 n$ and  $x_{i_{k'}}-x_{j_k+1} \leq \frac{ 3 {\rm ln}^4 n}{\sqrt n}$. Indeed, by Gustavsson, on the set of probability $1 - \varepsilon_n$, we have that on $[-1, 1]$, $|x_i - x_i^{exp}| \leq \frac{{\rm ln}^4 n}{\sqrt n}$, $\forall i$ and therefore if we assume the contrary, if all $3 {\rm ln}^4 n$ e.v. are $\leq \frac{\varepsilon}{\sqrt n}$ away from the nearest e.v. then $x_j - x_i \leq \frac{\varepsilon (j - i)}{\sqrt n}$ and since $x_j^{exp} - x_i^{exp} = \frac{j-i}{\sqrt n}$ we obtain $|x_j - x_j^{exp}| + |x_i - x_i^{exp}| \geq \frac{(1-\varepsilon)(j-i)}{\sqrt n}$.  Therefore for $\varepsilon \leq \frac{2}{3}$ we must have $j - i \leq 3 {\rm ln}^4 n.$)

Let $S$ be a particular realization of $x_i$ $i = 1, \ldots n$, we shall call $S$ a "configuration". We construct a configuration $S'$ from configuration $S$ by following the next step by step process: 

{\bf a)} $\forall k= 1, \ldots , p_1 n^{\frac{1}{2}-\frac{\varepsilon}{2}}$ we expand $[x_{j_k}, x_{j_k+1}]$ to $[x'_{j_k}, x'_{j_k+1}]$ in the following way. Let $\alpha_{j_k}=(x_{j_k+1}-x_{j_k}) n^{\frac{1}{2}+\varepsilon}$ (then $\alpha_{j_k} < 1$). We shift $x_{j_k+1}$ to the right to $x'_{j_k+1}$ and leave $x_{j_k}$ in place so that $x_{j_k}=x'_{j_k}$ and $$ x'_{j_k+1} - x'_{j_k} = \frac{\varepsilon^2}{2 {\rm ln}^8 n n^{\frac{1}{2}}} + \frac{\alpha_{j_k} \varepsilon^2}{2 {\rm ln}^8 n n^{\frac{1}{2}}},$$ i.e. $$x'_{j_k+1} = x_{j_k +1} + \frac{\varepsilon^2 (1 + \alpha_{j_k})}{2 {\rm ln}^8 n n^{\frac{1}{2}}} +\frac{ \alpha_{j_k}}{ n^{\frac{1}{2}+\varepsilon}}.$$
Now shift all $x_i$ with $i \in (j_k + 1, i_{k'})$ by exactly the same amount that $x_{j_k+1}$ was shifted i.e. $\frac{\varepsilon^2 (1 + \alpha_{j_k})}{2 {\rm ln}^8 n n^{\frac{1}{2}}} +\frac{ \alpha_{j_k}}{ n^{\frac{1}{2}+\varepsilon}}$ and leave in place all $x_i$ with $i \geq i_{k'} +1.$

We thus obtain a new configuration with $[x'_{j_k}, x'_{j_k +1}]$ of length $\frac{\varepsilon^2 (1 + \alpha_{j_k})}{2 {\rm ln}^8 n n^{\frac{1}{2}}}$ and $[x'_{i_k}, x'_{i_k +1}]$  of length $x_{i_k+1}-x_{i_k} - \frac{\varepsilon^2 (1 + \alpha_{j_k})}{2 {\rm ln}^8 n n^{\frac{1}{2}}} - \frac{\alpha_{j_k}}{n^{1/2+\varepsilon}}.$
Once we expand each interval $[x_{j_k}, x_{j_k+1}]$ $k = 1, \ldots p_1 n^{\frac{1}{2}-\frac{\varepsilon}{2}}$ by this procedure we obtain $S'$ from $S$.

Then we define the neighborhood of $S$, called $\delta(S)$ by letting each $x_i$ in $\delta(S)$ change by at most $(\frac{1}{n^{2+\delta}})$ (where $\delta > 0$). In some $S$ this can affect results of the shifts (for example, if $x_{j_k+1} - x_{j_k} \geq \frac{1}{n^{\frac{1}{2}+\varepsilon}} - \frac{1}{n^{2+\delta}}$), therefore in these cases we exclude the extra configurations from $\delta(S)$. Therefore, we construct $\delta(S)$ by taking all configurations $S$ such that $\forall x_j \in \delta(S)$, $\Delta x_j \leq \frac{1}{n^{2+\delta}}$ and if $x_{j_k+1} - x_{j_k} \leq \frac{1}{n^{\frac{1}{2}+\varepsilon}}$ for some $S$ in $\delta(S),$ then $x_{j_k+1} - x_{j_k} \leq \frac{1}{n^{\frac{1}{2}+\varepsilon}}$ $\forall S$ in $\delta(S).$ Similarly, if $x_{i_{k'+1}}-x_{i_{k'}} \geq \frac{\varepsilon}{n^{\frac{1}{2}}}$ for some $S$ in $\delta(S)$ then $x_{i_{k'+1}} - x_{i_{k'}} \geq \frac{\varepsilon}{n^{\frac{1}{2}}}$ $\forall S \in \delta(S).$

Then the mapping $S \rightarrow S'$ is continuous in $\delta(S)$ (for this paticular choice of $[x_{j_k}, x_{j_k+1}],$ $[x_{i_{k'}}, x_{i_{k'}+1}]$) and $\delta(S)$ maps to $\delta(S')$. We observe that due to the construction of the map $$ \int_{\delta(S')} \prod dx'_j \geq \int_{\delta(S)} \prod dx_j,$$ because the map $S \rightarrow S'$ expands distances.

{\bf b)} Now we compare $\prod |x_i - x_j|\, \vline_{\,S}$ and $\prod |x'_i - x'_j|\, \vline_{\,S'}$ where $S$ maps into $S'.$ 

For $x_i \in [ x_{j_k + 1}, x_{i_{k'}}]$ 
\begin{eqnarray}\nonumber
&& \prod_{j \geq i_{k'}+1} |x'_j - x'_i| = \prod_{j \geq i_{k'}+1} |x_j - x_i - (d' - d)| \cr &=&\prod_{j \geq i_{k'}+1} |x_j - x_i| \prod_{j \geq i_{k'}+1} |1 - \frac{d' - d}{x_j - x_i}| \cr 
&\approx&\prod_{j \geq i_{k'}+1} |x_j - x_i| {\rm exp}(- \sum_{j \geq i_{k'}+1} \frac{d' - d}{x_j - x_i}),
\end{eqnarray}
where $d'-d = \frac{1 + \alpha}{2} \frac{\varepsilon^2}{{\rm ln}^8 n n^{\frac{1}{2}}} - \frac{\alpha}{n^{\frac{1}{2}+\varepsilon}}$ i.e. equal to the amount of shift of $x_i$ since no shift of $x_j.$ 

Now we can calculate an estimate for the sum in the exponential
\begin{eqnarray} \nonumber 
\sum_{j \geq i_{k'}+1} \frac{1}{|x_j - x_i|} &=& \sum_{i_{k'}+1 \leq j \leq i_{k'} + 3 {\rm ln}^4 n} \frac{1}{|x_j - x_i|} + \sum_{i_{k'} + 3 {\rm ln}^4 n < j \leq n} \frac{1}{|x_j - x_i|} \cr &=& \sum_{i_{k'}+1 \leq j \leq i_{k'} + 3 {\rm ln}^4 n} \frac{1}{|x_j - x_i|} + \sum_{k \geq 1} \sum_{i_{k'} + 3 k {\rm ln}^4 n < j \leq i_{k'} + 3(k+1) {\rm ln}^4 n} \frac{1}{|x_j - x_i|} \qquad \cr &=& {\rm \alpha'_1} \frac{3 {\rm ln}^4 n}{|x_{i_{k'}+1} - x_{i_{k'}}|} + {\rm \alpha'_2} \sum_{k \geq 1} \frac{3 {\rm ln}^4 n}{|x_{i_{k'}+3 k {\rm ln}^4 n} - x_{i_{k'}}|} \cr &=& {\rm \alpha'_1} \frac{3 {\rm ln}^4 n}{|x_{i_{k'}+1} - x_{i_{k'}}|} + {\rm \alpha''_2} \sum_{k \geq 1} \frac{3 {\rm ln}^4 n \sqrt{n}}{(3k - 2) {\rm ln}^4 n} \cr  &=&  {\alpha''_1} \frac{3 {\rm ln}^4 n \sqrt{n}}{\varepsilon}(1 + \frac{\varepsilon}{{\rm ln}^8 n}) + \alpha''_2  3 {\rm ln} n \sqrt{n}\cr & =& \alpha''_1 \frac{3}{\varepsilon} ( 1 + \frac{\varepsilon}{{\rm ln}^8 n}) {\rm ln}^4 n \sqrt{n} + \alpha''_2 3 {\rm ln} n \sqrt{n},
\end{eqnarray} 
where $\alpha'_1, \ \alpha'_2, \ \alpha''_1, \ \alpha''_2 \in [0,1].$

We used in the estimate above that: $x_j \geq x_{i_{k'}+1},$ $x_i \leq x_{i_{k'}}$ therefore $|x_j - x_i| \geq |x_{i_{k'}+1} - x_{i_{k'}}|$ and $\frac{1}{|x_j - x_i|} \leq \frac{1}{|x_{i_{k'}+1} - x_{i_{k'}}|}$; also, if we have that $x_j \geq x_{i_{k'}+ 3 k {\rm ln}^4 n}$ and $x_i \leq x_{i_{k'}}$ implies $|x_j - x_i
| \geq |x_{i_{k'}+ 3 k {\rm ln}^4 n} - x_{i_{k'}}|$ and $\frac{1}{|x_j - x_i|} \leq \frac{1}{|x_{i_{k'}+3 k {\rm ln}^4 n} - x_{i_{k'}}|}.$
We also have that $|x_{i_{k'}+ 3 k {\rm ln}^4 n} - x_{i_{k'}}| \geq \frac{(3 k -2) {\rm 
ln}^4 n}{\sqrt n}$ and $\frac{1}{|x_j - x_i|} \leq \frac{\sqrt{n}}{(3 k - 2) {\rm ln}^4 n}.$ In addition, $|x_{i_{k'}+1} - x_{i_{k'}}| \geq \frac{\varepsilon}{\sqrt n} - \frac{\varepsilon^2}{{\rm ln}^8 n  n^{\frac{1}{2}}}$ and $\frac{1}{|x_{i_{k'}+1} - x_{i_{k'}}|} \leq \frac{\sqrt n}{\varepsilon}(1 + \frac{\varepsilon}{{\rm ln}^8 n}).$

Therefore, for the factor in the exponential we obtain
\begin{eqnarray}\nonumber
 \sum_{j \geq i_{k'}} \frac{d' - d}{|x_j - x_i|} = (\frac{1 + \alpha}{2} \frac{\varepsilon^2}{{\rm ln}^8 n  n^{\frac{1}{2}}} - \frac{\alpha}{n^{\frac{1}{2}+\varepsilon}})(\alpha''_1 \frac{3 {\rm ln}^4 n \sqrt{n}}{\varepsilon}(1 + \frac{\varepsilon}{{\rm ln}^8 n}) + \alpha''_2 3 {\rm ln} n \sqrt{n}) = \alpha' \frac{3 \varepsilon}{{\rm ln}^4 n}, 
\end{eqnarray}
 where $\alpha' \in [0, 1].$
And we can see that for $i \in [j_k+1,i_{k'}]$
\begin{equation}\label{estimate1}
\prod_{j \geq i_{k'}+1} |x_j' - x_i'| = \prod_{j \geq i_{k'}+1} |x_j - x_i| {\rm exp}(\alpha' \frac{3 \varepsilon}{{\rm ln}^4 n})
\end{equation}
This allows to derive from the formula ({\ref{estimate1}}) that
\begin{eqnarray}
\prod_{j_k + 1 \leq i \leq i_{k'}} \prod_{j \geq i_{k'}+1} |x_j' - x_i'|&=& 
\prod_{j_k + 1 \leq i \leq i_{k'}} \prod_{j \geq i_{k'}+1} |x_j - x_i|
\prod_{j_k + 1 \leq i \leq i_{k'}} {\rm exp}(\alpha'_i \frac{3 \varepsilon}{{\rm ln}^4 n})\qquad \cr &=&\prod_{j_k + 1 \leq i \leq i_{k'}} \prod_{j \geq i_{k'}+1} 
|x_j - x_i| {\rm exp}(\alpha' \frac{3 \varepsilon}{{\rm ln}^4 n} 3 
{\rm ln}^4 n) \cr &=&  \prod_{j_k + 1 \leq i \leq i_{k'}} \prod_{j \geq i_{k'} + 1}
|x_j - x_i| {\rm exp}(9 \varepsilon \alpha').
\end{eqnarray}
Similarly we estimate
\begin{eqnarray}\label{estimate2}
\prod_{j_k+1 \leq j \leq i_{k'}} \prod_{i \leq j_k} |x_j' - x_i'| &=&
\prod_{i \leq j_{k'}} \prod_{j_k+1 \leq j \leq i_{k'}} |x_j - x_i + d' - d| \cr &=&\prod_{i \leq j_{k'}} \prod_{j_{k}+1 \leq j \leq i_{k'}} |x_j - x_i| 
\prod_{i \leq j_{k'}} \prod_{j_k +1 \leq j \leq i_{k'}} |1 + \frac{d'-d}{x_j - x_i}| \qquad \cr & =& \prod_{i \leq j_{k'}} \prod_{j_{k}+1 \leq j \leq i_{k'}} |x_j - x_i| \beta, 
\end{eqnarray}
where $\beta \geq 1,$
 and that  
\begin{equation}\label{estimate3}
\prod_{i \leq j_k} \prod_{j \geq i_{k'} + 1} |x_j' - x_i'| = 
\prod_{i \leq j_k}
\prod_{j \geq i_{k'}+1} |x_j - x_i|.
\end{equation}	
Combining the estimates (\ref{estimate1}), ({\ref{estimate2}}), (\ref{estimate3}) we obtain 
\begin{equation}\nonumber
\prod_i \prod_j |x_j' - x_i'| = e^{-9 \alpha' \varepsilon} \beta \frac{|x_{j_k + 1}' - x_{j_k}'|}{|x_{j_k+1} - x_{j_k}|} \prod_i \prod_j |x_j - x_i|,
\end{equation}
the ratio is $$ \frac{|x_{j_k+1}' - x_{j_k}'|}{|x_{j_k+1}-x_{j_k}|} = \frac
{\frac{1+\alpha}{2} \frac{\varepsilon^2}{{\rm ln}^8 n n^{\frac{1}{2}}}}{\frac{\alpha}{n^{\frac{1}{2} + \varepsilon}}}= \frac{n^{\varepsilon} \varepsilon^2}{{\rm ln}^8 n} \frac{1 + \alpha}{2 \alpha}.$$

Thus, we find that for each expansion of $[x_{j_k}, x_{j_k + 1}]$ to $[x'_{j_k}, x'_{j_k+1}]$ the product changes by a factor
$$
\prod_i \prod_j |x_j' - x_i'| = e^{-9 \alpha'_k \varepsilon} \beta_k 
\frac{n^{\varepsilon} \varepsilon^2}{{\rm ln}^8 n} \frac{1 + \alpha_k}{2 \alpha_k} \prod_i\prod_j |x_j - x_i|.
$$
Therefore after expanding $p_1 n^{\frac{1}{2} - \frac{\varepsilon}{2}}$ intervals $[x_{j_k}, x_{j_k +1}]$ under the mapping of $S$ to $S'$ the resulting product changes by a factor
\begin{eqnarray}\label{estimate4} 
&\prod_i \prod_j |x_j' - x_i'| &= \prod_{k = 1}^{p_1 n^{\frac{1}{2} - \frac{\varepsilon}{2}}} \left( e^{-9 \alpha'_k \varepsilon} \beta_k \left(\frac{1 + \alpha_k}{2 \alpha_k}\right)\right) \left(\frac{n^{\varepsilon} \varepsilon^2}{{\rm ln}^8 n}\right)^{p_1 n^{\frac{1}{2} - \frac{\varepsilon}{2}}} \cr {\rm Therefore} \cr &\prod_i \prod_j |x'_j - x'_i|  &=
\beta \left(\frac{n^{\varepsilon} \varepsilon^2}{{\rm ln}^8 n}\right)^{p_1 n^{\frac{1}{2} - \frac{\varepsilon}{2}}} \prod_i \prod_j |x_j - x_i|, 
\end{eqnarray}
where we can always pick $\varepsilon >0$ small enough so that $\beta \geq 1.$
\vskip 0.1in
{\bf c)} 
\hskip 0.01in
Comparing $\sum_{i,j} f(x_{i,j})|_{\delta(S)} = z_0$ and $\sum_{i,j} f(x_{i,j}')|_{\delta(S')} = z,$
we can claim that
$$ z_0 = z + \sum_{i,j} f(x_{i,j}') x_{i,j}' \frac{C {\rm ln}^4 n \sqrt{p_1} \sqrt{C_n''}}{n^{\frac{23}{12}-\delta}}.$$
Indeed, when $x_{j_k +1}$ is shifted, we get a total shift of $$ d' - d =
\frac{1 + \alpha}{2} \frac{\varepsilon^2}{{\rm ln}^8 n n^{\frac{1}{2}}} -
\frac{\alpha}{n^{\frac{1}{2} + \varepsilon}}$$ for the eigenvalues
$x_{j_k+1}, \ldots, x_{i_{k'}}$ so at most $3 {\rm ln}^4 n$ eigenvalues.

To estimate the final distance between $S$ and $S'$ we get
$$
\sum_j (x_j' - x_j)^2 = \sum N_k (d' - d)^2 = {\tilde \alpha} 3 {\rm ln}^4 n
p_1 n^{\frac{1}{2}-\frac{\varepsilon}{2}} \left(\frac{\varepsilon^2}{{\rm ln}^8 n  n^{\frac{1}{2}}}\right)^2 = \frac{3 {\tilde \alpha} p_1 \varepsilon^4}{{\rm ln}^{12} n n^{\frac{1}{2} + \frac{\varepsilon}{2}}},$$
where ${\tilde \alpha} \in [0,1],$ $N_k \leq 3 {\rm ln}^4 n$ and $p_1 n^{\frac{1}{2} - \frac{\varepsilon}{2}}$ is the number of intervals $[x_{j_k+1}, x_{i_{k'}}]$ on which the shifts occur. This leads to the distance between $S$ and $S',$ denoted by $d(x, x')$ as $$
d(x, x') = \frac{\sqrt{3{ \tilde \alpha}} \varepsilon^2 \sqrt{p_1}}{{\rm ln}^6 n 
n^{\frac{1}{4}+\frac{\varepsilon}{4}}}.$$ 
By Lemma 9  on the set of measure $1 - \varepsilon_{K,n}$ over the distance
$d(x, x'),$ the distance between the hypersurface $\sum_{i,j} f(x_{i,j}) = z$
and the hypersphere $\sum_{i,j} x_{i,j}^2 = r^2$ intersecting at the point $x$ increases by at most $\frac{d(x,x')}{n^{\frac{2}{3}-\delta}}$ for any $\delta > 0$ and sufficiently large $n \geq n_{\delta}$. Therefore we obtain that  the radial distance between the hypersurface and the hypersphere, i.e. between points $x$ and $x'$, is at most 
\begin{equation}\label{dhyps}
\frac{d(x, x')}{n^{\frac{2}{3} - \delta}} = \frac{\sqrt{3{\tilde \alpha}} \varepsilon^2 \sqrt{p_1}}{{\rm ln}^6n  n^{\frac{11}{12}+\frac{\varepsilon}{4}- \delta }} = d.
\end{equation}
Let assume now that $(x_{i,j}^{(0)})_{i,j}$ be the point of the intersection of $\sum_{i,j} f(x_{i,j}) = z_0$ and $\sum_{i,j} x_{i,j}^2 = r_0^2$, and let $S$ be the eigenvalue configuration corresponding to the point $x^{(0)} = (x_{i,j}^{(0)})_{i,j},$ and $S'$ be the eigenvalue configuration corresponding to the point $x = (x_{i,j})_{i,j}$. 
Let also  $\sum_{i,j} f(x_{i,j}) = z$ to pass through $x.$

To estimate $z = \sum_{i,j} f(x_{i,j})$ we make the following construction: Suppose $x_{i,j}^{(0)}$ is on $S_r$ and $x_{i,j}$ is on $S'_r$. Let $x_{i,j}'$ be a point on the same radial line as $x_{i,j}$ such that $\sum_{i,j} (x_{i,j}')^2 = r_0^2$ (i.e. $\exists \lambda > 0$ such that $x_{i,j} = \lambda x_{i,j}'$ $\forall i, j$.) Let $x_{i,j}'^{(0)}$ be a point on the same radial line as $x_{i,j}$ such that $\sum_{i,j} f(x_{i,j}'^{(0)}) = z_0$ (i.e. $\exists \lambda' > 0$ such that $x_{i,j}'^{(0)} = \lambda' x_{i,j}$ for $\forall i,j.$)

Let $r = \sqrt{\sum_{i,j} (x_{i,j}')^2} = \sqrt{\sum_{i,j} (x_{i,j}^{(0)})^2}$, $r' = \sqrt{\sum_{i,j} (x_{i,j})^2}$ and $r_0' = \sqrt{ \sum_{i,j} {(x_{i,j}'^{(0)})^2}}$. 
Since $r'_0 = C n$ and by the argument in the equation (\ref {dhyps}), we have that
$$ \frac{r'_0}{r} = 1 + \frac{\sqrt{3 \alpha} \varepsilon^2 \sqrt{p_1}}
{{\rm ln}^6 n n^{\frac{23}{12}+\frac{\varepsilon}{4} - \delta}},$$ and  
$$ r'_0 = r + \frac{\sqrt{3 \alpha} \varepsilon^2 \sqrt{p_1} C 
n^{-\frac{11}{12} - \frac{\varepsilon}{4} + \delta}}{{\rm ln}^6 n}.$$
The distance between $S_r$ and $S'_{r}$ is obtained by a simple estimate, in which we denote by $I_J$ the set of $j$'s that are shifted by the map $ S \rightarrow S'$
\begin{equation}\nonumber
r'^2 = \sum_j (x_j + (d' - d)*1_{I_J})^2 = \sum_{i,j} x_{i,j}^2 + \sum_{j \in I_J} (d'-d)^2 + 2\sum_{j \in I_J} x_j (d' - d).
\end{equation}
Therefore
\begin{eqnarray}
r'^2 &=& r^2 + 3 \alpha' p_1 {\rm ln}^4 n n^{\frac{1}{2} - \frac{\varepsilon}{2}}
\frac{\varepsilon^4}{{\rm ln}^{16} n  n } + 6 \alpha'' {\rm ln}^4 n p_1 
\frac{n^{\frac{1}{2} - \frac{\varepsilon}{2}} \varepsilon^2}{{\rm ln}^8 n n^{\frac{1}{2}}} \cr
&=& r^2 + \frac{3 \alpha' p_1 \varepsilon^4}{{\rm ln}^{12}n n^{\frac{1}{2} +
\frac{\varepsilon}{2}}} + \frac{6 \alpha'' p_1 \varepsilon^2}{{\rm ln}^4 n 
n^{\frac{\varepsilon}{2}}}.
\end{eqnarray}
We can easily derive from this equation that 
$$
r' - r = \frac{3 \alpha'' p_1 \varepsilon^2}{{\rm ln}^4 n r 
n^{\frac{\varepsilon}{2}}} \approx \frac{3 \alpha'' p_1 \varepsilon^2}
{C {\rm ln}^4 n n^{1+\frac{\varepsilon}{2}}},
$$
and hence  
\begin{equation}\label{4zozestim}
r' = r_0' + \frac{\sqrt{3 {\tilde \alpha}} \varepsilon^2 \sqrt{p_1} 
C n^{-\frac{11}{12}-\frac{\varepsilon}{4}+\delta}}{{\rm ln}^6 n } =
r_0'(1 + \frac{\sqrt{3 {\tilde \alpha}} \varepsilon^2 \sqrt{p_1}}{{\rm ln}^6 n 
n^{\frac{23}{12} + \frac{\varepsilon}{4} -\delta}}).
\end{equation}
Using the equation (\ref{4zozestim}), we can estimate $z'-z_0$.
Since $x_{ij}'^{(0)} = x_{ij} \frac{r_0'}{r'}$ we have that
\begin{equation}\nonumber
z_0 = \sum_{ij} f({x'_{ij}}^{(0)}) = \sum_{i,j} f(x_{ij}) + \sum_{i,j} 
f'(x_{ij}) x_{ij} (\frac{r_0'}{r} - 1) + {\rm l.o.t.}
\end{equation}
Let $x_{ij}^{(0)}$ be in a set $C$ such that $$ \sum_{i,j} f'(x'_{ij}) x'_{ij} 
\geq {\rm E}(f'(x) x) \frac{n(n+1)}{2} + K n^{1 + \delta}$$ where $x'_{ij}$ is as in the construction described above, and let first demonstrate that $P(C) \leq 
\sqrt{\varepsilon_n}.$

{Note that in the construction of} $x'_{ij}$ the distance $d(x^{(0)}, x)$ and e.v. $x_j$ are 
fixed by the mapping $S \rightarrow S'$ so the only free variables are the angular part
or the direction of the vector $x_{ij}.$
We can see the set of $x'_{ij}$ as a projection of the hypersphere in angular variables with eigenvalues fixed, with the center at $x^{(0)}$ and the radius $d(x^{(0)}, x)$ to $S_{r_0}.$
So, for each $x^{(0)}$ we can consider a "bad set", i.e. the set of directions of $x^{(0)}$ to $x',$ such that for a particular direction
\begin{eqnarray}\nonumber
\sum_{i,j} f'(x_{ij}) x'_{ij} \geq {\rm E} (f'(x) x) \frac{n(n+1)}{2}+K n^{1+\delta},
\end{eqnarray}
for some fixed $\delta > 0,$ $K >> 1$ and $\forall n \geq n_{\delta, K}.$
Denote this "bad set" of $x'$ by $\Lambda^c_{x^{(0)}}.$
Then $P(x' \in \Lambda^c_{x^{(0)}} | x^{(0)}) \geq \sqrt{\varepsilon_n}$ can happen only on the set $C$ of $x^{(0)}$ of probability $P(C) \leq \sqrt{\varepsilon_n}.$ Indeed,
$$ P (x' \; is \; s.t. \; \sum_{i,j} f'(x_{ij}') x_{ij}' \geq {\rm E}(f'(x) x) \frac{n(n+1)}{2} + K n^{1+\delta}) \leq \varepsilon_n$$ by the Central Limit Theorem.

Therefore,
\begin{eqnarray}\label{pleqsqrte}
&& \varepsilon_n  \geq \int P(x \in \Lambda^c_{x^{(0)}}) P(d x^{(0)}) = \cr &=&
\int_{P(\Lambda^c_{x^{(0)}}| x^{(0)}) \leq \sqrt{\varepsilon_n}} 
P(x \in \Lambda^c_{x^{(0)}}| x^{(0)})  P(d x^{(0)}) + 
\int_{P(\Lambda^c_{x^{(0)}}| x^{(0)}) >  \sqrt{\varepsilon_n}}
P(x \in \Lambda^c_{x^{(0)}}| x^{(0)}) P(d x^{(0)}) \cr 
&\geq& \int_{P(\Lambda^c_{x^{(0)}}| x^{(0)}) \geq \sqrt{\varepsilon_n}} 
\sqrt{\varepsilon_n} P(d x^{(0)}).
\end{eqnarray}    
Thus 
$$ P (x^{(0)} \in C) \leq \sqrt{\varepsilon_n}.$$
Now we use this  and the relation  ${x_{ij}^{(0)}}' = x_{ij}' \frac{r_0'}{r'}$ to find that 
$$
z_0 = \sum_{ij} f(x_{ij}' \frac{r_0'}{r'}) = \sum_{ij} f(x_{ij}') + \sum_{ij} f'({x}_{ij}') x_{ij}' (\frac{r_0'}{r'}-1) + ({\rm l.o.t.}).
$$
Considering that from the estimates above we obtain 
$$
\frac{r_0'}{r} = 1 - \frac{C_n'\sqrt{p_1} {\rm ln}^{6} n}{n^{\frac{\varepsilon}{4} +\frac{23}{12} - \delta}},
$$
and  using the fact that by the Central Limit Theorem on the set of probability $1- \varepsilon_n$, $\sum_{ij} f'(x_{ij}') x_{ij}' = n^2 {\rm E} (f'(x) x)$,
we derive that 
$$ 
z_0 = z +  \frac{C'_n \sqrt{p_1} {\rm ln}^{4} n}{n^{\frac{\varepsilon}{4} -\frac{1}{12} - \delta}},
$$
and thus estimating  $z-z_0$.

{\bf d:}
The important part of the estimate is to get all the preimages of $S'$, therefore consider $\delta(S')$ to be a neighborhood of $S'$ where each eigenvalue changes by at most $\frac{1}{n^{2+\delta}}$ and find what $\delta(S)$ map into it.

We start with all $S $ that map into a particular $S'.$ 
For a given $S,$ we define $\delta_S(S)$ so that $\forall x_{ij} \in \delta_S(S)$ $x_{ij}$ changes by at most $\frac{1}{n^{2+\delta}}$ and the inequalities 
$$ |x_{j+1}- x_j| \leq \frac{1}{n^{\frac{1}{2}+\varepsilon}}, \
|x_{j+1} - x_j| \geq \frac{1}{n^{\frac{1}{2}}}$$ remain true everywhere in
 $\delta_S(S)$ for each $j$, (i.e. if for a particular $j'$ we have that for some ${\tilde S} \in \delta_S(S),$ $|x_{j'+1} - x_{j'}| \leq \frac{1}{n^{\frac{1}{2} + \varepsilon}}$ then $ \forall S \in \delta_S(S)$ we have $|x_{j'+1} - x_{j'}
| \leq \frac{1}{n^{\frac{1}{2}+\varepsilon}}.$)

 Now we let $\delta_S(S')$ to be the image of $\delta_S(S)$ under the map $S \rightarrow S'.$
We define the neighborhood of $S'$ corresponding to all the preimages of $S'$ to be   $$\delta(S') = \cap_{S \ {\rm preimages\ of}\ S'} \delta_S (S').$$
Clearly, the intersection is not empty.

We defined $\delta(S)$ to be a preimage of $\delta(S')$ for a particular map $\delta_S(S) \rightarrow \delta_S (S').$

We note that the preimage $S$ of $S'$ (and therefore $\delta(S)$ of $\delta(S')$) is defined uniquely once all the intervals $[x'_{j_k}, x'_{j_k+1}]$ and $[x'_{i_{k'}}, x'_{i_{k'}+1}]$  are specified in $S'.$

Indeed, if $d(x'_{j_k}, x'_{j_k+1})$ is given, then $d(x_{j_k}, x_{j_k+1})$ is
determined by 
$$ d(x'_{j_k}, x'_{j_k+1}) = \frac{1 + n^{\frac{1}{2}+\varepsilon} d(x_{j_k}, 
x_{j_k+1})}{2} \frac{\varepsilon^2}{{\rm ln}^8 n n^{\frac{1}{2}}}.$$

Now, if $d(x'_{i_{k'}}, x'_{i_{k'}+1})$ is given, we can find $d(x_{i_{k'}},
x_{i_{k'}+1})$ by taking $$ d(x_{i_{k'}}, x_{i_{k'}+1}) = d(x'_{i_{k'}},
x'_{i_{k'}+1}) - (d(x'_{j_k}, x'_{j_k+1}) - d(x_{j_k}, x_{j_k+1})).$$

Now specifying $[x'_{j_k}, x'_{j_k+1}]$ and $[x_{i_{k'}}, x_{i_{k'}+1}]$ out
of $2 \sqrt{n}$ possible choices for $[x_i, x_{i+1}]$ on $(-1,1)$ requires 
at most $$ N_{p_1} = \frac{(2 \sqrt{n})!}{(p_1 n^{\frac{1}{2} - \frac{\varepsilon}{2}})! (p_1 n^{\frac{1}{2}- \frac{\varepsilon}{2}})! (2 \sqrt{n} - 
2 p_1 n^{\frac{1}{2} - \frac{\varepsilon}{2}})!}$$ possibilities.  

Notice, that given $S'$, we should be able to rule out some of the possibilities
because of the lengths of $[x'_{j_k}, x'_{j_k+1}]$ and $[x'_{i_{k'}}, x'_{i_{k'+1}}].$

Therefore each $\delta(S')$ has at most $N_{p_1}$ preimages. We can now estimate $P(\delta(S))$ from the above information 
$$ P(\delta(S)) = P(S) {\rm Vol}(\delta(S)),$$ where $P(S)$ -- the probability density, can be assumed to be constant on $\delta(S)$ due to the small size of $\delta(S).$ 

We are going to show that 
$$ 
{\rm Vol} (\delta(S')) \geq \sum_{preimages} {\rm Vol} (\delta(S)).$$
Therefore,
\begin{eqnarray}\nonumber
&& P(\delta(S')) = P(S') {\rm Vol}(\delta(S')) \geq  \sum_{premages \ of \
\delta(S')} P(S') {\rm Vol}(\delta(S)) = \cr &=& \sum_{\delta(S): preimages \ of \ 
\delta(S')} \int_{\delta(S)} e^{-\sum_{i,j} f(x_{ij}')} \prod_{i,j} |x_i -
x_j| \prod_i d x_i \cr &=& \sum_{\delta(S): preimages \ of \ \delta(S')} \int_
{\delta(S)} e^{-\sum_{i,j} f(x_{ij}) + O_n} \prod_{i,j} |x_i - x_j| \prod_i 
d x_i, 
\end{eqnarray}
where $O_n = \sum_{i,j} f(x_{ij}') - \sum_{i,j} f(x_{ij}).$

In order for the argument to go through, we shall need to show that
\begin{equation}\label{condition1}
 \sum_{\delta(S): preimages \ of \ \delta(S')} e^{O_n} {\rm Vol} (\delta(S)) 
\leq {\rm Vol}(\delta(S')).
\end{equation}
Instead of ({\ref {condition1}}) it is sufficient to show that
\begin{equation}\label{condition2}
\frac{(2 \sqrt{n})! e^{O_n} {\rm Vol}(\delta(S))}
{((p_1 n^{\frac{1}{2}-\frac{\varepsilon}{2}})!)^2 (2 \sqrt{n} - 2 p_1 
n^{\frac{1}{2}-\frac{\varepsilon}{2}})!} \leq {\rm Vol} (\delta(S')).
\end{equation}
From ({\ref{estimate4}}) we have that
\begin{eqnarray}\nonumber
{\rm Vol} (\delta(S')) &=& \prod_{i,j} |x'_j - x'_i| \int_{\delta(S')}
\prod_i d x'_i * {\rm Angular \  part} \cr &=&
\beta (\frac{\varepsilon^2 n^{\varepsilon}}{{\rm ln}^8 n })^{p_1 n^{\frac{1}{2}
- \frac{\varepsilon}{2}}} \prod_{i,j} |x_j - x_i| 
\int_{\delta(S)} \prod_i d x_i * {\rm Angular \ part} * \frac
{\int_{\delta(S')} \prod_i d x'_i}{\int_{\delta(S)} \prod_i d x_i} =  \cr
&=& \beta (\frac{\varepsilon^2 n^{\varepsilon}}{{\rm ln}^8 n})^{p_1 
n^{\frac{1}{2} - \frac{\varepsilon}{2}}} {\rm Vol} (\delta(S)) * \left(
\frac{\int_{\delta(S')} \prod_i d x'_i}{\int_{\delta(S)} \prod_i d x_i}\right).
\end{eqnarray} 
Therefore (\ref{condition2}) becomes, since $O_n = C_n \sqrt{p_1} {\rm ln}^4 n 
n^{\frac{1}{12} + \delta - \frac{\varepsilon}{4}}$
\begin{equation}\nonumber
\frac{(2 \sqrt{n} )! {\rm exp}(C_n \sqrt{p_1} {\rm ln}^4 n n^{\frac{1}{12} + 
\delta- \frac{\varepsilon}{4}})}{((p_1 n^{\frac{1}{2} - \frac{\varepsilon}{2}})!)^2 
(2 \sqrt{n} - 2 p_1 n^{\frac{1}{2}-\frac{\varepsilon}{2}})!} \leq
\beta \left(\frac{\varepsilon^2 n^{\varepsilon}}{{\rm ln}^8 n}\right)^{p_1
n^{\frac{1}{2} - \frac{\varepsilon}{2}}} 
\left(\frac{\int_{\delta(S')} \prod_i d x'_i}{\int_{\delta(S)} \prod_i dx_i}
\right).
\end{equation}
 Due to the expanding nature of the map $S \rightarrow S'$  we know that
$$ 
\frac{\int_{\delta(S')} \prod_i d x'_i}{\int_{\delta(S)} \prod_i d x_i} \geq 1.
$$
This justifies replacing sufficient condition ({\ref {condition2}}) with ({\ref {condition3}}):
for all $p_1 \geq {\bar p}$, all $n \geq n_0$
\begin{equation}\label{condition3}
\frac{(2 \sqrt{n})! {\rm exp}(C_n \sqrt{p_1} {\rm ln}^4 n n^{\frac{1}{12} +
\delta- \frac{\varepsilon}{4}})}{((p_1 n^{\frac{1}{2}-\frac{\varepsilon}{2}})!)^2 
(2 \sqrt{n} - 2 p_1 n^{\frac{1}{2} - \frac{\varepsilon}{2}})!} \leq
\left(\frac{\varepsilon^2 n^{\varepsilon}}{{\rm ln}^8 n}\right)^{p_1
n^{\frac{1}{2} - \frac{\varepsilon}{2}}}.
\end{equation}
We simplify the left hand side of ({\ref {condition3}}) using Stirling's formula
$$ \frac{(2 \sqrt{n})!}{(2 \sqrt{n} - 2 p_1 n^{\frac{1}{2} - \frac{\varepsilon}{2}})!} = (2 \sqrt{n})^{2 p_1 n^{\frac{1}{2} - \frac{\varepsilon}{2}}}
{\rm exp}(- p_1 n^{- \frac{\varepsilon}{2}}(p_1 n^{\frac{1}{2} - \frac{\varepsilon}{2} } - \frac{1}{2})).$$
We obtain
\begin{eqnarray}\nonumber
&&{ \frac{(2 \sqrt{n})!  {\rm exp}(C_n \sqrt{p_1} {\rm ln}^4 n n^{\frac{1}{12}+ \delta-\frac{\varepsilon}{4}})}
{((p_1 n^{\frac{1}{2} - \frac{\varepsilon}{2}})!)^2 (2 \sqrt{n} - 2 p_1 
n^{\frac{1}{2} - \frac{\varepsilon}{2}})!}   }\cr
&=&
\frac{(2 \sqrt{n})^{2 p_1 n^{\frac{1}{2} - \frac{\varepsilon}{2}}}}
{(2 \pi p_1 n^{\frac{1}{2}-\frac{\varepsilon}{2}})
(p_1 n^{\frac{1}{2} - \frac{\varepsilon}{2}})^{2 p_1 n^{\frac{1}{2} - \frac{\varepsilon}{2}}}}   
\frac{{\rm exp} (- p_1 n^{-\frac{\varepsilon}{2}}(p_1 n^{\frac{1}{2} - \frac{\varepsilon}{2}} - \frac{1}{2})) } 
{{\rm exp}(- 2 p_1 n^{\frac{1}{2}-\frac{\varepsilon}{2}})   }
e^{C_n \sqrt{p_1} {\rm ln}^4 n n^{\frac{1}{12} + \delta - \frac{\varepsilon}{4}}}  \cr
&=&{\frac{1}{\left(2 \pi p_1 n^{\frac{1}{2} - \frac{\varepsilon}{2}}\right) }
\left(\frac{2 \sqrt{n}}{p_1 n^{\frac{1}{2} - \frac{\varepsilon}{2}}}\right)
^{2 p_1 n^{\frac{1}{2}-\frac{\varepsilon}{2}}}
 e^{2 p_1 n^{\frac{1}{2} - \frac{\varepsilon}{2}} - p_1^2 n^{\frac{1}{2} - \varepsilon} + {\frac{1}{2}} p_1 n^{-\frac{\varepsilon}{2}}}
e^{C_n \sqrt{p_1} {\rm ln}^4 n n^{\frac{1}{12} + \delta- \frac{\varepsilon}{4}}} 
}\end{eqnarray} 
The latter expression can easily be simplified further, so that  
 ({\ref {condition3}}) becomes
\begin{eqnarray}\nonumber 
&& \frac{1}{2 \pi p_1 n^{\frac{1}{2} - \frac{\varepsilon}{2}}}
\left(\frac{2 n^{\frac{\varepsilon}{2}} e}{p_1}\right)^{2 p_1 n^{\frac{1}{2}-
\frac{\varepsilon}{2}}} 
e^{-p_1^2 n^{\frac{1}{2} - \varepsilon} +
\frac{p_1 n^{-\frac{\varepsilon}{2}}}{2} + C_n \sqrt{p_1} {\rm ln}^4 n n^{\frac{1}{12} + \delta- \frac{\varepsilon}{4}}} \cr 
 &\leq &
\left(\frac{\varepsilon^2 n^{\varepsilon}}{{\rm ln}^8 n}\right)^{p_1 n^{\frac{1}{2} - \frac{\varepsilon}{2}}}.
\end{eqnarray}
\noindent And can be rewritten as
\begin{eqnarray} \label{condition4}   
\frac{1}{2  \pi p_1 
n^{\frac{1}{2} - \frac{\varepsilon}{2}}}  
\left(\frac{\varepsilon^2 n^{\varepsilon}}{{\rm ln}^8 n}\right)^{-p_1 
n^{\frac{1}{2} - \frac{\varepsilon}{2}}} \left( \frac{2 n^{\frac{\varepsilon}{2} 
}e}{p_1}\right)^{2 p_1 n^{\frac{1}{2} - \frac{\varepsilon}{2}}} \cr \qquad
 e^{- p_1^2 n^{\frac{1}{2} - \varepsilon} + C_n \sqrt{p_1} {\rm ln}^4 n^{\frac{1}{12}+\delta-\frac{\varepsilon}{4}}+ \frac{1}{2} p_1 n^{-\frac{\varepsilon}{2}}} \leq 1. 
\end{eqnarray}
The left hand side of ({\ref{condition4}}) can be simplified as 
\begin{eqnarray} \nonumber
&&{\frac{1}{2 \pi p_1 n^{\frac{1}{2}-\frac{\varepsilon}{2}}}}
\left( \frac{{\rm ln}^4 n}{\varepsilon n^{\frac{\varepsilon}{2}} }
\frac{2 e n^{\frac{\varepsilon}{2}} e^{-\frac{1}{2} p_1 n^{-\frac{\varepsilon}{2}}}}{p_1}\right)^{2 p_1 n^{\frac{1}{2} - \frac{\varepsilon}{2}}} 
e^{\frac{1}{2} p_1 n^{-\frac{\varepsilon}{2}}+ C_n \sqrt{p_1} {\rm ln}^4 n  
n^{\frac{1}{12} + \delta- \frac{\varepsilon}{4}}}  \cr 
&=&{\frac{1}{2 \pi p_1 n^{\frac{1}{2}-\frac{\varepsilon}{2}}}
\left(\frac{2 e {\rm ln}^4 n}{\varepsilon p_1}
 e^{-\frac{1}{2} p_1 n^{-\frac
{\varepsilon}{2}}} \right)^{2 p_1 n^{\frac{1}{2} - \frac{\varepsilon}{2}}}
e^{\frac{1}{2} p_1 n^{-\frac{\varepsilon}{2}} + C_n \sqrt{p_1} {\rm ln}^4 n 
n^{\frac{1}{12}+ \delta- \frac{\varepsilon}{4}}}}.   \nonumber
\end{eqnarray} 
\noindent and ({\ref {condition4}}) becomes
\begin{equation}\label{condition5}
\frac{1}{2 \pi p_1 n^{\frac{1}{2}-\frac{\varepsilon}{2}}}
\left(\frac{2 e {\rm ln}^4 n}{\varepsilon p_1} e^{-\frac{1}{2} p_1 n^{-\frac{
\varepsilon}{2}}}\right)^{2 p_1 n^{\frac{1}{2}-\frac{\varepsilon}{2}}} e^{\frac{1}{2}p_1 n^{-\frac{\varepsilon}{2}} + C_n 
\sqrt{p_1} {\rm ln}^4 n n^{\frac{1}{12}+\delta- \frac{\varepsilon}{4}}} 
 \leq 1.
\end{equation}
From ({\ref{condition5}}) and the discussion above we obtain the sufficient condition to have 
$$ P(\delta S') \geq \sum_{{\rm preimages}} P(\delta S)$$ is satisfied provided that
\begin{equation}\label{condition6}
\frac{1}{2 \pi n^{\frac{1}{2}- \frac{\varepsilon}{2}}} \sum_{p_1 \geq {\bar p}} ( p_1)^{-1}
\left(\frac{2 e {\rm ln}^4 n}{\varepsilon p_1} e^{-\frac{1}{2} p_1 n^{-\frac{\varepsilon}{2}}}\right)^{2 p_1 n^{\frac{1}{2} - \frac{\varepsilon}{2}}} e^{\frac{1}{2} p_1 n^{-\frac{\varepsilon}{2}} + C_n \sqrt{p_1} {\rm ln}^4 n 
n^{\frac{1}{12} + \delta- \frac{\varepsilon}{4}}} \leq \varepsilon_n.
\end{equation}
Now we are going to prove (\ref{condition6}).
 Consider $${\bar p} = \frac{2 e {\rm ln}^4 n}{\varepsilon}, \ 
p_1 = \frac{2 e {\rm ln}^4 n}{\varepsilon} + K, \ K \geq 1.$$
Then we show that the inequality ({\ref {condition6}}) holds.
Indeed, substituting $p{_1}, {\bar p}$ into the left hand side of ({\ref {condition6}}) we obtain
\begin{eqnarray}\label{lhscondition6}
&&\sum_{K \geq 1} \left(2 \pi n^{\frac{1}{2} - \frac{\varepsilon}{2}}\right)^{-1} \left( \frac{2 e 
{\rm ln}^4 n}{\varepsilon} + K\right)^{-1} \left(\frac{2 e {\rm ln}^4 n}{\varepsilon}\left(
\frac{2 e {\rm ln}^4 n}{\varepsilon} +K\right)^{-1}  
\right. \nonumber \cr && \left. \;
*\; e^{-\frac{1}{2}(\frac{2}{\varepsilon} 
e {\rm ln}^4 n + K) n^{-\varepsilon/2}}\right)^{(\frac{4}{\varepsilon} e{\rm ln}^4 n + 2K) n^{1/2 - \varepsilon/2}}
e^{(\frac{e {\rm ln}^4 n}{\varepsilon} + \frac{K}{2}) n^{-\varepsilon/2} 
+ C_n \sqrt{\frac{2}{\varepsilon}e {\rm ln}^4 n + K} {\rm ln}^4 n n^{\frac{1}{12} + \delta- \frac{\varepsilon}{4}}}  \nonumber  \cr 
 &=&
(2 \pi n^{\frac{1}{2}-\frac{\varepsilon}{2}})^{-1} \sum_{K \geq 1}(\frac{2 e {\rm ln}^4 n}{\varepsilon} + K)^{-1} \left((1 + \frac{K \varepsilon}{2 e {\rm ln^4 n}})^{-1} e^{-\frac{1}{\varepsilon} e {\rm ln}^4 n n^{-\frac{\varepsilon}{2}}- \frac{K}{2} n^{-\varepsilon/2}}\right)
^{(\frac{4}{\varepsilon} e {\rm ln}^4 n + 2 K)n^{\frac{1}{2} - \frac{\varepsilon}{2}}} \nonumber \cr 
&&*\; {\rm exp}(\frac{1}{\varepsilon} e {\rm ln}^4 n n^{-\varepsilon/2} +
\frac{1}{2} K n^{-\frac{\varepsilon}{2}} + C_n \sqrt{\frac{2 e}{\varepsilon}} 
{\rm ln}^6 n \sqrt{1 + \frac{K \varepsilon}{2 e {\rm ln}^4 n}}
{n^{\frac{1}{12}+\delta - \frac{\varepsilon}{4}}}) \nonumber  \cr  
&\leq& 
(2 \pi n^{\frac{1}{2} - \frac{\varepsilon}{2}})^{-1}
  \sum_{K \geq 1} (\frac{2}{\varepsilon} e {\rm ln}^4 n + K)^{-1}\nonumber  \cr 
 &&*\;{\rm exp}(- \frac{K \varepsilon}{2 e {\rm ln}^4 n}
(\frac{4 e {\rm ln}^4 n}{\varepsilon} n^{\frac{1}{2} - \frac{\varepsilon}{2}} + 
2 K n^{\frac{1}{2} - \frac{\varepsilon}{2}}) - \frac{4 e^2 {\rm ln}^8 n}{\varepsilon^2} n^{\frac{1}{2} - \varepsilon} - \frac{2 K e}{\varepsilon} {\rm ln}^4 n  n^{\frac{1}{2} - \varepsilon})\nonumber \cr
&&*\; 
{\rm exp} (-\frac{2 K e {\rm ln}^4 n}{\varepsilon} n^{\frac{1}{2} - \varepsilon} -  K^2 n^{\frac{1}{2} - \varepsilon} + \frac{e {\rm ln}^4 n n^{-\frac{\varepsilon}{2}}}{\varepsilon}  + \frac{K}{2} n^{-\frac{\varepsilon}{2}} + C'_n
\frac{\sqrt{2 e}}{\varepsilon} {\rm ln}^6 n \sqrt{1 + \frac{K \varepsilon}{2 e {\rm ln}^4 n}
} n^{\frac{1}{12} + \delta- \frac{\varepsilon}{4}})  \nonumber   \cr  
&=&
(\frac{4 \pi e}{\varepsilon} {\rm ln}^4 n n^{\frac{1}{2} - \frac{\varepsilon}{2}})^{-1}
\sum_{K \geq 1} (1 + \frac{K \varepsilon}{2 e {\rm ln}^4  n})^{-1}
  \cr
 &&*\;   {\rm exp}
[-2 K n^{\frac{1}{2} - \frac{\varepsilon}{2}} - \frac{K^2 \varepsilon}
{e {\rm ln}^4 n} n^{\frac{1}{2}-\frac{\varepsilon}{2}} - n^{\frac{1}{2} - 
\varepsilon}( \frac{4 e^2 {\rm ln}^8 n}{\varepsilon^2} + K^2 + 
\frac{4 K e}{\varepsilon} {\rm ln}^4 n)]  \cr
 &&*\;  
{\rm exp}( (\frac{e {\rm ln}^4 n}{\varepsilon} + \frac{K}{2}) n^{-\frac
{\varepsilon}{2}} + C_n' \frac{\sqrt{2e}}{\varepsilon} {\rm ln}^6 n \sqrt{1 + 
\frac{K \varepsilon} {4 e {\rm ln}^4 n}} n^{\frac{1}{12} + \delta- \frac{\varepsilon}{4}}). \nonumber 
\end{eqnarray}
\vskip 0.05in
\noindent{It is clear that for all $n \geq n_0 >> 1$ and $K \geq 1$ we have that} 
\begin{equation}\nonumber
-(2K + \frac{K^2 \varepsilon}{e {\rm ln}^4 n}) n^{\frac{1}{2} - \frac{\varepsilon}{2}} + C_n\frac{ \sqrt{2 e}}{\varepsilon} {\rm ln}^6 n \sqrt{1 + \frac{K \varepsilon}{
4 e {\rm ln}^4 n}} n^{\frac{1}{12} +\delta- \frac{\varepsilon}{4}} \leq - (K + \frac{K^2 \varepsilon}
{2 e {\rm ln}^4 n}) n^{\frac{1}{2} - \frac{\varepsilon}{2}}.
\end{equation}
This implies the sum in the formula ({\ref {condition6}}) is bounded for all $n \geq n_0$ by
\begin{equation}\nonumber
(\frac{4 e \pi}{\varepsilon} {\rm ln}^4 n\ n^{\frac{1}{2} - \frac{\varepsilon}{2}})^{-1}
\sum_{K \geq 1} {\rm exp}(-K n^{\frac{1}{2}-\frac{\varepsilon}{2}} - \frac{K^2 \varepsilon}{2 e
{\rm ln}^4 n} n^{\frac{1}{2} - \frac{\varepsilon}{2}}) 
<< (\frac{4 e \pi}{\varepsilon})^{-1} e^{-n^{\frac{1}{2} - \frac{\varepsilon}{2}}}({\rm ln}^4 n\ n^{\frac{1}{2}-\frac{\varepsilon}{2}})^{-1}.
\end{equation}
Therefore for $p_1 \geq C {\rm ln}^5 n$, probability that there are more than $p_1 \,n^{\frac{1}{2}-\frac{\varepsilon}{2}}$ pairs of eigenvalues such that $|x_{j+1} -x_j| \leq \frac{1}{n^{\frac{!}{2}+\varepsilon}}$ is
\begin{eqnarray}\nonumber
&&\sum_{p_1 \geq C {\rm ln}^5 n} P(\delta S) = \sum_{p_1 \geq C {\rm ln}^5 n}
\sum_{\delta S'} \sum_{preimages \ of \ \delta S'} P(\delta S)\big|_{\delta S
\rightarrow \delta S'} \cr &\leq& \sum_{p_1 \geq C {\rm ln}^5 n} 
\frac{ ( 2 \sqrt{n})!}{((p_1 n^{\frac{1}{2} - \frac{\varepsilon}{2}})!)^2 (2 \sqrt{n} - 2 p_1 
n^{\frac{1}{2} - \frac{\varepsilon}{2}})!}\sum_{\delta S'} P(\delta S) \big|_{
\delta S \rightarrow \delta S'} \cr  
&=& \sum_{p_1 \geq C {\rm ln}^5 n}
\sum_{\delta S} \frac{(2 \sqrt{n})! \int_{\delta S} e^{-\sum_{i,j} f(x_{ij})}
\prod_{i,j} |x_i - x_j| \prod_i d x_i}{((p_1 n^{\frac{1}{2} - 
\frac{\varepsilon}{2}})!)^2 (2 \sqrt{n} - 2 p_1 n^{\frac{1}{2} - 
\frac{\varepsilon}{2}})!}\big|_{\delta S \rightarrow \delta S'} \cr
&\leq & \sum_{p_1 \geq C {\rm ln}^5 n} \frac{(2 \sqrt{n})! \sum_{\delta S'} \int_{\delta S'} 
e^{-\sum_{i,j} f(x_{ij}') + C_n \sqrt{p_1} {\rm ln}^4 n 
n^{\frac{1}{12}+\delta}} (\frac{\varepsilon n^{\varepsilon}}{{\rm ln}^8 n})^
{-p_1 n^{\frac{1}{2}-\varepsilon}}}{((p_1 n^{\frac{1}{2} - \frac{\varepsilon}{2}})!)^2 
(2 \sqrt{n} - 2 p_1 n^{\frac{1}{2} - \frac{\varepsilon}{2}})!}
\prod_{i,j} |x_i'- x_j'| \prod_j d x_j' \cr & \leq & \sum_{K \geq 1}
{\rm exp} (-K n^{\frac{1}{2}-\frac{\varepsilon}{2}} - \frac{K^2 \varepsilon}
{2 e {\rm ln}^4 n} n^{\frac{1}{2}-\frac{\varepsilon}{2}}) \sum_{\delta S'}
\int_{\delta S'} e^{-\sum_{i,j} f(x'_{ij}) } \prod_{i,j} |x_i' - x_j'| \prod_j 
d x_j'   \cr & \leq & 2 e^{-n^{\frac{1}{2}-\frac{\varepsilon}{2}}} 
\sum_{\delta S'} P(\delta S') \leq 2 e^{-n^{\frac{1}{2} - \frac{\varepsilon}{2}}}.
\end{eqnarray}
Therefore we obtain:
\begin{eqnarray}
&&{\rm Probability}\; \left(\,{\rm  that}\;\;\right.
{\rm number \;\; of \;\; pairs} \;\;\; [x_j, x_{j+1}]\;\;\;{\rm such \;\; that}\;  \cr 
&&\;\;\;\;|x_{j+1}-x_j|\;\; \leq \;\; n^{-\frac{1}{2} - \varepsilon}  \left.
\;\;\;{\rm greater \;\; than}\;\;\;\;C\; {\rm ln}^5 n \;
n^{\frac{1}{2}-\frac{\varepsilon}{2}} \right)  \cr   
&&\leq  \varepsilon_n \Rightarrow_{n \rightarrow \infty} 0. 
\end{eqnarray}   
\vskip 0.1in 
The Lemma is proved.
\vskip 0.1 in
\noindent{\bf Lemma~8c:}
\vskip 0.01 in
{\it The radial distance between the surface $\sum_{ij} f(x_{ij}) = z$ and $\sum_{ij} x_{ij}^2 = r^2$ at the distance $\lambda$ along the hypersphere from an intersection point is less than $\frac{\lambda}{n^{\frac{2}{3} - \delta}}$ on the set of probability $1 - \varepsilon_{K,n}$.}
 \vskip 0.1in
 \noindent{\bf Proof:}
 \hskip 0.1in
{Consider the surface}
$$\sum_{ij} f(x_{ij})= - {\rm{ln}z}. $$
If we expand around $x^{(0)}_{ij}$, supposing that $\sum_{ij}(x_{ij}-x^{(0)}_{ij})^2 = O (n^{\frac{2}{3}} ) $, we obtain
\begin{equation}\label{1*}
\sum_{ij} f(x^{(0)}_{ij})+\sum_{ij} f^{'}(x^{(0)}_{ij})(x_{ij}-x^{(0)}_{ij}) 
+\sum_{ij} {\frac{1}{2}} f^{''}(x^{(0)}_{ij})(x_{ij}-x^{(0)}_{ij})^2+O(\varepsilon_n)=\sum_{ij} f(x^{(0)}_{ij}).
\end{equation}
We can rewrite the previous equation as
\begin{eqnarray}\label{2*}
\sum_{ij} a_{ij}(x_{ij} - \Delta_{ij})^2 = C,
\end{eqnarray}
where we are letting $a_{ij}={\frac{1}{2}} f^{''} (x_{ij}), \, \Delta_{ij} =- {\frac{f^{'}(x^{(0)}_{ij})} {f^{''}(x^{(0)}_{ij})}} + x^{(0)}_{ij} $ and $C={\frac{1}{2}}{\frac{f^{'}(x^{(0)}_{ij})^2}{f^{''}(x^{(0)}_{ij})}}$.
Denote by $\sum^{'}_{ij}$ the sum over all $\{ij\}$ such that $a_{ij}>0$ and by $\sum^{''}_{ij}$ the rest of the sum. Then we can rewrite (\ref{2*}) as
\begin{eqnarray}\label{3*}
\sum^{'}_{ij} {a_{ij}(x_{ij}-\Delta_{ij})^2}=C+\sum^{''}_{ij} {a_{ij}(x_{ij}-\Delta_{ij})^2}.
\end{eqnarray}
Consider a hypersphere $\sum_{ij} x^2_{ij} = \sum_{ij} (x^{(0)}_{ij})^2 = r^2_0 $. Rewrite the equation of this hypersphere as
\begin{eqnarray}\label{4*}
\sum^{'}_{ij} {x^2_{ij}} = r^2_0 - \sum^{''}_{ij} {x^2_{ij}} .
\end{eqnarray}
As can be easily observed, the hypersurfaces satisfying the equations (\ref{3*}) and (\ref{4*}) intersect at the point $\{x^{(0)}_{ij}\}$.
We would like to find out the upper bound on how the distance between these surfaces increases as the distance $\lambda$ from the distance of intersection increases.
In the equation (\ref{3*}), consider the set $S_c$ of $x_{ij}$ such that $\sum^{''}_{ij} a_{ij}(x_{ij}-\Delta_{ij})^2=\rm{const}$ . Then on this set, (\ref{3*}) is  an equation of a hypersphere.

Consider $\{x_{ij}\}$ in the set $S_c$. Then on this set we can consider a crossection passing through the point $\{x^{(0)}_{ij} \}$, the point of intersection of the ellipse and the sphere.
First suppose that the crossection is in the 2 dimensional plane. It is easy to derive in two dimensions that the distance between the ellipse and the sphere within distance $x$ from the intersection along the sphere is less than $c_1 x + c_2 x^2$. Now suppose that the crossection is the 3 dimensional hyperplane. Then we can parametrize this hyperplane by 2D planes all passing through 
$\{ x^{(0)}_{ij} \}$, and in each 2D crossection, we have that the distance 
 between the ellipse and the sphere is less than $c_1 x +c_2 x^2$, therefore
 the distance between the hyperellipse and the hypersphere in the 3D 
 crossection is less than $c_1 x +c_2 x^2$. 
 Similarly, consider a crossection by a 4D hyperplane passing through 
 $\{x^{(0)}_{ij} \}$. It can be parametrized 
 by the crossection by 3D hyperplanes passing through $\{x^{(0)}_{ij} \}$ in
 such a way that the distance in the 4D crossection is the minimum 
 of the distances in the 3D crossections. Therefore we derive that in any 
 4D crossection the distance between the hyperellipse and hypersphere 
 increases at most by  $c_1 x + c_2 x^2$ with the distance $x$ along 
 the sphere from the point of intersection.
Proceeding by induction, we get that the distance in any $N$ dimensional crossection of
 hyperellipse and a hypersphere increases at most as fast as  $c_1 x +c_2 x^2$ with 
the distance $x$ along the sphere from the point of intersection. 
Therefore the distance between the hypersphere and the hyperellipse increases at most as 
fast as $c_1 x +c_2 x^2$, if $x$ is the distance along the hypersphere from the point 
of intersection $\{x^{(0)}_{ij} \}$. 

Consider the point $\{x^{(0)}_{ij}\}$ and all the points distance $O(n^{\frac{2}{3}})$ from this point. 
Only on a set of $\{x^{(0)}_{ij}\}$ of probability measure $\varepsilon_n$ 
it is possible to have that at all the points distance $O(n^{\frac{2}{3} })$, 
the distance between the surface $\sum_{ij}{f(x_{ij})} = - {\rm{ln}} z$ 
and the surface 
$\sum_{ij}  {x^{2}_{ij}}=r^2 $ is greater than $O(K)$ 
for some large $K$. 
(That is because by the Central Limit Theorem, for the points on the surface $\sum_{i,j} f(x_{i,j}) = z$ in the set of probability $1 - \varepsilon_n$, $\sum_{i,j} x_{i,j}^2 = \frac{n(n+1)}{8} \pm Kn$. Therefore over the distance along the sphere of $O(n^{\frac{2}{3} - \delta})$ the distance between the surface and the sphere increases by $O(K)$.)

For a set of $x^{(0)}_{ij}$ of measure 
$ 1- {\frac{1}{2}} \varepsilon_n $ we have that the distance between 
the surfaces $\sum_{ij} f(x_{ij})= - {\rm{ln}}z$ and $\sum_{ij} {x^{2}_{ij}}=r^2$,
intersecting at $\{x^{(0)}_{ij} \}$, at a distance $O(n^{\frac{2}{3} } )$ 
from the point of intersection is with probability 
$1 - {\frac{1}{2}} \varepsilon_n$ less than some large constant $K$. 
Let the distance between  $\sum_{ij} { f(x_{ij})} = \sum_{ij}{ f(x^{(0)}_{ij})}$ and 
$ \sum_{ij} {x^2_{ij}} = \sum_{ij} {(x^{(0)}_{ij})^2}$ at the distance $O({n^{\frac{2}{3}}})$ from $x^{(0)}_{ij}$ in the direction ${\vec{ \rho}}$ be denoted by $d({\vec{ \rho}})$. 
Consider the set $S_{\sqrt{\varepsilon_n}} = \{(x^{(0)}_{ij},{\vec{ \rho}})$ such that $d({\bf {\vec{ \rho}}}) > K \}$. Then by the Central Limit Theorem probability of $S_{\sqrt{\varepsilon_n}}$ is less that $\sqrt{\varepsilon_n}$.
Indeed, since
\begin{eqnarray} 
P(d(\vec{ \rho}) > K)\,=\,
\int{P(d({\vec{ \rho}})} > K|x^{(0)}_{ij},{\vec{ \rho}})\, f(x^{(0)}_{ij},{\vec{ \rho}}) \, \prod_{ij}{dx_{ij} d{\vec{ \rho}}} \leq{\varepsilon_n},
\end{eqnarray}
we obtain that the set of $(x^{(0)}_{ij},{\vec{ \rho}})$ such that 
$P(d({\vec{ \rho}}) > K|x^{(0)}_{ij},{\vec{ \rho}}) > 
\sqrt{\varepsilon_n}$ has probability less than $\sqrt{\varepsilon_n}$.
Therefore the total probability 
 of a set of $x^{(0)}_{ij}$, such 
that given the $x^{(0)}_{ij}$, we have that 
on a set of ${\vec{ \rho}}$ of measure at least ${\varepsilon_n}^\frac{1}{4}$, 
the distance $d({\vec{ \rho}})$ is $>$ than $K$, must be less than ${\varepsilon_n}^{1/4}$.
We have that for given the $x^{(0)}_{ij}$ (outside of the set of probability $\varepsilon^{1/4}_n$) in the set of ${\vec{ \rho}}$ of measure $\geq {1-{\varepsilon_n}^{1/4}}$, $d({\vec{ \rho}})\leq K $.
This implies that  
 we can choose a particular direction ${\vec{ \rho}}$ from the directions in the set of measure
 ${1-{\varepsilon_n}^{1/4}}$ such that $d({\vec{ \rho}}) \leq K $.
By scaling arguments discussed above, we have that for the same $x_{ij}$ and ${\vec{ \rho}}$, at distance $\lambda$ from $x^{(0)}_{ij}$, the distance between the surface
$$ \sum_{ij}\, f(x_{ij} ) = \sum_{ij}\, f(x^{(0)}_{ij} )\;\;\;{\rm and} \;\;\; 
\sum_{ij}\, x^2_{ij} = \sum_{ij}\, (x^{(0)}_{ij} )^2  $$
is less than $\frac{ \lambda K} {{n}^{\frac{2}{3}}} $.
\vskip 0.01in
\noindent {Lemma is proved.}
\vskip 0.1in
\noindent {\bf Lemma~9:}
\vskip 0.01 in
\noindent{\it Consider the cylinders $C_k$ such that in each cylinder as $x_i$ shifts  by $\lambda$, where $|\lambda| \leq \frac{N}{\sqrt n}$, each $x_j$ shifts by $\lambda\left(1 - \frac{|i-j|}{N}\right)$ for $|j-i| \leq N$,  and stays the same for $|j-i| > N$. 
In these cylinders
$$\frac{{\rm Vol}(\frac{b \sqrt{{\rm ln} n}}{\sqrt n} \leq x_i \leq \frac{c \sqrt{{\rm ln} n}}{\sqrt n}, V_K, C_k)}{{\rm Vol}(V_K, C_k)} = \frac{{\rm Vol}(\frac{b \sqrt{{\rm ln} n}}{\sqrt n} \leq x_i \leq \frac{c \sqrt{{\rm ln} n}}{\sqrt n}, S_K, C_k)}{{\rm Vol}(S_K, C_k)}(1+\varepsilon_n).
$$
}
\vskip 0.1in
\noindent {\bf Proof:}
\hskip 0.05in
In such a cylinder $C_k$ distance to a point on the other side of $\left[\prod_{jk} g(x_{jk}) = z \cap C_k \right]$ is
$$
\Delta x = \sqrt{\sum_j (x_j - x_j^{(0)})^2 } = \lambda \sqrt N \leq \frac{N^{\frac{3}{2}}}{\sqrt n}.
$$
By Lemma~8c, the distance between the sphere and the surface $\prod_{ij} g(x_{ij}) = z$ intersecting on one side of the cylinder is less than $\frac{\Delta x}{n^{\frac{2}{3} - \delta}} \leq \frac{N^{\frac{3}{2}}}{n^{\frac{7}{6} - \delta}}.$

The ratio of the volume of the intersection of a small cone with $\prod_{jk} g(x_{jk}) =z$ and with $\sum_{jk} x_{jk}^2 = r^2$ is 
$$
\left(\frac{r'}{r}\right)^{\frac{n(n+1)}{2}} = 1 + O\left(\frac{N^{\frac{3}{2}}}{n^{\frac{1}{6} - \delta}}\right).
$$
This $1 + \varepsilon_n$ for all $N << n^{\frac{1}{9} - \frac{2}{3} \delta}$.
Therefore for each small cone the volume of the intersection with $\prod_{jk} g(x_{jk}) = z$ is the same as the volume of the intersection with the surface $\sum_{jk} x_{jk}^2 = r^2$. 
Lemma~9 is  proved.

\vskip 0.1in
\noindent{\bf Lemma~10:}
\vskip 0.01 in
{\it
\begin{eqnarray}\label{10a3}
&&  \frac{{\rm Vol}(\frac{b~ \sqrt{{\rm ln} n}}{\sqrt n} \frac{r'}{r} \leq x_i \leq \frac{c~ \sqrt{{\rm ln} n}}{\sqrt n} \frac{r'}{r}, \prod_{ij} g(x_{ij}) = z', V_m, C_m)}{{\rm Vol}(\prod_{i,j} g(x_{ij}) = z', V_m, C_m)}\cr && =\frac{ {\rm Vol} (\frac{b \sqrt{{\rm ln} n}}{\sqrt n} \leq x_i \leq \frac{c \sqrt{{\rm ln} n}}{\sqrt n}, \prod_{ij} g(x_{ij}) = z, V_m, C_m)}{{\rm Vol}(\prod_{i,j} g(x_{ij}) = z, V_m, C_m)} (1 + \varepsilon_n),
\end{eqnarray}
and also
\begin{eqnarray}\label{10a1}
&& \frac{{\rm Vol} (\frac{b \sqrt{{\rm ln} n}}{\sqrt n} \leq x_i \leq \frac{c \sqrt{{\rm ln} n}}{\sqrt n}, \prod_{ij} g(x_{ij}) = z, V_m, C_m)}{{\rm Vol} (\prod_{ij} g(x_{ij}) = z, V_m, C_m)} \cr 
&& = \frac{{\rm Vol} (\frac{b \sqrt{{\rm ln} n}}{\sqrt n} \leq x_i \leq \frac{c \sqrt{{\rm ln}n}}{\sqrt n}, V_m, C_m)}{{\rm Vol} (V_m, C_m)} (1 + \varepsilon_n).\qquad 
\end{eqnarray}
}
\vskip 0.1in
\noindent{\bf Proof:}
\hskip 0.1in
Consider a cone $C_k$ and in this cone choose a cross-section $C_k \cap \{\lambda = {\rm const} \}$ perpendicular to the long direction of $C_k \cap S_r$.
Also consider two surfaces $\prod_{ij} g(x_{ij}) = z$ and $\prod_{ij} g(x_{ij}) = z'$ in the same perpendicular cross-section. Let $x_{ij}^{(0)}$ be a point in the center of the surface $\prod_{ij} g(x_{ij}) = z$ in this cross-section ($x_{ij}^{(0)}$ is chosen later). Then $x_{ij}^{(0)} \frac{r'}{r}$ is the corresponding point on the same radial line on the surface $\prod_{ij} g(x'_{ij}) = z'$. Let $x_{ij}$ be a point on the surface $\prod_{ij} g(x_{ij}) = z$.
Now perform a Taylor expansion of  $\sum_{i,j} f\left(x_{ij} \frac{r'}{r}\right)$ around the point $x_{ij}$ and then of the remainder terms around the point $x_{ij}^{(0)}$. (Note that the point $x_{ij} \frac{r'}{r}$ is not necessarily on $\prod g(x_{ij}) = z'.$)
By doing so we obtain
\begin{eqnarray}\label{lmm10prf}
&&\sum_{i,j} f\left(x_{ij} \frac{r'}{r}\right) = \sum_{i,j} f(x_{ij}) + \sum_{i,j} f'(x_{ij}) x_{ij}\left(\frac{r'}{r} -1\right) + \sum_{i,j} f''(x_{ij}) (x_{ij})^2 \left(\frac{r'}{r} - 1\right)^2  \cr  &+& O(\frac{1}{n})  = \sum_{i,j} f(x_{ij}) + \sum_{i,j} f'(x_{ij}^{(0)}) x_{ij}^{(0)} \left(\frac{r'}{r} - 1\right) + \sum_{i,j} f''(x_{ij}^{(0)}) (x_{ij}^{(0)})^2 \left(\frac{r'}{r} -1\right)^2 \cr &+& \sum_{i,j} f''(x_{ij}^{(0)}) x_{ij}^{(0)} (x_{ij} - x_{ij}^{(0)}) \left(\frac{r'}{r} -1\right) + \sum_{i,j} f'(x_{ij}^{(0)}) (x_{ij} - x_{ij}^{(0)}) \left(\frac{r'}{r} -1\right) +O(\frac{1}{n}). \qquad\quad
\end{eqnarray}
Considering that $x_{ij}^{(0)}$ is fixed, on the subset of measure $(1- \varepsilon_n)P(\sum_{i,j} f(x_{ij}) = z \cap C_k \cap \{ \lambda = {\rm const}\})$ of the set of $x_{ij}$ such that $ \{ \sum_{i,j} f(x_{ij}) = z \cap C_k \cap \{ \lambda = {\rm const} \} \} $ we have that by the Central Limit Theorem 
$$
\sum_{i, j} f''(x_{ij}^{(0)}) x_{ij}^{(0)} (x_{ij} - x_{ij}^{(0)}) = O(\sqrt {\sum_{i,j} (f''(x_{ij}^{(0)}) x_{ij}^{(0)} (x_{ij} - x_{ij}^{(0)}))^2}) = O\left(\frac{\delta  {\rm ln} n \,{n^\frac{1}{2}}}{n^{\varepsilon}}\right).
$$
Similarly 
\begin{eqnarray} \nonumber
&&\sum_{i,j} f'(x_{ij}^{(0)}) (x_{ij} - x_{ij}^{(0)})
 \cr
&=& O(\sqrt { \sum_{i,j} (f'(x_{ij}^{(0)}) (x_{ij} - x_{ij}^{(0)}))^2 }) = O\left(\frac{\delta {\rm ln} n  \,n^\frac{1}{2}}{n^{\varepsilon}}\right).
\end{eqnarray}

The idea of the proof is the following:
Consider a particular cone $C_k$ and a point $x_{ij}^{(0)} \in C_k \cap S_K$. 
In a small neighborhood around this point the surface $\sum_{ij} f(x_{ij}) = z$ is approximately a plane. The $\frac{n(n-1)}{2} - 1$ projections $(x_{ij} - x_{ij}^{(0)})^P$ of the coordinates $x_{ij} - x_{ij}^{(0)}$ onto this plane (omitting $x_{nn} - x_{nn}^{(0)}$) are independent random variables. 
Since in the cross section ${\lambda = {\rm const}}$ each $|\Delta x_j| \leq \frac{\delta}{n^{\frac{1}{2} + \varepsilon}}$, then 
$\sqrt{\sum_{i,j} (\Delta x_{ij})^2}
 = \sqrt{\sum_j (\Delta x_j)^2}
\leq \,\frac{\delta}{n^ \varepsilon}.$

Consider a probability measure conditioned on the set $\{ \sum_{i,j} f(x_{ij}) = z \cap C_k \cap \{\lambda ={\rm const}\}\}$. The set $C_k \cap \{ \lambda = {\rm const} \}$ is a rectangular set, and therefore $x_{ij} - x_{ij}^{(0)}$ (for all $i,j$ except for $x_{n-1, n} - x_{n-1,n}^{(0)}$) are independent on this set and therefore are independent with respect to the probability measure conditioned on this set. By the arguments above we obtain that $x_{ij} - x_{ij}^{(0)}$ (for all $i,j$ except for $x_{n-1,n} - x_{n-1,n}^{(0)}$ and $x_{n,n} - x_{n,n}^{(0)}$) are independent on the set $\{ \sum_{i,j} f(x_{ij}) = z \cap C_k \cap \{ \lambda = {\rm const} \} \}$ and therefore are independent with respect to the probability measure conditioned on this set.

Choose $x_{ij}^{(0)}$ so that ${\rm E}_{{\rm conditional}}(x_{ij} - x_{ij}^{(0)}) = 0$ (where the expectation $E_{{\rm conditional}}$ is taken with respect to the conditional measure on the set $\{ \sum_{i,j} f(x_{ij}) = z \} \cap C_k \cap \{ \lambda = {\rm const} \} $). 
Then by the Central Limit Theorem we obtain:
$$
\sum_{ij} f'(x_{ij}^{(0)}) (x_{ij} - x_{ij}^{(0)})^P = \sqrt{\sum_{ij} {\rm Var} 
(f'(x_{ij}^{(0)}) (x_{ij} - x_{ij}^{(0)})^P)}\  \phi,
$$
where $\phi$ is a standard normal random variable.

Since by Lemma 11, $f'(x_{ij}) \leq O({\rm ln} n)$ we obtain $$\sqrt{\sum_{i,j} {\rm Var}(f'(x_{ij}^{(0)}) (x_{ij} - x_{ij}^{(0)})^P)} \leq O({\rm ln} n) \sqrt{ \sum_{i,j} {\rm E} (x_{ij} - x_{ij}^{(0)})^2 } \leq \frac{O( \delta{\rm ln} n)}{n^{\varepsilon}}.$$ 
(The estimate of $\sum_{ij} f''(x_{ij}^{(0)})x_{ij}^{(0)} (x_{ij} - x_{ij}^{(0)}) = O(\frac{{\delta \rm ln} n}{n^{\varepsilon}})$ is proved similarly.) Also the cases when $C_k \cap \{\lambda = {\rm const} \}$ is a rotated rectangular set or a parallelepiped give similar results.
Indeed, for a $C_k \cap \{\lambda = {\rm const} \}$ a rotated parallelepiped, consider the coordinates ${\tilde x}_{ij} = f'(x_{ij}^{(0)})x_{ij}^{(0)} (x_{ij} - x_{ij}^{(0)})$, in these coordinates $C_k \cap \{ \lambda = {\rm const} \}$ is still a parallelepiped. Then consider the rotated coordinates $z_{ij}$, parallel to the axis of the parallelepiped $C_k \cap \{ \lambda = {\rm const} \}$. The coordinates $z_{ij}$ (exclluding $z_{nn}$) are independent on the parallelepiped, and $E_{{\rm conditional}} z_{ij} = 0$, so all the arguments above apply to $z_{ij}$, and we obtain that on the set of probability $(1 - \varepsilon_n)P(C_k \cap \{\sum_{i,j} f(x_{ij}) = z\} \cap \{ \lambda = {\rm const} \})$ we have that
$$
\sum_{i,j} z_{ij} = \sqrt {\sum_{i,j} {\rm Var} z_{ij}} \phi. 
$$
Since $\sqrt {\sum_{i,j} {\rm Var} z_{ij}} \leq {\rm const} \sqrt{\sum_{i,j} z_{ij}^2}$ and since 
$\sum_{i, j} z_{ij}^2 = \sum_{i,j} {\tilde x}_{ij}^2$, we obtain that $$
\sum_{i,j} z_{ij} = O\left(\frac{\delta {\rm ln} n}{n^{\varepsilon}}\right).$$
Denoting  for simplicity   $X_k = {\tilde x}_{ij}$ and $Z_k = z_{ij}$, where $k = 1, \ldots {\frac{n(n+1)}{2}}$, we obtain 
$$ X_i = \sum_{j} \alpha_{ij} Z_j,$$
for the rotation matrix $(\alpha_{ij})$: $\alpha^{\tau} \alpha = \alpha \alpha^{\tau} = I$. 
Therefore
$$
\sum_i X_i = \sum_j Z_j \sum_i \alpha_{ij} = \sqrt{\sum_j {\rm Var} Z_j^2 (\sum_i \alpha_{ij})^2 } \phi \leq n~ {\rm max}Z_j~ \phi.
$$
In the cross-section $\{\lambda = {\rm const}\}$ each $|\Delta z_{ij} |\leq \frac{\delta}{n^{\frac{1}{2} + \varepsilon}}$.
Thus we obtain that on the subset of probability $(1 - \varepsilon_n) P(\{\sum_{ij} f(x_{ij}) = z\} \cap C_k \cap \{ \lambda = {\rm const} \})$ of the set $\{\sum_{i,j} f(x_{ij}) = z\} \cap C_k \cap \{ \lambda = {\rm const} \}$
$$
|\sum_{i,j} f'(x_{ij}^{(0)}) x_{ij}^{(0)}(x_{ij}-x_{ij}^{(0)})| = |\sum_{i,j} {\tilde x}_{ij}| \leq {\rm const}~ n ~{\rm max} |\Delta z_{ij}| \leq \delta ~{\rm ln} n~ n^{\frac{1}{2} - \varepsilon} .
$$
Now, taking into account that $\frac{r'}{r} - 1 = O\left(\frac{1}{n}\right)$,
we can estimate that
\begin{eqnarray}
\sum_{i,j} f''(x_{ij}^{(0)}) x_{ij}^{(0)} (x_{ij} - x_{ij}^{(0)}) \left(\frac{r'}{r} -1\right) + \sum_{i,j} f'(x_{ij}^{(0)}) (x_{ij} - x_{ij}^{(0)}) \left(\frac{r'}{r} -1\right) = \frac{O(\delta {\rm ln} n)}{n^{\frac{1}{2} + \varepsilon}},\nonumber
\end{eqnarray}
Therefore using that $$\sum_{i,j} f(x_{ij}^{(0)}) = \sum_{i,j} f(x_{ij})$$ and  
\begin{eqnarray}\label{lmm10prf2}
&& \sum_{i,j} f(x_{ij}^{(0)}) + \sum_{i,j} f'(x_{ij}^{(0)}) x_{ij}^{(0)}\left(\frac{r'}{r} -1\right) + \sum_{i,j} f''(x_{ij}^{(0)}) (x_{ij}^{(0)})^2 \left(\frac{r'}{r} - 1\right)^2 + O(\frac{1}{n})\cr && = \sum_{i,j} f\left(x_{ij}^{(0)}\frac{r'}{r}\right) = \sum_{i,j} f(x_{ij}'),
\end{eqnarray}
 and taking into consideration that the  remainder term, $O(\frac{1}{n})$, is the same in both terms up to $O(\frac{\delta {\rm ln} n}{n^{1 + \varepsilon}})$, we come to a more precise estimate
$$
\sum_{ij} f\left(x_{ij} \frac{r'}{r}\right) - f(x_{ij}') = \frac{O(\delta {\rm ln} n)}{n^{\frac{1}{2}+ \varepsilon}}.
$$

If $R$ and $R'$ are radial distance to the points $x_{ij} \frac{r'}{r}$ and to the point on the same radial line on the surface $f(x_{ij}') = {\rm ln} z'$, we obtain that
$$
\sum_{i,j} f\left(x_{ij}\frac{r'}{r}\right) =
\sum_{i,j} f\left(x_{ij}\frac{r'}{r}\,\frac{R'}{R}\right)  + \sum_{i,j} f'\left(x_{ij} \frac{r'}{r}\right) x_{ij} \frac{r'}{r} \left(\frac{R'}{R} -1\right)+{\rm l.o.t.}  
$$
Therefore
$$
\sum_{i,j} f'\left(x_{ij} \frac{r'}{r}\right) x_{ij} \frac{r'}{r} \left(\frac{R'}{R} -1\right) = \frac{O(\delta {\rm ln} n)}{n^{\frac{1}{2}+\varepsilon}}.
$$
Since by the Central Limit Theorem
$$
\sum_{i,j} f'\left(x_{ij} \frac{r'}{r}\right) x_{ij} \frac{r'}{r} = O(n^2),
$$
we obtain that on the set of probability $(1 - \varepsilon_n)P(C_k \cap \{\lambda = {\rm const}\})$
$$ 
\frac{R'}{R} = 1 + O\left(\frac{\delta {\rm ln} n}{n^{\frac{5}{2} + \varepsilon}}\right).
$$

Therefore we obtain that the distance between the point $x_{ij} \frac{r'}{r}$ and the point $x_{ij}\frac{r'}{r} \frac{R'}{R}$ on the same radial line on the surface $\sum_{ij} f(x'_{ij}) = z'$ is $$x_{ij} \frac{r'}{r} \left(\frac{R'}{R} -1\right)$$ in the $x_{ij}$ direction. Since typical $x_{ij} - x_{ij}^{(0)}$ is of the order on $\frac{1}{n^{1 + \varepsilon}}$ in any direction perpendicular to the long  direction of the cylinder, the volume of the $\prod_{ij} g(x_{ij}) = z \cap C_k \cap \{\lambda = {\rm const}\}$ is the same for different $z$ up to a factor $\left(\frac{r'}{r}\right)^{\frac{n(n+1)}{2}}(1 + O(\frac{1}{n^{\frac{1}{2}+\varepsilon}}))$ (on a set of probability $(1- \varepsilon_n)P(C_k \cap \{\lambda = {\rm const} \} )$:
$$
\frac{{\rm Vol}(\prod_{i,j} g(x'_{ij}) = z', C_k, \{\lambda = {\rm const}\} )}{{\rm Vol}(\prod_{i,j} g(x_{ij}) = z, C_k, \{ \lambda = {\rm const} \} )} = \left(\frac{r'}{r}\right)^{\frac{n(n+1)}{2}} \left(1 + O\left(\frac{1}{n}\right)\right).
$$

 Therefore on a set of probability $(1 - \varepsilon_n) P(C_k)$:
$$
\frac{{\rm Vol}(C_k \cap \prod_{ij} g(x_{ij}') = z' \cap \{\lambda = {\rm const} \})}{{\rm Vol}(C_k \cap \prod_{ij} g(x_{ij}) = z \cap \{ \lambda = {\rm const} \})} = \frac{{\rm Vol}(C_k \cap \prod_{ij} g(x'_{ij}) = z')}{{\rm Vol}(C_k \cap \prod_{ij} g(x_{ij}) = z)}.
$$

If we take the cones $K_v$ to be so narrow that $K_v \cap \sum_{ij} x_{ij}^2 = \frac{n(n+1)}{8}$ contains either no points such that $\frac{b \sqrt{{\rm ln} n}}{\sqrt n} \leq x_i \leq \frac{c \sqrt{{\rm ln} n}}{\sqrt n}$ or contains only the points such that $\frac{b \sqrt{{\rm ln} n}}{\sqrt n} \leq x_i \leq \frac{c \sqrt{{\rm ln} n}}{\sqrt n}$, then
$$
\frac{{\rm Vol}(\prod_{i,j} g(x'_{ij}) = z', K_v)}{{\rm Vol}(\prod_{i,j} g(x_{ij}) = z, K_v)} = \left(\frac{r'}{r}\right)^{\frac{n(n+1)}{2}} \left( 1 +O\left( \frac{1}{n}\right)\right).
$$
Now, $r'$ and $r$ were considered for a particular value of $\lambda$. They can potentially depend on the value of $\lambda$ we chose. 
To estimate the ratio 
$
\frac{{\rm Vol}(\frac{b{\rm ln} n}{\sqrt n} \leq x_i \leq \frac{c{\rm ln} n}{\sqrt n}, C_m)}{{\rm Vol}(C_m)}$
we need to show that $\left(\frac{r'}{r}\right)^{\frac {n\,(n+1)}{2}}$ stays the same for different $\lambda$.

We return to the argument in the beginning of the lemma, but now we choose $x_{ij}^{(0)}$ at  $\lambda=0 $, and take the line
$ x'_{j} = x_{j} + \lambda\,\left( 1 -\frac{|j'-i|}{N}\right), $
and we take also a point  $x_{ij}$   on this line with $x_j, \xi_{i,j}$ fixed 
and only $\lambda$ changing. This is the ``long'' direction of the cone. Now we write 
the formula ({\ref{lmm10prf}}) again for this $x{^{(0)}_{ij}}', x^{(0)}_{ij}, x_{ij}$ in this new cross-section.
\begin{eqnarray}\label{lmm10prfad1}
&&\sum_{i,j} f\left(x_{ij} \frac{r'}{r}\right) = \sum f(x_{ij}) + \sum  f'(x_{ij}^{(0)}) x_{ij}^{(0)}\left(\frac{r'}{r} -1\right) + \sum_{i,j} f''(x_{ij}^{(0)}) (x_{ij}^{(0)})^2 \left(\frac{r'}{r} - 1\right)^2  \cr  &+& 
\sum_{i,j} f''(x_{ij} ^{(0)}) x_{ij}^{(0)} (x_{ij} - x_{ij}^{(0)}) \left(\frac{r'}{r} -1\right) + \sum_{i,j} f'(x_{ij}^{(0)}) (x_{ij} - x_{ij}^{(0)}) \left(\frac{r'}{r} -1\right) +O(\frac{1}{n}). \quad\quad
\end{eqnarray}
We estimate as before
$$\sum_{i,j} f''(x_{ij}^{(0)}) x_{ij}^{(0)} (x_{ij} - x_{ij}^{(0)}) = O \left(\sqrt{
\sum \left( f''(x_{ij}^{(0)}) x_{ij}^{(0)} (x_{ij} - x_{ij}^{(0)})\right)^2}\right).$$

Since each $|\Delta x_j| \leq \lambda \leq \frac {N \delta}{{n}^{\frac{1}{2}+\varepsilon}}$ we obtain that each ${\Delta x_{ij}} \leq \frac {N \delta}{{n}^{\frac{1}{2}+\varepsilon}}\sum{|\xi_{ki}||\xi_{kj}|}
\leq \frac {N \delta}{{n}^{\frac{1}{2}+\varepsilon}} $.
And as above we derive that
$$\sum f''(x_{ij}^{(0)}) x_{ij}^{(0)} (x_{ij} - x_{ij}^{(0)})\left(\frac{r'}{r}-1\right) = O \left(\frac{\delta\,N\,{\rm ln}\,n}{n^{\frac{1}{2}+ \varepsilon}}\right).$$
Similarly we obtain that 
$$\sum f'(x_{ij}^{(0)}) (x_{ij} - x_{ij}^{(0)})\left(\frac{r'}{r}-1\right) = O \left(\frac{\delta\,N\,{\rm ln}\,n}{n^{\frac{1}{2}+\varepsilon}}\right).$$
We can rewrite (\ref{lmm10prf2}) for these $x_{i,j}^{(0)},\,x_{i,j}$ as (since $x_{ij}^{(0)} \frac{r'}{r}$ is the point on the surface $\sum_{ij} f(x_{ij}) = z'$) 
\begin{eqnarray}
&&\sum f(x'_{ij}) = \sum f(x_{ij}^{(0)} \frac{r'}{r}) \cr 
&=&\sum f(x_{ij}^{(0)}) + \sum f'(x_{ij}^{(0)}) x_{ij}^{(0)}\left(\frac{r'}{r} -1\right) + \sum f''(x_{ij}^{(0)}) (x_{ij}^{(0)})^2 \left(\frac{r'}{r} - 1\right)^2 + O(\frac{1}{n}) \nonumber,
\end{eqnarray}
to obtain that 
$$ \sum f(x_{ij} \frac{r'}{r})- \sum f(x'_{ij})= O\left(\frac{\delta\,N\,{\rm ln}\,n}{n^{\frac{1}{2}+\varepsilon}}\right).$$ 
As before, if $R$ and $R'$ are radial distances to the points $x_{ij}\,\frac{r'}{r}$ and the points on the same radial line on the surface
$f(x'_{ij})= {\rm ln}\,z'$ we obtain that 
\begin{eqnarray}
&&\sum f(x_{ij}\frac{r'}{r}) = \sum f(x_{ij} \frac{r'}{r}\,\frac{R'}{R})+ \sum f'(x_{ij}\frac{r'}{r} )\,{x_{ij}}\,\frac{r'}{r}\, \left(\frac{R'}{R}-1\right) \cr 
&=&\sum f(x'_{ij}) + \sum f'(x_{ij}\frac{r'}{r} )\, x_{ij} \,\frac{r'}{r}\, \left(\frac{R'}{R} -1\right). 
\end{eqnarray}
Therefore
$$\sum f'(x_{ij}\,\frac{r'}{r}) \, x_{ij} \,\frac{r'}{r} \, \left(\frac{R'}{R} -1\right) = O\left(\frac{\delta\,N\,{\rm ln}\,n}{n^{\frac{1}{2}+\varepsilon}}\right), $$ 
and as before we obtain that
$$\frac{R'}{R} \, = 1\,+\, O\left(\frac{\delta\,N\,{\rm ln}\,n}{n^{\frac{5}{2}+\varepsilon}}\right), $$ 
and  also that the radial distance between the point $x_{ij}\frac{r'}{r}$ and $ x_{ij}\frac{r'}{r}\,\frac{R'}{R}$ is  
$$\Delta\,=\, R'\,-\,R\,=\,
 O\left(\frac{\delta\,N\,{\rm ln}\,n}{n^{\frac{3}{2}+\varepsilon }}\right).$$

Let $r(\lambda)$ be the distance to $x_{ij}$ at some value of $\lambda$ and $R'$ is the distance to $x'_{ij}$ on the same radial line. Then
\begin{eqnarray}{\nonumber}
&&\frac{{\rm Vol}(\prod_{i,j} g(x_{ij}) = z', C_k, \{\lambda = {\rm const}\} )}{{\rm Vol}(\prod_{i,j} g(x_{ij}) = z, C_k, \{ \lambda = {\rm const} \} )} = \left(\frac{R'}{r(\lambda)}\right)^{\frac{n(n+1)}{2}} \left(1 + O\left(\frac{1}{n}\right)\right)
\cr &=&\left(\frac{\frac{r'}{r}\,r(\lambda)+R'-R}{r(\lambda)}\right)^{\frac{n(n+1)}{2}} \left(1 + O\left(\frac{1}{n}\right)\right) 
=\left(\frac{r'}{r}\,+ \frac{R'-R}{r(\lambda)}\right)^{\frac{n(n+1)}{2}} \left(1 + O\left(\frac{1}{n}\right)\right)\cr
&=& \left(\frac{r'}{r}+ 
O\left(\frac{N\,{\rm ln}\,n}{n^{\frac{5}{2} }}\right)^{\frac{n(n+1)}{2}}\right) \left(1 + O\left(\frac{1}{n}\right)\right) = \left(\frac{r'}{r}\right)^{\frac{n(n+1)}{2}} \left(1 + O\left(\frac{1}{n}\right)\right).
\end{eqnarray}
From this we derive
\begin{eqnarray}\label{10a2}
&&\frac{{\rm Vol}(\frac{b \sqrt{{\rm ln} n}}{\sqrt n} \frac{r'}{r} \leq x_i \leq \frac{c \sqrt{{\rm ln} n}}{\sqrt n }\frac{r'}{r}, \prod_{i,j} g(x_{ij}) = z', C_k )}
{{\rm Vol}(\prod_{i,j} g(x_{ij}) = z', C_k)}\cr && =
\frac{{\rm Vol}(\frac{b \sqrt{{\rm ln} n}}{\sqrt n} \leq x_i \leq \frac{c \sqrt{{\rm ln} n}}{\sqrt n}, \prod_{i,j} g(x_{ij}) = z, C_k)}{{\rm Vol}(\prod_{i,j} g(x_{ij}) = z, C_k)} (1 + \varepsilon_n).
\end{eqnarray}
Using the arguments similar to those in Lemma 2 we obtain
\begin{eqnarray}\label{10a4}
&&\frac{{\rm Vol}(\frac{b \sqrt{{\rm ln} n}}{\sqrt n} \leq x_i \leq \frac{c \sqrt{{\rm ln} n}}{\sqrt n}, \prod_{i,j} g(x_{ij}) = z, C_k)}{{\rm Vol}(\prod_{i,j} g(x_{ij}) = z, C_k)} \cr & =& \frac{{\rm Vol}(\frac{b \sqrt{{\rm ln} n}}{\sqrt n} \leq x_i \leq \frac{c \sqrt{{\rm ln} n}}{\sqrt n}, V_k, C_k)}{{\rm Vol}(V_k, C_k)} (1 + \varepsilon_n).\qquad
\end{eqnarray}

\vskip 0.1 in
\noindent{\bf Lemma~11.}
\vskip 0.01 in
\noindent{\it Let  $f(x)$ be such that for all $\Lambda$ large enough ${\rm max}_{[0, \Lambda]} f(x) \leq C {\rm max}_{\Lambda, \Lambda +1} f(x),$ 
and such that $|f'(x)|, |f''(x)|, |f'''(x)| \leq {\rm max}(C |f(x)|, C) $.}

{\it Then, on the set of $P_g$ measure $1 - \varepsilon_n$
$$
    {\rm max} (|f'(x)|, |f''(x_{ij})|, |f'''(x_{ij})|)\leq O({\rm ln}~ n), ~\forall~i,j = 1, \ldots, n.
$$}
\vskip 0.01in
\noindent{\bf Proof:}
\hskip 0.1in
{Let $\Lambda$ be a cutoff such that} 
$$
P( |\xi_{ij}| \leq \Lambda, i,j = 1, \ldots, n) \geq 1 - \varepsilon_n.
$$
We now restrict our consideration to $x_{ij}$ such that $|x_{ij}| \leq \Lambda$ for all $i,j$. $\Lambda$ can be chosen so that it works both for a given distribution $g$ and for a Gaussian.
Writing that $g(x) = e^{-f(x)}$, we obtain
$$
(1 - \int_{\Lambda}^{\infty} e^{-f(x)} dx)^{n^2} \geq 1 - \varepsilon_n.
$$
This implies
$$
\int_{\Lambda}^{\infty} e^{-f(x)} dx \leq \frac{\varepsilon_n}{n^2}.
$$
Therefore
$$
\frac{\varepsilon_n}{n^2} \geq \int_{\Lambda}^{\Lambda+1} e^{-f(x)} dx \geq e^{-{\rm max}_{\Lambda \leq x \leq \Lambda+1} f(x)}.
$$
Thus
$$
{\rm max}_{\Lambda \leq x \leq \Lambda +1} f(x) \leq {\rm ln}~ (\frac{n^2}{\varepsilon_n}).
$$
Assuming that ${\rm max}_{[0, \Lambda]} f(x) \leq {\rm max}_{[\Lambda, \Lambda+1]} f(x)$ and that $|f'(x)|, |f''(x)|, |f'''(x)| \leq C |f(x)| + C$ we obtain that on the set of measure $1 - \varepsilon_n$
$$
{\rm max}_{ij}|f'(x_{ij})|, |f''(x_{ij})| , |f'''(x_{ij})| \leq O({\rm ln}~  n).
$$
Thus the Lemma is proved.

\vskip 0.25in
\noindent{\bf Acknowledgements:} 
\vskip 0.01 in
\noindent{The author would like to thank P. Sarnak} for 
bringing this longstanding open problem to the author's attention, the referee 
of the first version of the manuscript submitted in October 2005 for his 
comments and for introducing the author to the paper \cite{Gu} by Gustavsson, 
Jan Wehr for his comments regarding the first version of the manuscript 
and Gustavsson's paper \cite{Gu}.


\vskip 0.25in
\begin{thebibliography}{1}
\vskip 0.1in
\bibitem{A1} L. Arnold, ``On the asymptotic distribution of the eigenvalues of
random matrices'', J. Math. Anal. Appl., {\bf 20}, 262-268 (1967).

\bibitem{A2} L. Arnold, ``On Wigner's semicircle law for the eigenvalues of
random matrices'',  Z. Wahrscheinlichkeitstheorie Verw. Gebiete, {\bf
  19}, 191-198 (1971).

\bibitem{B} Z.D. Bai, ``Convergence rate of expected spectral distributions of large
random matrices. I. Wigner matrices,'' Ann. Probab., {\bf 21}, 625-648
(1993).

\bibitem{BDJ} J.Baik, P.Deift, K.Johansson, On variance of the length of the longest increasing subsequence of random permutations, {\it J. Amer.Math.Society.}

\bibitem{BDMMZ} J. Baik, P. Deift, K. McLaughlin, P. Miller, X. Zhou, Optimal tail estimates for directed last passage site percolation with geometric random variables, arXiv:math.PR/0112162v1 16 Dec. 2001.










\bibitem{BMS} A.Boutet de Monvel and M.V. Shcherbina, ``On the norm of random
matrices'', Mat. Zametki, {\bf 57}, No. 5, 688-698 (1995).











\bibitem{BZ} E. Brezin, A. Zee, Universality of correlations between
  eigenvalues of large random matrices, {\it Nuclear Phys. B}, {\bf
  402}, 613-627 (1993).








\bibitem{CL} O. Costin and J.L. Lebowitz, ``Gaussian fluctuations in random
matrices'', Phys. Rev. Lett., {\bf 75}, No. 1, 69-72 (1995).





\bibitem{DIZ} P. Deift, A. Its, X. Zhou, A Riemann-Hilbert approach
  to asymptotic problems arising in the theory of random matrix
  models, and also in the theory of integrable statistical mechanics,
  {\rm Ann. of Math.}, {\bf 146}, 149-235 (1997).
\bibitem{BI} P. Bleher, A. Its, Semiclassical asymptotics of
  orthogonal polynomials, Rieman-Hilbert problem, and universality in
  the matrix model, {\it Ann. of Math.}, {\bf 150}, no. 1, 185-266 (1999).

\bibitem{DKMVZ} P. Deift, T. Kriecherbauer, K.T.- R. McLaughlin,
  S. Venakides, X. Zhou, Uniform asymptotics for polynomials
  orthogonal with respect to varying exponential weights and
  applications to universality questions in random matrix theory, {\it
  Comm. Pure Appl. Math.}, {\bf 52}, no. 11, 1335-1425 (1999). 











\bibitem{Fl} Feller, W.  (1971): {\it An introduction to Probability
Theory and Its Applications}, v. 2 (Sect. XVI.4).
(2nd ed., Wiley series in Probability and Mathematical Statistics).


 


\bibitem{F} P.J. Forrester, ``The spectrum edge of random matrix ensembles'',
Nucl. Phys. B, {\bf 402}, 709-728 (1994).

\bibitem{FK} Z. Fu\"redi and J.Komloz, ``The eigenvalues of random symmetric
matrices'', Combinatorica, {\bf 1}, No. 3, 233-241 (1981).
    

\bibitem{G} V.L. Girko, The spectral theory of random matrices,
Nauka, Moscow, 1988.

\bibitem{Gu} J. Gustavsson, Gaussian Fluctuations of Eigenvalues in the GUE, arXiv:math.PR/0401076 v3 10 May 2004.

\bibitem{Jo1} K. Johannson, Universality of local eigenvalues in certain Hermitian Wigner matrices.

\bibitem{Jo2} K. Johannson, Shape fluctuations and random matrices.


\bibitem{KS} N. Katz and P. Sarnak, ``Zeros of zeta function and symmetry'', Bull. of AMS, {\bf 36}, No. 1, 1-26 (1999).

\bibitem{KKPS} A.M.Khorunzhy, B.A. Khoruzhenko, L.A. Pastur, and
M.V. Shcherbina, ``The large n-limit in statistical mechanics and the
spectral theory of disordered systems,'' in: Phase Transitions and
Critical Phenomena, Vol. 15, (C.Domb and J.L. Lebowitz, eds.),
Academic Press, London (1992).

\bibitem{KKP} A.M.Khorunzhy, B.A. Khoruzhenko, and L.A. Pastur, ``Asymptotic
properties of large random matrices with independent entries'',
J. Math. Phys., {\bf 37}, 5033-5059 (1996).

\bibitem{K} A.Khorunzhy, ``On smoothed density of states for Wigner random
matrices'', Random Oper. Stochastic Equations, {\bf 5}, No. 2, 147-162
(1997).








\bibitem{M} V.A. Marchenko and L.A. Pastur, The distribution of the
eigenvalues in some ensembles of random matrices'', Mat.Sb., {\bf 72},
507-536 (1967).

\bibitem{Meh} M.L. Mehta, Random matrices, Academic Press, New York, 1991.


\bibitem{P1} L.A. Pastur, ``On the spectrum of random matrices'',
Teor.Mat.Fiz., {\bf 10}, 102-112 (1972).

\bibitem{P2} L.A. Pastur, ``The spectra of random self-adjoint operators'',
Usp. Mat. Nauk, {\bf 28}, 3-64 (1973).

\bibitem{PS} L.A. Pastur, M. Shcherbina, Universality of the local
  eigenvalue statistics for a class of unitary invariant random matrix
  ensembles, {\it J. Stat. Phys.}, {\bf 86}, 109-147 (1997).




\bibitem{Ok} A. Okunkov, Random matrices and random permutations.




 
\bibitem{Ru} A. Ruzmaikina, "Generalizing results on edge
distribution of eigenvalues of Wigner random matrices to slowly
decaying distributions of entries."

\bibitem{Ruz} A. Ruzmaikina, ``Universality of the edge distribution of eigenvalues of Wigner random matrices with polynomially decaying distributions of entries.'' 
\bibitem{RuAi} A. Ruzmaikina, M. Aizenman, Characterization of invariant measures at the leading edge for competing particle systems, accepted {\it Annals of Probability}.

\bibitem{R1} A. Ruzmaikina, Characterization of invariant measures at the leading edge for particle systems on a line.





\bibitem{SS1} Ya. Sinai and A. Soshnikov, ``Central limit theorem for traces of
large symmetric matrices with independent matrix elements'',
Bol. Soc. Brazil.Mat., {\bf 29}, No. 1, 1-24 (1998).

\bibitem{SS} Ya. Sinai and A. Soshnikov, ``A refinement of Wigner's semicircle
law in a neighborhood of the spectrum edge for random symmetric
matrices'', Funct. Anal. and Appl., {\bf 32}, No. 2 (1998).

\bibitem{S} A. Soshnikov, ``Universality at the edge of the spectrum in Wigner
random matrices'', Comm. Math. Phys. 207, No. 3, 697-733 (1999).

\bibitem{TW1} C.Tracy and H.Widom, ``On orthogonal and symplectic matrix
ensembles'', Commun. Math.Phys., {\bf 177}, 727-754 (1996).

\bibitem{TW2} C.Tracy and H.Widom, ``Level-spacing distribution and Airy
kernel'', Commun. Math. Phys., {\bf 159}, 151-174 (1994).


\bibitem{Wa} K.W. Wachter, ``The strong limits of random matrix spectra for
sample matrices of independent elements'', Ann.Probab., {\bf 6}, No.1,
1-18 (1978).


\bibitem{W1} E.Wigner, ``Characteristic vectors of bordered matrices with
infinite dimensions'', Ann.Math., {\bf 62}, 548-564 (1955).

\bibitem{W2} E.Wigner, ``On the distrbution of the roots of certain symmetric
matrices'', Ann.Math. {\bf 67}, 325-328 (1958).

\end{thebibliography}
\end{document}